\documentclass[12pt,iopams,setstack,epsfig,epsf,float]{iopart}
\usepackage{cite}
\usepackage{epsf}
\usepackage{epsfig}
\usepackage{float}
\usepackage{amssymb}

\def\ep{{\epsilon}}
\def\cg{{\cal G}}
\def\cd{{\cal D}}
\def\tg{{\tilde{\cal G}}}
\def\k{{{\bf k}}}
\def\om{{\omega}}
\def\q{{{\bf q}}}

\def\g{{\bf{g}}}

\def\q{{\bf{q}}}

\def\P{{\bf{P}}}

\def\beq{\begin{equation}}
\def\eeq{\end{equation}}
\def\beqa{\begin{eqnarray}}
\def\eeqa{\end{eqnarray}}

\def\g0{{\gamma_0}}
\def\ii{{\mbox{i}}}
\def\sgn{{\mbox{sgn}}}
\def\imp{{\mbox{\scriptsize imp}}}

\def\Im{{\mbox{Im}}}
\def\H{{\mbox{H}}}
\def\me{{\mbox{e}}}
\def\dna{{\downarrow}}
\def\upa{{\uparrow}}

\eqnobysec
\begin{document}
\bibliographystyle{plain}
\input epsf

\title[Dynamics and transport properties of Kondo Insulators]{Dynamics and transport properties 
of Kondo insulators}

\author{N S Vidhyadhiraja\dag, Victoria E Smith\dag, 
David E Logan\dag \\and H R Krishnamurthy\ddag}

\address{\dag\ University of Oxford, Physical and Theoretical Chemistry
Laboratory,\\ South Parks Rd, Oxford OX1~3QZ, UK}
\address{
\ddag\ Department of Physics, IISc, Bangalore 560~012; and 
JNCASR,\\ Jakkur, Bangalore 506~064, India}

\begin{abstract}

  A many-body theory of paramagnetic Kondo insulators is described, 
focusing specifically on single-particle dynamics, scattering rates,
d.c.\ transport and optical conductivities. This is achieved by
development of a non-perturbative local moment approach
to the symmetric periodic Anderson model within the framework
of dynamical mean-field theory. Our natural focus is the strong 
coupling, Kondo lattice regime; in particular 
the resultant `universal' scaling behaviour in terms of the
single, exponentially small 
low-energy scale characteristic of the problem. Dynamics/transport on
all relevant ($\omega, T$) scales are considered, from the gapped/activated 
behaviour characteristic of the low-temperature insulator
through to explicit connection
to single-impurity physics at high $\omega$ and/or $T$; and for optical
conductivities emphasis is given to the nature of the optical
gap, the temperature scale responsible for its destruction, and the consequent
clear distinction between indirect and direct gap scales.
Using scaling, explicit comparison is also made to experimental results for
d.c.\ transport and optical conductivites of 
$Ce_{3}Bi_{4}Pt_{3}$, $SmB_{6}$ and
$YbB_{12}$. Good agreement is found, even quantitatively; and a mutually
consistent picture of transport and optics results.

\end{abstract}


\submitto{\JPCM}

\section{Introduction.}
\label{sec:intro}

  In the field of strongly correlated electrons, lanthanide- or
actinide-based heavy electron materials constitute a longstanding
challenge to experimentalists and theorists alike [1,2]. The majority
of such systems, heavy fermions, are of course metallic;
whether they be paramagnetic or ordered, Fermi- or non-Fermi
liquids. Among them however resides a class of materials with
insulating ground states: the so-called Kondo insulators, containing
a large variety of compounds reviewed {\it e.g.\ }in [3-7], and well known
examples including $Ce_{3}Bi_{4}Pt_{3}$, $SmB_{6}$, $YbB_{12}$,
$CeRhAs$, $FeSi$. This mainly cubic class of paramagnetic
systems exhibit narrow-gap insulating/semiconducting behaviour at 
low temperatures, while their `high' temperature behaviour is
largely indistinguishable from metallic heavy fermions, amounting in 
essence to a lattice of $f$ ions that scatter conduction electrons 
independently via the Kondo effect. The insulating gap has long been argued 
(see {\it e.g.\ }[3]) to arise from hybridization between essentially localized $f$-levels 
and a broad conduction band, the essential physics involving a flat $f$-band 
crossing one conduction band such that there are exactly 
two electrons per unit cell (`half-filling'); albeit that the resultant
hybridization gap is not of course a simple one-electron entity, being strongly
renormalized by many-body interactions that reflect the localized and hence
correlated nature of the $f$-levels. As such, the Kondo insulators provide [3] a
realisation of the simplest, canonical model for understanding heavy electron
systems [1,2]: the half-filled periodic Anderson model (PAM), in which each
lattice site contains a non-degenerate, correlated $f$-level hybridizing locally
with a non-interacting conduction band, and which represents the natural lattice
generalization of the single-impurity Anderson model (AIM) [2].

In the present paper we consider the half-filled, symmetric PAM within 
the powerful framework of dynamical mean-field theory (DMFT, 
reviewed in [8-11]).  Formally exact in the large-dimensional limit, 
DMFT provides a tangible approximation
in finite dimensions, whereby electron dynamics become essentially 
local but remain wholly non-trivial [8-11].
Our basic aims here are to provide a many-body 
description of dynamical and transport properties of paramagnetic 
Kondo insulators,
specifically single-particle dynamics, dynamical conductivities and static
electrical transport; and to develop the theory to the point where quantitative
comparison with experiment can be made.

   These goals are of course easier stated than achieved, and the PAM has been 
studied  extensively within DMFT via a wide range of techniques.
Numerical methods include the numerical renormalization group (NRG) [12,13], 
quantum Monte Carlo (QMC)
[14-17] and exact diagonalization [18]; while theoretical approaches include 
perturbation theory in the interaction strength [19,20], iterated perturbation 
theory [21,22], the lattice non-crossing approximation [23,24] and the simpler 
average $t$-matrix approximation [25], large-$N$ mean-field theory [26,27], 
and the Gutzwiller approach [28,29].  NRG aside however, the above techniques 
suffer in general from well recognised limitations;  whether it be an 
inability to handle large interactions and hence recover the 
exponentially small scales that are the hallmark of strongly correlated 
systems, failure to recover Fermi liquid behaviour at low-energies, 
unrealistic confinement to the lowest energies, and so on. Within DMFT
all correlated lattice-fermion models reduce to an effective quantum impurity 
hybridizing self-consistently with the
surrounding fermionic bath [8-11], {\it i.e.\ }to an effective, self-consistent
AIM. Techniques for the latter thus underpin the former.
Motivated in part by this we have been 
developing a `local moment approach' (LMA) to quantum impurity 
models (AIMs) [30-35], the main emphasis of
which is on dynamics and transport. Intrinsically non-perturbative
and able to capture the spin-fluctuation physics characteristic of
the strongly correlated Kondo regime, the LMA encompasses all interaction
strengths $U$ and recovers simple perturbative behaviour in weak coupling [30].
Dynamics on all energy scales are handled, and
the low-energy dictates of Fermi liquid behaviour satisfied (although the
approach is not confined to Fermi liquid ground states, 
see {\it e.g.\ }[33,34]).
Results for dynamics arising from the LMA have been shown [30-32,34] to 
give very good agreement with NRG calculations; and, for static magnetic
properties, with exact results from the Bethe ansatz [35].

  In a recent paper [36] we have further developed the local moment approach
to encompass $T=0$ single-particle dynamics/spectra of the symmetric PAM.
While plain perturbative behaviour is again recovered in weak coupling, the
natural focus of [36] was on the strong coupling ({\it i.e.\ }large-$U$) 
Kondo lattice regime.
At sufficiently
low energies in the vicinity of the Fermi level, the LMA recovers correctly
the `insulating Fermi liquid' behaviour [36] that reflects adiabatic 
continuity in $U$ to the non-interacting limit of the simple 
hybridization-gap insulator [2,10]. This is 
manifest in preservation of the single-particle gap, now characterised by a 
renormalized  gap scale $\Delta_{g}$ which is reduced from its non-interacting 
counterpart by the quasiparticle weight $Z$ that embodies many-body 
interactions.
In agreement with consensus [12-15,24,28,29], strong coupling dynamics
were found [36] to be characterized by the \it single \rm low-energy 
gap scale $\Delta_{g}$; which
is exponentially small in strong coupling [36] (reflecting its dependence
on $Z$), thus leading to a clear separation between low- and high-energy
scales. In consequence, `low'-$\omega$ dynamics exhibit scaling: being dependent
solely and universally upon $\omega/\Delta_{g}$, with \it no \rm dependence 
on the `bare' high-energy parameters ($U$ etc) that enter the PAM Hamiltonian.
The simplest manifestation of scaling is that at sufficiently
low energies $\omega/\Delta_{g}$
the spectral behaviour amounts [2,10] to a quasiparticle renormalization
of the non-interacting hybridization-gap insulator; which is of course the
justification for renormalized band structure ideas. By itself however such 
quasiparticle behaviour gives rather a crude caricature of the scaling spectra, 
for it is confined to the immediate vicinity of the 
Fermi level $\omega =0$ [36]; 
beyond which and on scales on the order of $\Delta_{g}$ itself, 
non-trivial dynamics rapidly sets in,
embodied in long, slowly varying spectral tails that
reflect genuine many-body 
scattering/lifetime effects. As we show in the present paper, 
it is in fact this that dominates
both dynamics and transport properties for {\it all} temperatures.

  The paper is organised as follows.
  The model and basic underlying theory is discussed in \S2; formulated for an
arbitrary host lattice, and including relevant aspects of the LMA in general
(\S2.1) as well as the specific class of diagrams contributing to the associated
dynamical self-energies $\Sigma_{\sigma}(\omega;T)$ that we employ here in practice.
\S 3 deals briefly with some basic formal results for conduction electron 
scattering rates, electrical transport and optical conductivities. 
Here and thereafter we 
consider explicitly and together two canonical host lattices [8-11], the 
hypercubic and Bethe lattices; our aim throughout being to emphasise both the 
differences and, more importantly, similarities between these two 
representative cases. Results arising from the LMA are then presented 
systematically in \S s 4ff. Our primary emphasis is again the strong coupling 
Kondo lattice  regime of the PAM, this being both where the theoretical 
difficulties lie and the regime generally applicable to small-gap Kondo insulators. 
By the same token we focus largely, albeit not exclusively, on the scaling behaviour
of dynamics and/or transport, now depending universally on
$\tilde{\omega} = \omega/\Delta_{g}$ \it 
and \rm $ \tilde{T} = T/\Delta_{g}$. 
This is important for many reasons, not least because the lack of scale separation 
inherent to some previous approaches has we believe led to a number of misconceptions
in the literature, particularly in regard  to the scales relevant to the $T$- and 
$\omega$-dependence of the conductivity.
In \S's 4,5, and considering \it all \rm $\tilde{\omega}$ and $\tilde{T}$ scales,
single-particle dynamics, conduction electron scattering rates and d.c. transport
are considered; including explicit connection to single impurity physics at high 
frequencies and/or temperatures (`Kondo logs' etc).
Optical conductivities $\sigma(\omega;T)$ are considered in \S6, with particular 
emphasis given to the nature of the optical gap and the clear separation between 
indirect and direct gap scales in both the $\omega$-dependence and
thermal evolution of $\sigma(\omega;T)$.

  In \S 7 we turn to experiment, considering three prototypical materials for
which extensive and reliable data is available [3-7], 
{\it viz} $Ce_{3}Bi_{4}Pt_{3}$,
$SmB_{6}$ and $YbB_{12}$; our aim being direct comparison between the present
theory and experimental results for both d.c. transport and optical 
conductivities. That may be achieved in a minimalist fashion, employing directly
the scaling behaviour discussed in \S s 4-6; which requires neither multiparameter 
fits nor in general a specification of the bare model parameters. Good agreement 
between theory and experiment is found, even quantitatively; with many of the 
characteristic features arising theoretically being directly apparent in experiment, 
and a mutually consistent description of transport and optics thereby arising.
The paper concludes with a brief summary.

\section{Model and theory.}
\label{sec:model}

  In standard notation, the Hamiltonian for the PAM is given by:

\beq
\hspace{-1cm}
\hat{H}=-t\sum_{(i,j),\sigma} c^\dag_{i\sigma} c^{\phantom{\dag}}_{j\sigma}
+\sum_{i,\sigma} (\ep_f + 
\case{U}{2} f^\dag_{i\,-\sigma}f^{\phantom{\dagger}}_{i\,-\sigma})
f^\dag_{i\sigma}f^{\phantom{\dag}}_{i\sigma}
+V\sum_{i,\sigma} (f^\dag_{i\sigma} c^{\phantom{\dag}}_{i\sigma} +\mbox{h.c.})
\label{eq:model}
\eeq
The first term describes the uncorrelated conduction($c$) band with
nearest neighbour hopping $t_{ij}=t$, rescaled as $t\propto t_*/\surd Z_c$
in the large dimensional limit where the coordination number $Z_c \rightarrow
\infty$ [8-11]
(with $t_*$ the basic unit of energy).
 The second term refers to the
$f$-levels with site energies $\ep_f$ and on-site repulsion $U$; such that
$\ep_f=-\case{U}{2}$ for the particle-hole (p-h) symmetric PAM considered 
here, for which $n_f=\sum_\sigma <\! f^\dag_{i\sigma}
f^{\phantom{\dag}}_{i\sigma}\! >=1$ and $n_c=\sum_\sigma <\! c^\dag_{i\sigma}
c^{\phantom{\dag}}_{i\sigma}\! >=1$ (for all $U$) as appropriate
to the Kondo insulating state.
The final term in equation \eref{eq:model} describes $c/f$-level
hybridization via the local matrix element $V$, whence the model is 
characterized by two independent dimensionless parameters, {\it viz}
$U/t_*$ and $V/t_*$.

  Our natural focus is on local single-particle dynamics, embodied in the
retarded Green functions $G^f_{ii}(\om)\;(\leftrightarrow -\ii\theta(t) 
<\{ f_{i\sigma}(t), f^\dag_{i\sigma}\}>)$ and likewise $G^c_{ii}(\om)$
for the $c$-levels, with corresponding local spectra $D^\nu_{ii}(\om)
=-\case{1}{\pi}\Im\, G^\nu_{ii}(\om)$ (and $\nu=c$ or $f$). A knowledge
of local dynamics and their thermal evolution is in turn sufficient
within DMFT [8-11]
to determine optical and 
transport properties, as detailed in $\S$\ref{sec:transp}.

  We begin with some brief remarks on the trivial limit $V=0$, where
(equation \eref{eq:model}) the $f$-levels decouple from the free
conduction band. The latter is specified by its local density of
states $\rho_0(\ep)=-\case{1}{\pi}\Im\, g_0(\ep)$, and it proves useful in
the following to denote by $\H(z)$ the Hilbert transform
\beq
\H(z)=\int^\infty_{-\infty} d\ep \; \frac{\rho_0(\ep)}{z-\ep}
\label{eq:ht}
\eeq
for arbitrary complex $z$. The free $c$-electron (local) propagator $g_0(\om)$
is itself given simply by
\numparts
\beqa
g_0(\om)&=& \H(\om^+)  \\
&=&  \left[ \om^+ - S_0(\om) \right]^{-1}
\label{eq:s0def}
\eeqa
\endnumparts
with $\om^+=\om + \ii 0^+$ here and throughout; where (as used below) 
equation \eref{eq:s0def} defines the Feenberg self-energy 
$S_0(\om)$ [37,38],
with $S_0(\om)\equiv S[g_0]$ a functional of $g_0$ (since $g_0=\H(S+1/g_0)$).
While our subsequent discussion holds for an arbitrary conduction band
$\rho_0(\ep)$, specific results will be given in $\S$'s 4{\it ff} 
for the Bethe lattice (BL) and hypercubic lattice (HCL); for which
within DMFT the $\rho_0(\ep)$ are respectively a semi-ellipse and an 
unbounded Gaussian, given explicitly by [8-11] :
\numparts
\beqa
\rho_0(\ep)= \frac{2}{\pi t_*} \left[ 1-(\ep/t_*)^2\right]^\frac{1}{2}
\hspace{1cm} :\; |\ep| \leq t_* \hspace{1.2cm}  \mbox{BL}  \label{eq:r0bl} \\
\rho_0(\ep)= \frac{1}{\surd{\pi} t_*} \exp\left(-[\ep/t_*]^2\right)
\hspace{3.5cm}  \mbox{HCL } \label{eq:r0hc} 
\eeqa
\endnumparts

As noted in $\S$\ref{sec:intro} we are interested in the homogeneous
paramagnetic phase of the PAM, for which the $G^\nu_{ii}(\om)\equiv
G^\nu(\om)$ are site-independent. The major simplifying feature of DMFT
is that the $f$-electron self-energy is 
site-diagonal [8-11],
and from
straightforward application of Feenberg renormalized perturbation 
theory [37,38]
the $G^\nu(\om)$ are given by:
\numparts
\beqa
G^c(\om)&=& \left[ \om^+ - \frac{V^2}{\om^+ - \Sigma_f(\om;T)}
-S(\om) \right]^{-1} 
\label{eq:gc} \\
G^f(\om)&=& \left[ \om^+ - \Sigma_f(\om;T)  - \frac{V^2}{ \om^+ 
-S(\om)} \right]^{-1} 
\label{eq:gf}
\eeqa
\endnumparts
Here $\Sigma_f(\om;T)$ is the conventional single self-energy (defined
to exclude the trivial Hartree contribution which identically cancels
$\ep_f=-\case{U}{2}$), such that $\Sigma_f(\om;T)=\Sigma^R_f(\om;T)
-\ii\,\Sigma^I_f(\om;T)$ with $\Sigma^I_f(\om;T)\geq 0$ for all $(\om,T)$;
and with p-h symmetry reflected in 
\beq
\Sigma_f(\om;T) = -\left[\Sigma_f(-\om;T)\right]^*
\eeq
together with $S(\om)=-\left[S(-\om)\right]^*, G^\nu(\om)=
-\left[ G^\nu(-\om)\right]^*$ and hence $D^\nu(\om)=
D^\nu(-\om)$ for the spectra. Equation \eref{eq:gf} embodies the 
connection to an effective impurity model that is inherent to DMFT [8-11],
for it may be cast in the `single impurity' form $G^f(\om)=
\left[ \om^+ - \Sigma_f(\om;T)  - \Delta_{\mbox{\scriptsize eff}}(\om)
\right]^{-1}$;
with an effective hybridization $\Delta_{\mbox{\scriptsize eff}}(\om) =
V^2\left[ \om^+ -S(\om) \right]^{-1}$ which, in contrast to that for a 
pure Anderson impurity model (AIM) and by virtue of its dependence on
the Feenberg self-energy $S(\om)$, depends implicitly on coupling to the 
other sites in the correlated lattice and as such must thus be 
self-consistently determined. Specifically, the Feenberg self-energy 
$S(\om)\equiv S[G^c]$ is precisely the same functional of $G^c(\om)$ as
it is of $g_0(\om)$ in the $V=0$ limit ({\it e.g.\ }$S=\case{1}{4}t_*^2 G^c$ 
for the BL).  In consequence, $G^c$ is given using equation \eref{eq:gc}
by
\beq
G^c(\om)=\H(\gamma) \label{eq:gchg} 
\eeq
where
\beq
\gamma(\om)=\om^+ - \frac{V^2}{\om^+ - \Sigma_f(\om;T)}
\label{eq:gamma}
\eeq
(and we add in passing that in physical terms $\gamma_I(\om)=
\Im\,\gamma(\om)$ gives the conduction electron scattering rate which
will be considered further in $\S$\ref{sec:transp}). 

It is this that,
for an arbitrary $\rho_0(\ep)$, prescribes the conventional `single
self-energy' route to the propagators $G^\nu(\om)$: given 
$\Sigma_f(\om;T)$, and hence $\gamma(\om)$, $G^c(\om)=\H(\gamma)$ follows
directly by Hilbert transformation; $S(\om)$ follows (from equation 
\eref{eq:gc}) as 
\beq
S(\om)=\gamma - \frac{1}{\H(\gamma)}
\label{eq:som}
\eeq
and $G^f(\om)$ then follows in turn from equation \eref{eq:gf}. In practice
of course the problem must be solved iteratively and self-consistently,
because the approximate $ \Sigma_f(\om;T)$ employed will itself in general be a functional of self-consistently determined propagators. Self-consistent
second order perturbation theory in $U$ [20]
provides a direct example; as too does iterated perturbation theory 
(IPT) [21]
where $ \Sigma_f(\om;T)$ is constructed
from the second order (in $U$) skeleton diagram, employing host/medium
$f$-electron propagators $\cg(\om)=\left[(G^f(\om))^{-1} + \Sigma_f(\om;T)
\right]^{-1}$ given from equation \eref{eq:gf} by
\beq
\cg(\om)=\left[ \om^+ - \frac{V^2}{\om^+ - S(\om)}\right]^{-1}
\label{eq:hostg}
\eeq
and thus dependent on $S(\om)$ itself. Results arising from IPT will
be discussed in $\S$'s 4-6.

\subsection{Local Moment Approach.}
\label{ssec:lma}

In the conventional route to dynamics sketched above, the success of any
particular theory is naturally determined by the approximation employed 
for the single self-energy $\Sigma_f$. Therein lie well known 
difficulties [2],
notably the inability of conventional perturbation theory to handle strong
interactions in general, and to recover exponentially small scales that are
the hallmark of strongly correlated behaviour; together with the divergences
that plague perturbation theory in $U$ [2]
if one attempts to perform essentially
standard diagrammatic resummations ({\it e.g.\ }RPA) of the sort one 
intuitively expects should be required to capture the regime of strong 
electron correlations. For these reasons the LMA [30-36]
eschews an approach based
directly on the single $\Sigma_f$, and focuses instead on a two-self-energy
description  that is a natural consequence  of the mean-field approach from
which it starts.

  There are three essential elements to the 
LMA.
(i) First that $f$-electron
local  moments (`$\mu$'), viewed as the initial effect of interactions, 
are introduced explicitly from the outset. The starting point is thus broken 
symmetry static mean-field  (MF, {\it i.e} 
unrestricted Hartree-Fock); containing
two degenerate, local symmetry broken MF states,
corresponding to $\mu=\pm|\mu|$. Grossly deficient by itself (see [36]),
MF nevertheless provides a starting point for a non-perturbative many-body
approach, to which end (ii) the LMA employs the two-self-energy description
that is a natural consequence of the underlying two local saddle points. 
The associated self-energies are built diagrammatically from the underlying MF
propagators, and include in particular a non-perturbative class of diagrams
($\S$\ref{sec:se} and \fref{fig:se} below) that capture the spin-flip dynamics
essential to describe the strongly correlated regime. (iii) The final key
idea behind the LMA at $T=0$ is that of symmetry restoration [30-36]:
self-consistent
restoration of the broken symmetry endemic at pure MF level, and hence
recovery on the lowest energy scales of the Fermi liquid/quasiparticle
behaviour that reflects adiabatic continuity in $U$ to the non-interacting
limit.

As detailed in [36]
the $G^\nu(\om)$, which  are as they must be
rotationally invariant, are expressed formally as ({\it cf} equations (2.5))
\beq
G^\nu(\om)= \case{1}{2}\left[G^\nu_\upa(\om) +
G^\nu_\dna(\om)\right] 
\label{eq:sumg}
\eeq
where
\numparts
\beqa
G^c_\sigma(\om)&=& \left[\om^+ - \frac{V^2}{\om^+ - 
\tilde{\Sigma}_\sigma(\om;T)} -S(\om) \right]^{-1} 
\label{eq:gcsig} \\
G^f_\sigma(\om)&=&\left[\om^+ - \tilde{\Sigma}_\sigma(\om;T) 
- \frac{V^2}{\om^+ - S(\om)} \right]^{-1}
\label{eq:gfsig}
\eeqa
\endnumparts
(and $\sigma=\upa/\dna$ or $+/-$); and with the $f$-electron 
self-energies separated as 
\beq
\tilde{\Sigma}_\sigma(\om;T) = -\case{\sigma}{2}U|\bar{\mu}| + 
\Sigma_\sigma(\om;T) \,.
\label{eq:sepsigma}
\eeq
The first term here represents the purely static Fock bubble diagram that
alone is retained at pure MF level (with $|\bar{\mu}|$ given explicitly
by equation \eref{eq:mubar} below). The second term $\Sigma_\sigma(\om;T)
=\Sigma^R_\sigma(\om;T) - \ii \Sigma^I_\sigma(\om;T)$, is the all 
important dynamical contribution mentioned above, with p-h symmetry reflected 
in 
\beq
\Sigma_\sigma(\om;T) = -\left[\Sigma_{-\sigma}(-\om;T)\right]^*
\label{eq:sigsym}
\eeq
(such that $G^\nu_\sigma(\om)=-\left[G^\nu_{-\sigma}(-\om)\right]^*$
and hence $G^\nu(\om)=-\left[G^\nu(-\om)\right]^*$).

Equations (\ref{eq:sumg},12) are the
two-self-energy counterparts of the single self-energy equations 
(2.5). And for an arbitrary conduction band 
$\rho_0(\ep)$ and given $\{\tilde{\Sigma}_\sigma(\om;T)\}$,
they may likewise be solved straightforwardly ({\it cf} the above 
discussion of equations (\ref{eq:gchg}-9)): defining 
\beq
\tilde{\gamma}_\sigma(\om)=\om^+ - \frac{V^2}{\om^+ - 
\tilde{\Sigma}_\sigma(\om;T)}
\eeq
such that $G^c(\om)=\case{1}{2} \sum_\sigma\left[\tilde{\gamma}_\sigma
-S\right]^{-1}$ (equations (\ref{eq:sumg},12a)), and comparing
to  $G^c(\om)= \left[\gamma -S\right]^{-1}$ (equations (\ref{eq:gc},8)),
the $\tilde{\gamma}_\sigma$'s are related to the single
$\gamma(\om)$ (equation \eref{eq:gamma}) by
\beq
\gamma(\om)=\case{1}{2}\left[\tilde{\gamma}_\upa(\om) + 
\tilde{\gamma}_\dna(\om)
\right] + \frac{\left[ \frac{1}{2}\left(\tilde{\gamma}_\upa(\om) -
\tilde{\gamma}_\dna(\om)\right)\right]^2}{S(\om)-\frac{1}{2}\left[
\tilde{\gamma}_\upa(\om)+\tilde{\gamma}_\dna(\om)\right]}  \,.
\label{eq:sumgamma}
\eeq
Given $\tilde{\Sigma}_\sigma(\om;T)$ and hence $\tilde{\gamma}_\sigma(\om)$,
this equation together with $S(\om)=\gamma-1/\H(\gamma)$ (equation 
\eref{eq:som}) may be solved iteratively for $S(\om)$ (employing an initial
`startup' $S$, say $S=\frac{1}{4} t_*^2g_0(\om)$); which procedure is both
straightforward and numerically fast. And with $S(\om)$ then known the 
$G^\nu(\om)$ follow directly  from equations (\ref{eq:sumg},12).
In particular, the underlying MF propagators may be
obtained from this procedure in one shot, the static MF self-energies being 
given by $\tilde{\Sigma}_\sigma\equiv \tilde{\Sigma}^0_\sigma=-\sigma x$
with $x=\case{1}{2}U|\mu|$. For any given $x$, the MF propagators 
$g^\nu_\sigma(\om)$ and  hence spectra $d^\nu_\sigma(\om)
\equiv d^\nu_\sigma(\om;x)$ thus follow; and at pure MF level the local
moment $|\mu|$ is then found from solution of $|\mu|=|\bar{\mu}|$, where
the Fock bubble $|\bar{\mu}|\equiv |\bar{\mu}(x)|$ appearing generally in
equation \eref{eq:sepsigma} is given by
\beq
|\bar{\mu}| = \int^\infty_{-\infty} d\om\,\left[ d^f_\upa(\om) -
d^f_\dna(\om)\right] f(\om;T)
\label{eq:mubar}
\eeq
with $ f(\om;T)=\left[ \me^{\om/T} + 1 \right]^{-1}$ the Fermi function.

  The single self-energy $\Sigma_f(\om;T)$ likewise follows as a byproduct 
of the above procedure, since solution of equations \eref{eq:som}, 
\eref{eq:sumgamma} given $\{\tilde{\Sigma}_\sigma(\om;T)\}$ 
determines both $S(\om)$ {\it and} $\gamma(\om)$, whence
(see equation \eref{eq:gamma}) $\Sigma_f(\om;T)=\om^+ - V^2\left[
\om^+ - \gamma(\om)\right]^{-1}$ thus follows; which relation may be recast 
equivalently as 
\beq
\fl \hspace{1cm}
\Sigma_f(\om;T) = \case{1}{2}\left[\tilde{\Sigma}_\upa(\om;T) +
\tilde{\Sigma}_\dna(\om;T)\right] + \frac{\left[ \frac{1}{2}\left(
\tilde{\Sigma}_\upa(\om;T) - \tilde{\Sigma}_\dna(\om;T)\right)\right]^2}
{\cg^{-1}(\om) - \frac{1}{2}\left[\tilde{\Sigma}_\upa(\om;T)
+\tilde{\Sigma}_\dna(\om;T)\right]}
\label{eq:ssig}
\eeq
where $\cg(\om)=\left[\om^+ - V^2\left(\om^+ - S(\om)\right)^{-1}\right]^{-1}$
is precisely the host/medium $f$-propagator equation \eref{eq:hostg}.
The resultant  conventional single self-energy may thus be obtained
directly, given the two self-energies $\tilde{\Sigma}_\sigma(\om;T)$
 equation \eref{eq:sepsigma} (although not of course vice versa); and
 the particular class of diagrams contributing to the dynamical 
$ \Sigma_\sigma(\om;T)$ that we retain in practice are specified in 
$\S$\ref{sec:se}.

 As mentioned above and discussed further in [36],
the final, central idea
behind  the $T=0$ LMA is self consistent restoration of the broken symmetry
inherent at MF level. This is embodied  mathematically in the symmetry
restoration (SR) condition $\tilde{\Sigma}_\upa(\om=0;T=0)=
\tilde{\Sigma}_\dna(\om=0;T=0)$ at the Fermi level $\om=0$; and hence
$\tilde{\Sigma}_\sigma(0;0)=0$ (for either $\sigma$) for the p-h
symmetric PAM here considered, {\it i.e.\ }
\beq
\tilde{\Sigma}^R_\upa(0;0)= \Sigma^R_\upa(0;0) -  \case{1}{2} U|\bar{\mu}|
=0
\label{eq:sr}
\eeq
(using $\tilde{\Sigma}_\sigma(0;0)=\tilde{\Sigma}^R_\sigma(0;0)$). In
physical terms, satisfaction of SR ensures [36] that single-particle dynamics
on the lowest energy scales amount to a quasiparticle renormalization
of the non-interacting limit $U=0$, reflecting Fermi liquid behaviour
in the general sense of adiabatic continuity to that limit. For 
$U=0\;(=\Sigma_f)$ the non-interacting Green functions are denoted by
$g^\nu_0(\om;V^2)$, with spectra $d^\nu_0(\om;V^2)$ and the 
$V^2$-dependence explicit; such that (via equations (\ref{eq:ht},3,5,7,8))
$d^c_0(\om;V^2)=\rho_0(\om-
V^2/\om)$ and $d^f_0(\om;V^2) = \case{V^2}{\om^2} d^c_0(\om;V^2)$ with
$\rho_0(\ep)$ the free ($V=0$) conduction band density of states,
{\it e.g.\ }equations (2.4). For $U=0$ and all $V\neq 0$ the 
system is thus of course a simple hybridization gap insulator [39],
with a gap
$\Delta^0_g(V^2)$ that is hard for the generic case of a bounded $\rho_0(\ep)$
({\it e.g.\ }the BL, equation \eref{eq:r0bl}) and (strictly) soft 
for an unbounded 
$\rho_0(\ep)$ satisfying $\rho_0(\ep)\rightarrow 0$ as $|\ep|\rightarrow
\infty$ ({\it e.g.\ }the Gaussian characteristic of the HCL, equation
\eref{eq:r0hc}). If the SR condition equation \eref{eq:sr} is satisfied
then (see [36])
for $U>0$  the leading, 
lowest-$\om$ behaviour of the full $T=0$ $G^\nu(\om)$ follows as
\numparts
\beqa
G^c(\om)&\sim& g^c_0(\om;ZV^2) \label{eq:rngc} \\
G^f(\om)&\sim& Z\, g^f_0(\om;ZV^2) \label{eq:rngf}
\eeqa
\endnumparts
where $Z=[ 1-(\partial\Sigma^R(\om;0)/\partial\om)_{\om=0}
]^{-1}$ is the quasiparticle weight (given equivalently by [36]
$Z=[ 1-(\partial\tilde{\Sigma}^R_\sigma(\om;0)/
\partial\om)_{\om=0}]^{-1}$); resulting in preservation of
the insulating gap, $\Delta_g=\Delta^0_g(ZV^2)$, that is generically
reduced from the non-interacting hybridization gap by the quasiparticle 
weight factor $Z$, with $Z\ll 1$ in strong coupling.

Equations (2.20) embody the quasiparticle behaviour
of the PAM on the lowest energy scales, {\it i.e.\ }the `insulating Fermi liquid'
behaviour that evolves continuously from the non-interacting limit. 
Imposition of SR equation \eref{eq:sr}, as a single condition {\it at}
the Fermi level $\om=0$, naturally underlies the LMA; and amounts in practice,
as detailed in [36],
to a self-consistent determination of the
local moment
$|\mu|$ (superseding the pure MF condition $|{\mu}|= |\bar{\mu}(x)|$,
see equation \eref{eq:mubar}).

\subsection{Dynamical self-energies.}
\label{sec:se}

\begin{figure}[t]
\epsfxsize=200pt
\centering
{\mbox{\epsffile{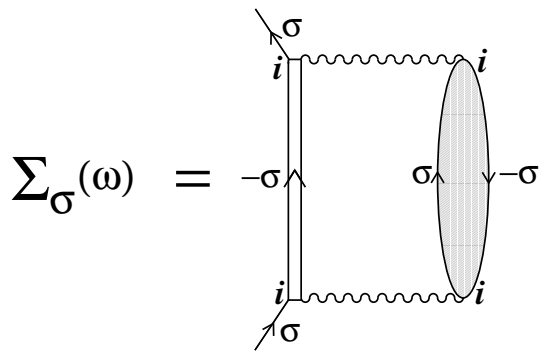}}}
\caption{Class of diagrams for the $f$-electron self-energies
$\Sigma_\sigma(\om)$ here retained in practice. The interaction
$U$ is denoted by a wavy line and the renormalized host/medium
propagator (see text) by a double line; the transverse spin
polarization propagator is shown hatched.}
\label{fig:se}
\end{figure}

   Our final task is to specify the class of diagrams retained in practice
for the dynamical $\Sigma_\sigma(\om;T)$'s (equation (2.13)).
These embody self-consistent dynamical coupling  of single-particle
excitations to low-energy transverse spin fluctuations and are precisely
as considered in [36]
for $T=0$, extended to finite-$T$ following arguments
identical to [32]
for the Anderson impurity model; the reader is thus 
referred to [32,36],
for full details. The diagrams are summarized in figure~\ref{fig:se}
where wavy lines denote the local interaction $U$, the double line 
propagator denotes the broken symmetry host/medium $f$-electron 
propagator $\tilde{\cg}_{-\sigma}(\om)$ specified below (equation 
\eref{eq:hostgspin}),
and the local $f$-level transverse spin polarization
propagator is shown as hatched; diagrammatic expansion of the resultant 
$\Sigma_\sigma$ in terms of MF propagators and dynamical self-energy 
insertions is discussed in [36,40].

   In physical terms the diagrams shown in figure~\ref{fig:se} describe
correlated
spin-flip scattering processes that are essential to capture in particular
the strong coupling Kondo lattice regime of the model: in which having say,
added, a $\sigma$-spin electron to a $-\sigma$-spin occupied $f$-level
on site $i$, the $-\sigma$-spin hops off the $f$-level thus generating
an on-site spin-flip ( embodied in the transverse spin polarization 
propagator); the $-\sigma$-spin electron then propagates through the 
lattice/host in a correlated fashion, interacting {\it fully}
with $f$-electrons on  {\it any} site $j\neq i$ (reflected in the
host/medium $\tilde{\cg}_{-\sigma}(\om)$); before returning to site $i$
at a later time whereupon the originally added $\sigma$-spin is removed 
(and which  process simultaneously restores the spin-flip on site $i$).

  The renormalized medium $f$-propagator $\tilde{\cg}_{-\sigma}(\om)$,
which embodies correlated propagation  of the $-\sigma$-spin electron through
the lattice, is given explicitly by [36]
({\it cf} equation \eref{eq:hostg}):
\beq
\tilde{\cg}_{-\sigma}(\om)=\left[ \om^+ - \case{\sigma}{2}U|\mu| -
\frac{V^2}{\om^+ - S(\om)} \right]^{-1}
\label{eq:hostgspin}
\eeq
As used below, $\tilde{\cg}_{\sigma}(\om)$ may be separated as 
$\tilde{\cg}_{\sigma}(\om)= \tilde{\cg}^+_{\sigma}(\om) +
\tilde{\cg}^-_{\sigma}(\om)$ into the one-sided retarded Hilbert 
transforms, given by
\beq
\tilde{\cg}^\pm_{\sigma}(\om)=\int^\infty_{-\infty} d\om_1\,
\frac{\tilde{\cd}_{\sigma}(\om_1)\theta(\pm\om_1)}{\om-\om_1+\ii 0^+}
\label{eq:onesht}
\eeq
with $\tilde{\cd}_{\sigma}(\om)=-\case{1}{\pi}\Im\tilde{\cg}_{\sigma}(\om)$
the corresponding spectral density (and $\theta(x)$ the unit step function).

  Specifically, the retarded LMA $\Sigma_\upa(\om;T) (=-[\Sigma_\dna
(-\om;T)]^*$ by p-h symmetry) is given by
\numparts
\beq
\fl
\hspace{0.8cm}\Sigma_\upa(\om;T)= U^2\int^\infty_{-\infty}\frac{d\om_1}{\pi}
\int^\infty_{-\infty}d\om_2\;\chi^{+-}(\om_1;T)\frac{
\tilde{\cd}_\dna(\om_2)}{\om+\om_1-\om_2 + \ii 0^+}\; h(\om_1;\om_2)
\label{eq:sigT}
\eeq
with
\beq
h(\om_1;\om_2)=\theta(\om_1)f(\om_2;T)+\theta(-\om_1)\left[1-
f(\om_2;T)\right]
\label{eq:hw12}
\eeq
\endnumparts
(which reflects the hard core boson character of the local $f$-level
spin flips [32],
whose statistics are dictated by the probability with
which fermions can hop from/to site $i$ to/from the surrounding host
lattice, as embodied in the Fermi functions). $\chi^{+-}(\om;T)\geq 0$
denotes the (local) spectral density of transverse spin excitations,
given by $\chi^{+-}(\om;T) = \sgn(\om)\:\Im\,\Pi^{+-}(\om;T)$ with 
$\Pi^{+-}(\om;T)$ the retarded, finite-$T$ polarization propagator. The 
latter is given at the simplest level, considered here, by an RPA-like
p-h ladder sum in the transverse spin channel, obtained by straightforward
analytical continuation of the imaginary time
\beq
\Pi^{+-}(\ii\,\om_m)=\, ^0\Pi^{+-}(\ii\,\om_m)\left[1-U\, ^0\Pi^{+-}(\ii\,\om_m)
\right]^{-1} \,.
\label{eq:pi_rpa}
\eeq
The bare polarization bubble diagram $^0\Pi^{+-}$
may itself be expressed either in terms
of the broken symmetry MF propagators $\{g^f_\sigma(\om;x)\}$ 
($\S$\ref{ssec:lma}), as shown explicitly in figure 1c of [36]
and with the 
resultant LMA referred to therein as LMA I; or in terms of the self-consistent
medium propagators $\{\tilde{\cg}_\sigma \}$, correspondingly referred 
to as LMA II.  In practice as shown in [36],
results for single-particle dynamics arising
from LMA I/II are very similar, and for that reason explicit results are
given in the present paper for LMA I alone. We also add that for $T=0$,
equations (2.23) reduce generally (via equation 
\eref{eq:onesht}) to
\beq
\fl
\hspace{1.0cm}\Sigma_\upa(\om;0)= U^2\int^\infty_{-\infty}\frac{d\om_1}{\pi}
\;\chi^{+-}(\om_1;0)\left[\theta(\om_1)\tg_\dna^-(\om_1+\om) +
\theta(-\om_1)\tg^+_\dna(\om_1+\om)\right]
\label{eq:sigT0}
\eeq
as employed (in time-ordered form) for $T=0$ in [36].

  The above considerations specify the LMA two-self-energies that we consider
in practice; $\tilde{\Sigma}_\sigma(\om;T)$ being given in its entirety
by equation \eref{eq:sepsigma}, $|\bar{\mu}|$ therein by equation 
\eref{eq:mubar} and the dynamical $\Sigma_\upa(\om;T)$ by equations
(2.23) ( with $\Sigma_\dna(\om;T)$ by p-h symmetry).
The problem is readily solved numerically. As explained in $\S$\ref{ssec:lma}
(following  equation \eref{eq:sumgamma}), for given $\{\tilde{\Sigma}_\sigma
\}$ equations (\ref{eq:som},16) may be solved 
straightforwardly for the Feenberg self-energy $S(\om)$ and the full 
$c/f$-electron propagators $G^\nu(\om)$. An iterative, self-consistent 
solution is naturally required, since the $\{\tilde{\Sigma}_\sigma\}$
are functionals of the renormalized medium $f$-propagators
$\{\tg_\sigma\}$, themselves given explicitly by equation \eref{eq:hostgspin}
and thus dependent on $S(\om)$. And since both $\tg_\sigma$ and the MF 
propagators $g^f_\sigma$ depend explicitly on $x=\case{1}{2}U|\mu|$, it
is numerically optimal to solve for fixed $x$, with $U$ determined,
as opposed to vice versa [36].

  The problem is first solved for $T=0$, ensuring that symmetry restoration
(equation \eref{eq:sr}) is satisfied at each iterative step. For any given
$x=\case{1}{2}U|\mu|$ solution of equation \eref{eq:sr} determines $U$,
and the local moment $|\mu|\equiv|\mu(T=0)|$ then follows directly. As
explained in [36]
this step generates a spin-flip resonance in 
$\chi^{+-}(\om;T=0)$ centred on a non-zero frequency $\om_m$. This is the 
low-energy scale characteristic of the Kondo lattice (and with
$\om_m \propto Z$, the quasiparticle weight); its origins within the LMA
thus stemming from SR, and its physical significance being that it sets
the timescale $\tau\sim h/\om_m$ for restoration of the locally broken symmetry
inherent at the crude level of pure MF. For $T>0$ the same iterative algorithm
may be employed, except that SR is no longer required. Temperature enters
the problem in two distinct ways: (a) explicitly, and centrally, via
the Fermi functions inherent in $h(\om_1;\om_2)$ and $\chi^{+-}(\om_1;T)$
(equation (2.23)); and (b) implicitly, via the 
$T$-dependence of the local moment $|\mu|\equiv|\mu(T)|$ in 
$x=\case{1}{2}U|\mu|$.
The latter may be encompassed via $|\mu(T)|=|\mu(0)| + \delta|\mu(T)|$,
with $|\mu(0)|$ the $T=0$ moment required to satisfy SR as above; and with
$\delta|\mu(T)|$ calculated in practice at MF level [32],
such that 
$\delta|\mu(0)|=0$. As for the Anderson impurity model [32],
we find 
however that the resultant $T$-dependence of $|\mu|$ is negligible for
essentially all $(U,V)$ (provided one is not concerned with physically 
irrelevant temperatures of the order of $U$), and we thus omit it from the
results shown explicitly in $\S$'s 4{\it ff}.
\section{Electrical transport and optical conductivity.}
\label{sec:transp}

  Within the large-dimensional framework of DMFT a knowledge of
single-particle dynamics, in particular the $(\om,T)$-dependences of
the $f$-electron self-energy $\Sigma_f(\om;T)$, enable $\q=0$ transport 
properties to be determined [8-11].
In this section
we specify some basic results for the conduction electron scattering
rate and $\om$-dependent electrical conductivity, for both the hypercubic 
and Bethe lattices. These are independent of the particular approximation
employed to determine $\Sigma_f(\om;T)$. But they naturally underlie the 
results obtained via the LMA that are given in $\S$'s~5{\it ff}.

   Equation~\eref{eq:gchg} for the conduction electron Green function,
$G^c(\om)=\H(\gamma)$ with 
$\gamma(\om)=\om^+ - V^2\left[\om^+ - \Sigma_f(\om;T)\right]^{-1}$, is
equivalently but more familiarly expressed as 
$G^c(\om)=N^{-1}\sum_\alpha[\om^+ - \ep_\alpha - \Sigma_c(\om;T)]^{-1}$.
Here $\ep_\alpha$ denote the states of the free ($V=0$) conduction band
with density of states (equation (2.4)) $\rho_0(\ep)=N^{-1}\sum_\alpha
\delta(\ep-\ep_\alpha)$ ({\it e.g.\ }$\ep_\alpha\equiv \ep_\k$ for a 
Bloch decomposible lattice); and $\Sigma_c(\om;T)\; (=\om^+ - \gamma(\om))$
is the purely local conduction electron self-energy, related to the
$f$-electron single self energy $\Sigma_f$ by:
\beq
\Sigma_c(\om;T)=V^2\left[\om^+ - \Sigma_f(\om;T)\right]^{-1}
\eeq
It will prove useful in the following to rewrite equation (2.7) as
\beq
G^c(\om)=\int^\infty_{-\infty} d\ep\,\rho_0(\ep)G^c(\ep;\om)
\hspace{0.5cm}\equiv\; <\!G^c(\ep;\om)\!>_\ep
\eeq
with the $\ep$-resolved propagator $G^c(\ep;\om)=[ \gamma(\om) - \ep]^{-1}
=[\om^+ - \ep - \Sigma_c(\om;T)]^{-1}$ and corresponding spectrum
$D_c(\ep;\om)=-\case{1}{\pi}\Im G^c(\ep;\om)$; and  where $<\!A(\ep)\!>_\ep
=\int d\ep\,\rho_0(\ep)\,A(\ep)$ defines the $\ep$-average of any $A(\ep)$.

   In particular, the conduction electron scattering rate $1/\tau(\om;T)$
($\hbar=1$) considered in $\S$4.1, is given by 
\beq
\frac{1}{\tau(\om;T)}=\gamma_I(\om;T)=-\Im \Sigma_c(\om;T)
\eeq
(with $\gamma_I=\Im\gamma$). It is conveniently expressed in the 
dimensionless form
$1/\tilde{\tau}(\om;T)=\tilde{\gamma_I}(\om;T)=\pi\rho_0\gamma_I(\om;T)$,
with $\rho_0=\rho_0(\ep=0)$; and is given in terms of the $f$-electron 
self-energy $\Sigma_f=\Sigma_f^R-\ii\,\Sigma^I_f$ by
\beq
\frac{1}{\tilde{\tau}(\om;T)}=\tilde{\gamma_I}(\om;T)
=\frac{\Delta_0^{-1}\Sigma^I_f(\om;T)}{[\om'-\Delta_0^{-1}
\Sigma_f^R(\om;T)]^2 + [\Delta_0^{-1}\Sigma^I_f(\om;T)]^2}
\eeq
where $\om'=\om/\Delta_0$ and $\Delta_0$ is defined by
\beq
\Delta_0=\pi V^2\rho_0 \,.
\eeq
Physically, $\Delta_0$ is the hybridization strength that would arise
for a pure Anderson impurity model (AIM), in which a single correlated 
$f$-level is locally coupled (via $V$) to the free metallic conduction band 
$\rho_0(\ep)$. And equation (3.4) will prove important in connecting
results for the PAM at large $\om$ and/or $T$, to those for the pure AIM
itself (and notwithstanding the fact that the ground state for the latter
is metallic, while that for the symmetric PAM is of course insulating);
see also equation (3.12) below.

 Calculation of the $\om$-dependent conductivity is in principle
 straightforward
in the large-dimensional limit of DMFT, since the strict absence of 
vertex corrections [41] to the ($\q =0$) current-current correlation
function means that only the lowest order conductivity bubble
diagram survives [8-11]. We denote the trace of the resultant conductivity
tensor by $\tilde{\sigma}(\om;T)$ ($\case{1}{3}$ of which, denoted by
$\sigma(\om;T)$, provides an
approximation to the isotropic conductivity for a $d=3$ dimensional system).
This may be cast in the form
\beq
\frac{\tilde{\sigma}(\om;T)}{\sigma_0}=F_\alpha(\om;T)
\eeq
where $\sigma_0=\case{\pi e^2 a^2}{\hbar}\case{N}{V}\simeq \case{\pi e^2}
{\hbar a}$  such that $\sigma_0 \sim 10^4-10^5\, [\Omega cm ]^{-1}$ for 
lattice constants $a$ in the physically realistic regime of $1-10$ \AA. 
The dimensionless $F_\alpha(\om;T)$ depends on the lattice type, and
is given explicitly for the hypercubic lattice (HCL) and Bethe lattice (BL)
by
\numparts
\beqa
\fl
\hspace{0.5cm}F_{HCL}(\om;T)&=\frac{t_*^2}{\om}\int^\infty_{-\infty} d\om_1 \;
[f(\om_1)-f(\om_1+\om)] \; <\! D_c(\ep;\om_1)D_c(\ep;\om_1+\om)\!>_\ep
\\
\fl
\hspace{0.5cm}F_{BL}(\om;T)&=\frac{t_*^2}{\om}\int^\infty_{-\infty} d\om_1 \;
[f(\om_1)-f(\om_1+\om)] \; 
<\! D_c(\ep;\om_1)\!>_\ep<\!D_c(\ep;\om_1+\om)\!>_\ep
\eeqa
\endnumparts
(with $f(\om)$ here the Fermi function). The exact result equation (3.7a) for
the HCL is of course well known (see {\it e.g.\ }[8-11]) and widely 
used even in 
studies employing the BL ({\it e.g.\ }[21]); although for the latter 
we emphasize
that it is equation (3.7b) which follows from direct 
analysis of
the conductivity bubble diagram. Equations (3.7a,b) correspond in an 
obvious physical sense to limiting forms of behaviour, from fully
correlated to uncorrelated averages of the $D_c(\ep;\om)$'s. Both will 
be considered in $\S$'s 5{\it ff}.

  Before proceeding we comment briefly on evaluation of $F_{HCL}(\om;T)$
itself, under a single approximation: namely that in
$<\!\!D_c(\ep;\om_1)D_c(\ep;\om_1+\om)\!\!>_\ep=\int^\infty_{-\infty} d\ep\,\rho_0(\ep)\,
 D_c(\ep;\om_1)D_c(\ep;\om_1+\om)$ entering equation (3.7a) the
$\ep$ dependence of $\rho_0(\ep)$ is neglected, $\rho_0(\ep)
\simeq \rho_0(0) \equiv \rho_0$. A straightforward integration over 
$\ep\in (-\infty,\infty)$ then
yields
\beq
\fl \hspace{0.2cm}
 <\! D_c(\ep;\om_1)D_c(\ep;\om_1+\om)\!>_\ep\;\simeq\:
 \frac{\rho_0}{\pi}\; \frac{[\gamma_I(\om_1+\om)+\gamma_I(\om_1)]}
{[\gamma_R(\om_1+\om)-\gamma_R(\om_1)]^2 + 
[\gamma_I(\om_1+\om)+\gamma_I(\om_1)]^2}
\eeq
(where $\gamma(\om)=\gamma_R(\om)+\ii\,\gamma_I(\om)$). Use of this in
equation (3.7a), relating $\gamma$ to the conduction electron self 
energy $\Sigma_c$ as above, gives
\beq
F_{HCL}(\om;T) \simeq\; \frac{-\rho_0 t_*^2}{\pi\om}\;\;\Im\:
\int^\infty_{-\infty}d\om_1 \;\frac{[f(\om_1)-f(\om_1+\om)]}
{\om-\Sigma_{c,r}(\om_1+\om)+\Sigma_{c,a}(\om_1)}
\eeq
where $r/a$ here denote retarded/advanced functions; and for the
particular case of the d.c limit $\om=0$, equation (3.8) with 
equations (3.4,7a) give:
\beq
\fl\hspace{0.5cm}
F_{HCL}(0;T) \simeq\; \frac{1}{2} \left[\rho_0 t_*\right]^2
\;\int^\infty_{-\infty}d\om\, \frac{-\partial f(\om)}{\partial \om}\;
\tilde{\tau}(\om;T)\;\equiv\; \frac{1}{2}\left[\rho_0 t_*\right]^2
<\!\tilde{\tau}\!>
\eeq
Equations (3.9,10) are likewise well known [25] and widely used; and the 
latter in particular, expressing the d.c conductivity in terms of
an averaged scattering time, is physically intuitive. We emphasize
nevertheless that they are approximate (granted even the legitimate
neglect of vertex corrections), being dependent on the `flat-band'
approximation $\rho_0(\ep)\simeq\rho_0 \,\forall\, \ep$ as above; and it is in fact
simple to show that this approximation by itself fails to account
for the existence of the Kondo insulating gap that is characteristic
of the symmetric PAM. That said however, one expects physically that
equation (3.10) should provide a good approximation to
the d.c conductivity at least for sufficiently high temperatures
dominated by incoherent scattering; the question here, considered briefly in
$\S$5, being how high is high?

  Finally, for explicit use in $\S$5, we consider the pure Anderson
impurity  model itself, denoting by $\rho_{\imp}(T)$ the change in resistivity
due to addition of the impurity to the non-interacting host, and 
$\rho'_{\imp}(T)=\rho_{\imp}(T)/\rho_{\imp}(0)$. This has the same
functional form as equation (3.10), {\it viz} (see {\it e.g.\ }[2]):
\beq
\frac{1}{\rho'_{\imp}(T)}\;=\;\int^\infty_{-\infty}d\om\, 
\frac{-\partial f(\om)}{\partial \om}\;
\tilde{\tau}_{\imp}(\om;T)
\eeq
The impurity scattering rate is given by [2] $1/\tilde{\tau}_{\imp}(\om;T)
=\pi\Delta_0 D_\imp(\om;T)$ with $D_\imp(\om;T)$ the impurity spectral
function and $\Delta_0$ the hybridization strength equation (3.5)
(such that $\pi\Delta_0 D_\imp(0;0)=1$ from the Friedel sum rule [2]).
Denoting the impurity single self-energy by $\Sigma(\om;T)
=\Sigma^R(\om;T)-\ii\,\Sigma_I(\om;T)$, the impurity transport rate is given
explicitly by
\beq
\frac{1}{\tilde{\tau}_{\imp}(\om;T)}\;=\;\frac{[1+\Delta_0^{-1}\Sigma^I
(\om;T)]}{[\om'-\Delta_0^{-1}\Sigma^R(\om;T)]^2 +
[1+\Delta_0^{-1}\Sigma^I(\om;T)]^2}
\eeq
with $\om'=\om/\Delta_0$; which should be compared to its counterpart
for the PAM, equation (3.4).

\section{ Single-particle dynamics. }

  We turn now to single-particle spectra resulting from the LMA specified
in \S2. Our focus for obvious physical reasons is the strong coupling (large-$U$)
Kondo lattice regime, wherein
universality arises; in particular the resultant scaling behaviour of dynamics,
and the thermal destruction of the Kondo (insulating) gap. Conduction electron
scattering rates $1/\tilde{\tau}(\omega; T)$ are also considered here (\S4.1) since
these are closely related to single-particle dynamics (as directly evident from
equation (3.4)).

   In strong coupling, as shown in [36] for $T=0$ and consistent with previous
work [12-15,24,28,29], the problem is characterised by a single 
low-energy scale; as embodied in $\Delta_{g}$ given by
\begin{equation}
\Delta_{g} = Z\frac{V^{2}}{t_{*}}
\end{equation}
with $Z= [1-(\partial\Sigma_{f}^{R}(\omega; 0)/\partial\omega)_{\omega =0}]^{-1}$
the quasiparticle weight.  This sets the scale for the Kondo gap,
which (\S2.1) is generically hard (as for the Bethe lattice), albeit strictly
soft for the hypercubic lattice; other embodiments of this low-energy scale, 
such as the spin-flip $\omega_{m}$ arising in $\chi^{+-}(\omega; 0)$ 
(\S2.2 and [36]) are equivalent to $\Delta_{g}$, being simply proportional
to it.
There are two distinct issues relating to the low-energy scale. First its
dependence on the `bare' high-energy parameters entering the Hamiltonian, {\it viz} the
$f$-electron Coulomb repulsion $U$, hybridization $V$ and bandwidth scale 
$t_{*}$ (or equivalently $\rho_{0}^{-1} \propto t_{*}$). In strong coupling it is
known, from NRG results in particular [13], that the gap scale becomes 
exponentially small; such behaviour is indeed found within the LMA
({\it viz} $ln\Delta_{g} \propto -U/(8V^{2}\rho_{0})$  as detailed in [36]).
This is important insofar as it guarantees a clean separation between 
low-energy and/or temperature scales on the order of the Kondo gap and multiples
of it; and high $\omega$ 
and/or $T$ on the order of the bare energy scales (Hubbard satellites in the 
$f$-electron spectra to cite an extreme). Failure to recover this
pristine separation of scales can obscure much relevant physics in the strong
coupling regime (we provide examples in the following sections). This may arise
either because strong coupling and/or low temperatures are difficult to access 
in a numerical approach ({\it e.g.\ }QMC); or because the approximate theory used leads
to an insufficiently small low-energy scale, for example algebraic rather than
exponential decay in $U$. Iterated perturbation theory (IPT) [21,22,10] 
provides an
example of the latter, and is discussed further in the following sections.

  Granted a clean separation of energy scales however, the precise dependence of
the gap scale $\Delta_{g}$ on bare parameters is subsidiary. Of primary importance
is that physical properties, whether dynamics or transport, exhibit universal
scaling behaviour on experimentally  relevant `low' $\omega$ and/or $T$ scales
on the order of the Kondo gap and (in principle arbitrary) multiples thereof. 
The scaling is of course in terms of the gap  $\Delta_{g}$ itself, and
is thus \it independent \rm of the bare model parameters; in contrast to the
correspondingly non-universal behaviour arising on truly high $\omega$ or $T$
scales characteristic of the bare parameters. Neither is such scaling of purely
theoretical interest, for an experimental gap of (say) $50K \sim 4 meV$ is tiny
on the $eV$ scale typical of bare parameters, and many multiples of it may arise
before non-universal scales are reached in practice. Moreover it is arguably
less preferable, as well as unnecessary in general, to have comparison to the
$\omega$ and/or $T$ dependence of experiment hinge on a particular choice of bare 
model parameters; as opposed to a knowledge solely of the experimental gap scale, 
which alone is required if comparison is made instead to the relevant scaling form
(\S7).

\begin{figure}[t]
\epsfxsize=400pt
\centering
\epsffile{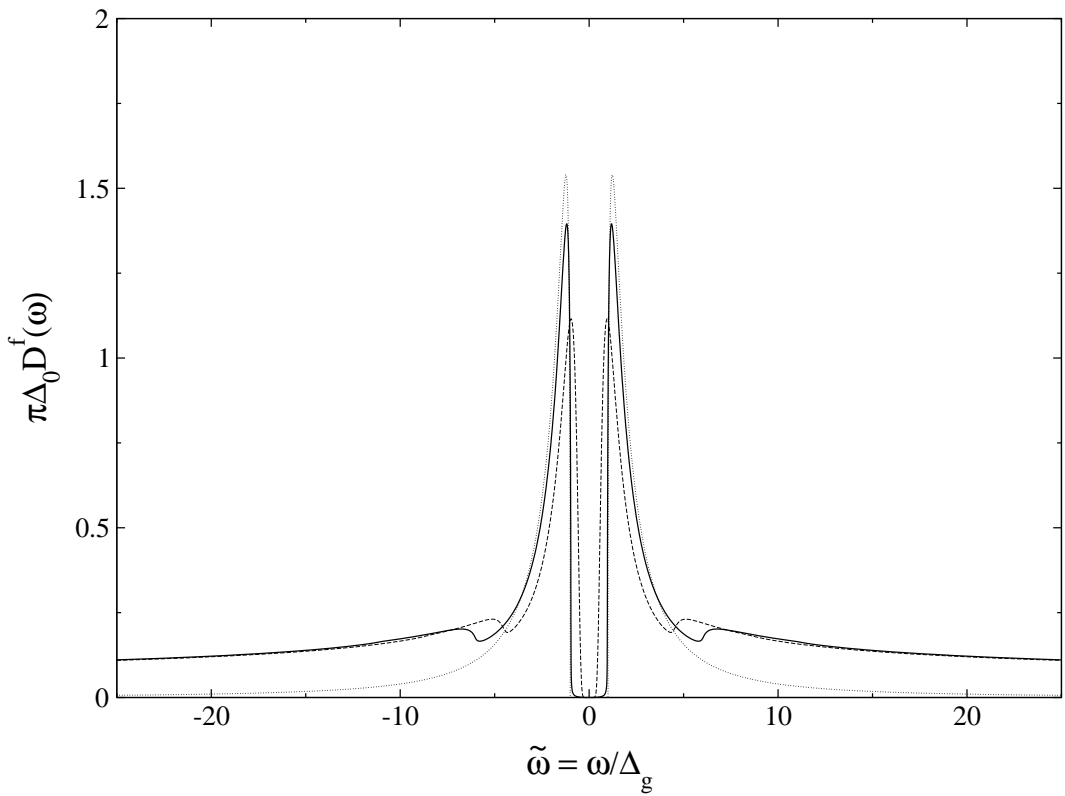}
\caption{$T=0$ scaling spectra $\pi \Delta_0 D^f(\om)$ {\it vs} $\tilde{\om}=
\om/\Delta_g$ for the BL (solid line) and the HCL (dashed line). The 
limiting low energy quasiparticle form is also shown (for the BL, dotted line).}
\end{figure}

  Figure 2 summarizes salient results for the $T=0$ $f$-electron 
spectrum [36]; showing $\pi\Delta_{0}D^{f}(\omega)$ (with $\Delta_{0} =
\pi V^{2}\rho_{0}$, equation (3.5)) as a function of $\tilde{\omega} = \omega/\Delta_{g}$
[42], and for both the BL (solid line) and HCL (dashed). This is 
the universal scaling form arising from the LMA [42], with no dependence whatever 
on any of the bare parameters $U$, $V$, $t_{*}$. It naturally depends on the lattice
type, but only weakly and on scales up to a few times the gap, beyond which the
two scaling spectra coincide. The Kondo insulating gap is directly
apparent in figure 2. For the BL we also show explicitly the 
limiting quasiparticle form,
whose recovery at sufficiently low-$\tilde{\omega}$ embodies adiabatic
continuity to the non-interacting limit (`insulating Fermi liquid' behaviour,
\S2.1 and [36]); given explicitly using equation (2.20b) by
$\pi\Delta_{0}D^{f}(\omega) \sim 
(4/|\tilde{\omega}|^{2})[1-1/\tilde{\omega}^{2}]^{\frac{1}{2}}$ for
$|\tilde{\omega}| >1$ (and zero for $|\tilde{\omega}| = |\omega|/\Delta_{g} < 1$,
the gap). For $|\tilde{\omega}| \gtrsim 3$ however the quasiparticle form
is simply inadequate: it decays rapidly as $\sim 1/|\tilde{\omega}|^{2}$ and fails 
to recover the long, slowly decaying tails evident in figure 2. The latter, which 
dominate the scaling spectra at (moderate to) large $\tilde{\omega}$ --- and 
in consequence transport properties at correspondingly `high' temperatures
(\S5) --- are logarithmically slow and discussed
further in the following sections. Universal scaling in terms of $\tilde{\omega}$
is not of course confined to the $f$-electron spectra: the $T=0$ $c$-electron
spectrum $t_{*}D^{c}(\omega)$ (or equivalently $D^{c}(\omega)/\rho_{0}$)
behaves likewise; and in consequence, as follows straightforwardly
using equations (2.5), $\Delta_{0}^{-1}\Sigma_{f}(\omega; 0)$ also scales universally
(which is why the dimensionless conduction electron scattering rate
has been defined as in equation (3.4)).

 For finite temperatures, what one expects in scaling terms is obvious; {\it viz} that 
the $f/c$ spectra and $f$-electron self-energy should now exhibit universal scaling
in terms of both $\tilde{\omega} = \omega/\Delta_{g}$ \it and \rm
$\tilde{T} = T/\Delta_{g}$. That such scaling correctly arises within the
LMA is demonstrated in figure 3 where, for fixed $\tilde{T} =2$, the $f$- and $c$-electron
scaling spectra are shown for three different interaction strengths
$\tilde{U} = U/t_{*} = 5.6, 6.1, 7.0$ and $V^{2}/t_{*}^{2} = 0.2$. The inset shows 
$\pi\Delta_{0}D^{f}(\omega)$ on an `absolute' scale, {\it i.e.}
\it vs \rm $\omega/t_{*}$, illustrating the exponential reduction in the gap
scale with increasing interaction strength. The main figures by contrast show 
the $f/c$-spectra \it vs \rm $\tilde{\omega} = \omega/\Delta_{g}$, from which
the scaling collapse is evident (and in practice sets in by rather
moderate interactions $\tilde{U} \sim 4$ or so).
\begin{figure}[h]
\epsfxsize=300pt
\centering
\epsffile{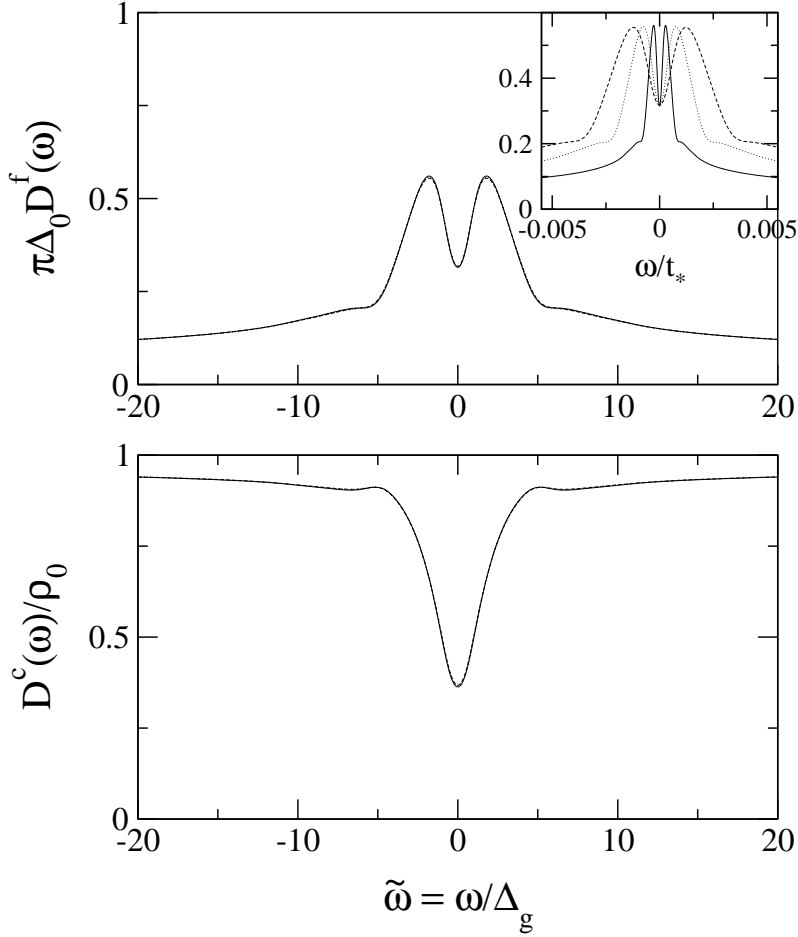}
\vspace{0.1cm}
\caption{Universal scaling spectra (BL) for fixed $\tilde{T}=T/\Delta_g=2:
\pi\Delta_0D^f(\om)$ and $D^c(\om)/\rho_0$ {\it vs} $\om/\Delta_g$ for 
$V^{2}/t_{*}^{2} = 0.2$ and $\tilde{U} = U/t_{*} = 5.6$ (dashed),
6.1 (dotted), and 7.0 (solid). Inset: corresponding $f$-spectra on an
absolute scale, {\it vs} $\om/t_*$.}
\label{fig:ftscsp}
\end{figure}

  The scaling illustrated above arises generically, and figure 4 shows the resultant
LMA scaling spectra (for the BL) for a range of scaled temperatures up to 
$\tilde{T} = 25$. The
thermal destruction/infill of the Kondo insulating gap is evident, occurring 
as expected physically for temperatures on the order of the gap $\Delta_{g}$, and
accompanied in the case of the $f$-electron spectra by a corresponding destruction
of the characteristic $T=0$ spectral `horns'. We also note, as evident from the
inset to the $f$-spectrum which shows the spectrum on an enlarged $\tilde{\omega}$
scale, that for any $\tilde{T}$ the high-frequency asymptotics of the scaling spectra
coincide with that for $\tilde{T}=0$; as likewise expected physically, and which
behaviour arises for frequencies $|\tilde{\omega}| \gg \tilde{T}$.
\begin{figure}[ht]
\epsfxsize=300pt
\centering
\epsffile{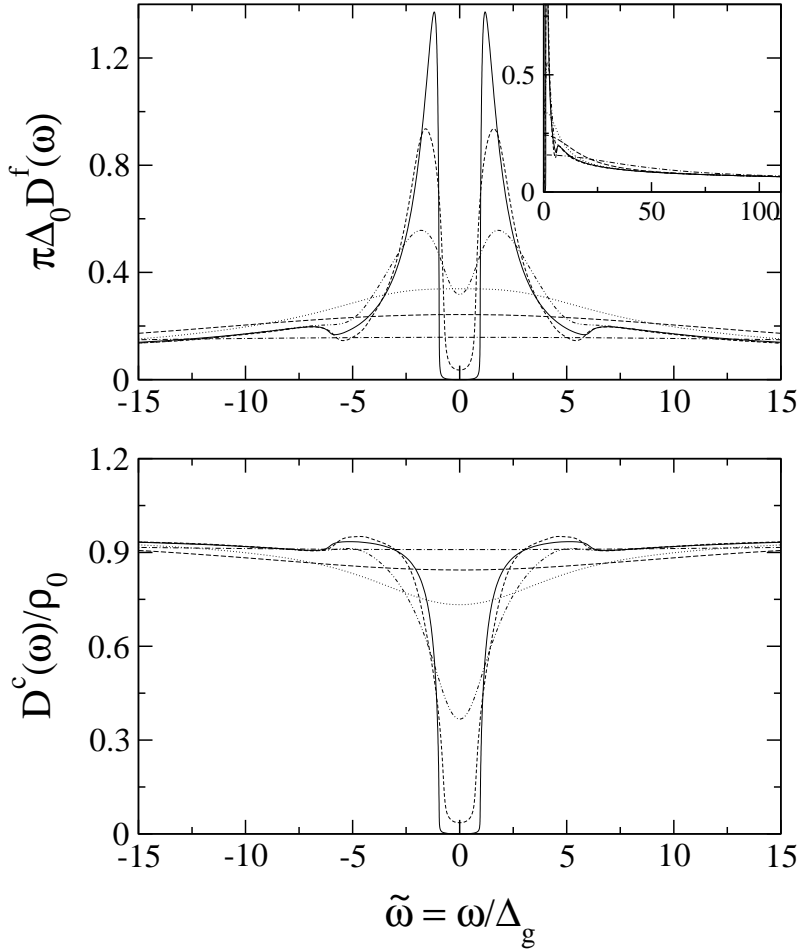}
\caption{Temperature dependence of the BL scaling spectra:
$\pi\Delta_0D^f(\om)$
and $D^c(\om)/\rho_0$ {\it vs} $\om/\Delta_g$ for 
temperatures $\tilde{T}=T/\Delta_g=0$ (solid), 1 (short dash),
2 (double point-dash), 5 (dotted), 10 (long-dash) and 25 (point-dash).
Inset: $f$-spectra on an enlarged scale, out to $\tilde{\om}=100$.}
\label{fig:ftevol}
\end{figure}

\begin{figure}[h]
\epsfxsize=280pt
\centering
\epsffile{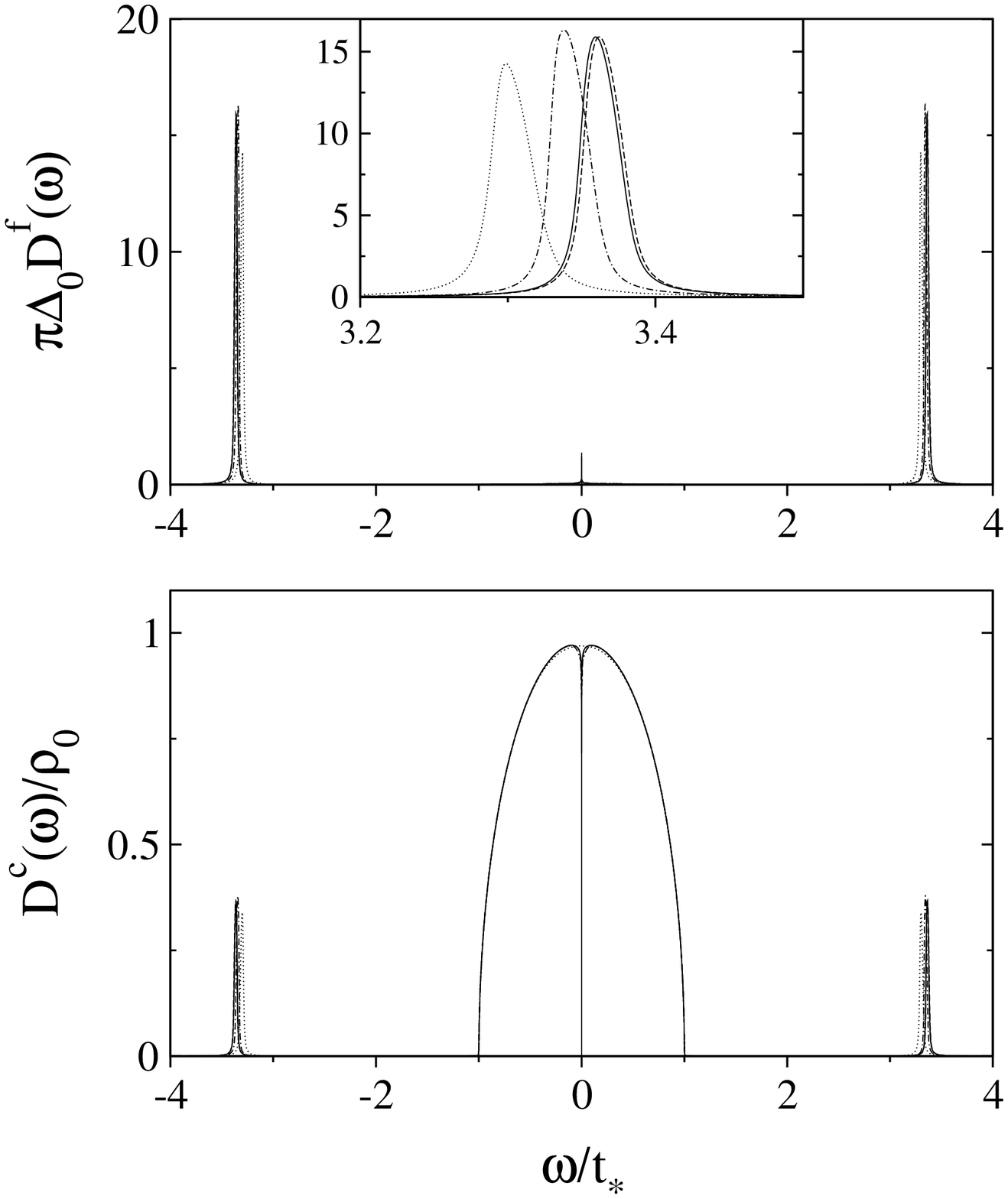}
\caption{Spectra on all scales (BL with $U/t_*$=6.1 and $V^2/t_*^2=0.2$):
$\pi\Delta_0D^f(\om)$ and $D^c(\om)/\rho_0$ {\it vs} $\om/t_*$ for
temperatures $\tilde{T}=T/\Delta_g=0$ (solid), 10 (dashed), 50 (point-dash)
and 450 (dotted). Inset: Hubbard satellites on a much reduced scale.}
\label{fig:allsc}
\end{figure}
  To give an `all scales' perspective on thermal evolution, figure 5 shows the $f$- and
$c$-electron spectra for the BL (specifically for $\tilde{U} = 6.1$ and
$V^{2}/t_{*}^{2}=0.2$) on an absolute energy scale, \it vs \rm $\omega/t_{*}$;
and for temperatures $\tilde{T} = T/\Delta_{g} = 0, 10, 50, 450$. The
$f$-spectrum is naturally dominated by the non-universal Hubbard satellites
at $|\omega| \sim \frac{U}{2}$ which carry almost all the spectral weight (and
are of course `projected out' of the scaling spectra);
the key low-energy universal spectral features shown in figures 2-4 being
nigh on invisible in figure 5 as expected, since their net spectral weight is of order
$Z \ll 1$. The $c$-spectrum, which shows weakly remnant Hubbard peaks, is
by contrast dominated by the envelope of the free conduction band (the $|\omega|
\leq t_{*}$ semiellipse for the BL); and again the thermal destruction of the
low-energy spectral gap is barely visible. We add moreover that until temperature
reaches essentially non-universal scales, the $T$-dependence of single-particle
dynamics is confined to the relevant low-energies illustrated in figures 3,4 (see
{\it e.g.\ }the inset to figure 5 where the Hubbard satellites are
enlarged, noting that $\tilde{T} =450$ here corresponds to $T \sim 0.2t_{*}$).

\vspace{0.3cm}

\subsection{Scattering rates. }

  The conduction electron scattering rates $\tau^{-1}(\omega; T)$ are now
considered. These are directly related to the $f$-electron self-energy 
$\Delta_{0}^{-1}\Sigma_{f}(\omega; T)$ as in equations (3.3,4); and determine the
conductivity via the dependence thereof (equation (3.7)) on the $c$-electron
spectrum $D_{c}(\epsilon; \omega)$ (or rather more directly via equation (3.10) in
the latter's regime of applicability, considered explicitly in \S5).

  Before proceeding we note that our conclusions regarding the scattering
rates differ significantly from the work of [21]. In order to explain typical
experimental conductivities for Kondo insulators, it was argued in [21] that
scattering rates in the vicinity of  the Fermi level should be on the order 
$\sim 0.1-1$ of the bandwidth $t_{*}$ (values some 2-3 orders of magnitude higher
than for a clean metal like $Cu$). Scattering rates calculated in [21] 
were however 
found to be some 1-2 orders of magnitude lower than required. 
While the authors of [21] note that this behaviour is very surprising, they 
attribute it to an intrinsic limitation of the model itself; 
rather than {\it e.g.\ }to a limitation of the 
approximate calculations employed, or to a possible
misidentification of the relevant temperature scales involved.
We find by contrast, as shown below, that scattering rates 
can certainly attain values on the order of the bandwidth $t_{*}$; and indeed
argue that rates of this order \it must \rm arise over a wide, experimentally
relevant temperature regime.
\begin{figure}[h]
\epsfxsize=400pt
\centering
\epsffile{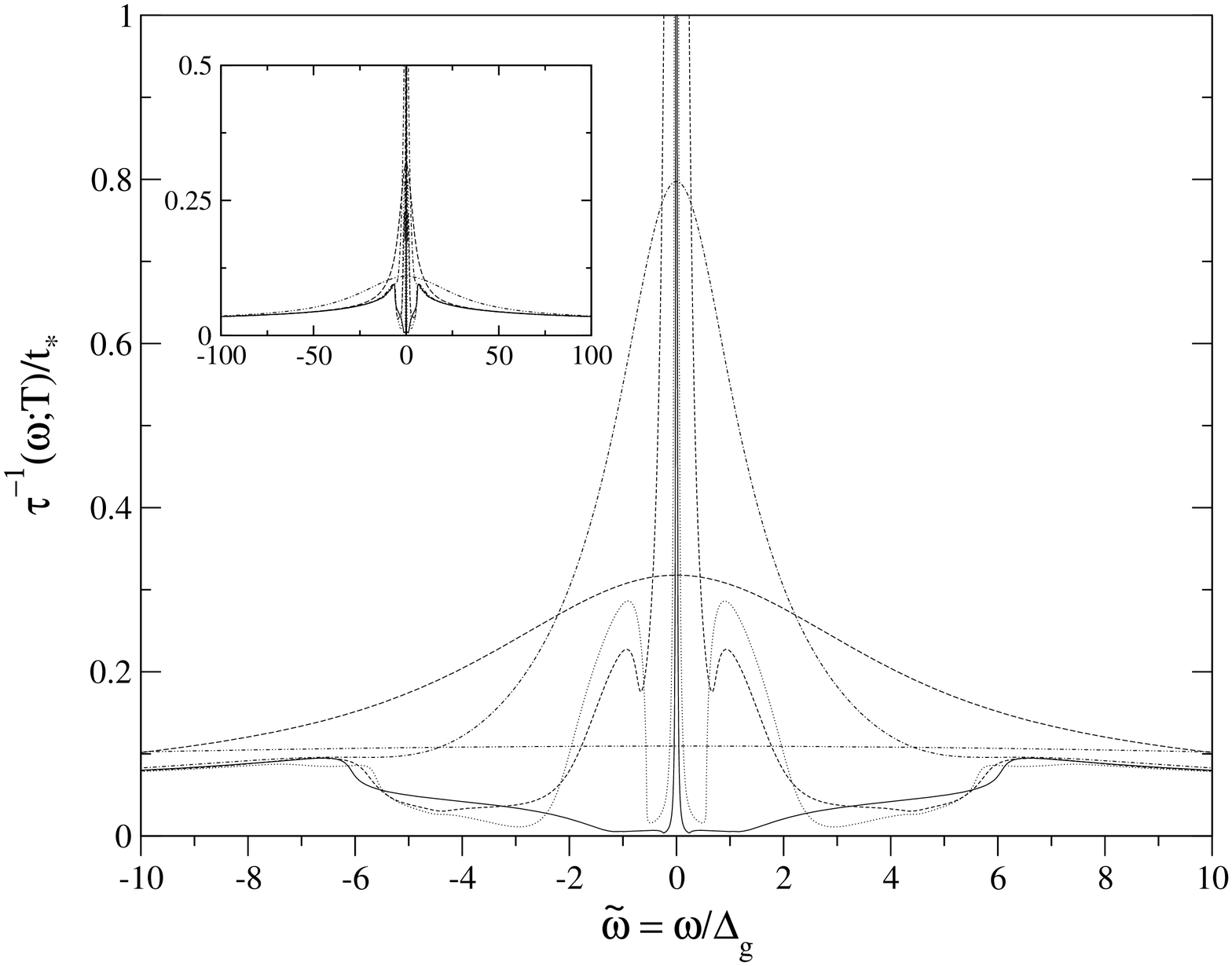}
\caption{Thermal evolution of the conduction-electron scattering
rates: $\tau^{-1}(\om;T)$ (in units of the BL bandwidth $t_*$)
{\it vs} $\tilde{\om}=\om/\Delta_g$ for temperatures $\tilde{T}
= T/\Delta_{g}$= 0.1 (solid), 0.5 (dotted), 1 (short dash), 2.5
 (point dash), 5 (long dash) and 20 (double point-dash).
 Inset: on an expanded scale, out to $\tilde{\om}=100$.}
\label{fig:scat_T}
\end{figure}

  The $\omega$- and $T$-dependences of the scattering rate arising in strong 
coupling are illustrated in figure 6: {\it viz} $\tau^{-1}(\omega;T)$
in units of the bandwidth $t_{*}$, \it vs \rm $\tilde{\omega} = \omega/\Delta_{g}$
and for a range of  temperatures $\tilde{T} = T/\Delta_{g}$ from 0.1 to 20; as
given explicitly via equation (3.4) (where the `bare' factor of $\omega^{'}
\equiv \tilde{\omega}\frac{\Delta_{g}}{\Delta_{0}}$ plays no role since
$\frac{\Delta_{g}}{\Delta_{0}} \propto Z$ is exponentially small in strong coupling).
The inset to figure 6 shows the same results on an expanded scale out to 
$\tilde{\omega} = 100$ (showing that for $|\tilde{\omega}| \gg \tilde{T}$ the `tail' 
behaviour reduces to that characteristic of $T=0$, the logarithmically slow
decay of which is considered explicitly below, 
figure 8). The only remaining $\omega$-dependence arises on non-universal (and 
essentially irrelevant) scales on the order of the Hubbard satellites 
$|\omega| \sim \frac{U}{2}$, where as expected physically the scattering rate 
is strongly peaked [21]; this is omitted from figure 6 for obvious reasons.

  The first point to note about $\tau^{-1}(\omega;T)$ is that for $T=0$
precisely it contains a $\delta$-function contribution at the Fermi level
$\omega =0$; specifically $\pi t_{*}\delta(\tilde{\omega})$ as follows 
generally via equation (3.4) using $\Sigma_{f}^{R}(\omega; 0) \sim -\frac{\omega}{Z}$
as $\omega \rightarrow 0$. For $T \neq 0$ this becomes the resonance evident
in figure 6, which naturally broadens with increasing temperature; and the \it only \rm 
thermal scale on which this can occur is the gap $\Delta_{g}$ --- the single 
low-energy scale characteristic of the system in strong coupling. This argument is
rather general. The scattering rate in the vicinity of the Fermi level,
$|\tilde{\omega}| \lesssim \tilde{T}$, in consequence diminishes with 
increasing $\tilde{T}$ from essentially arbitrarily large values (reflecting the
insulating nature of the $T=0$ state); and does so on temperature scales of the order
of the gap, $\Delta_{g}$. The results of figure 6 show moreover that for
temperatures $\tilde{T} = T/\Delta_{g}$ in the range $\sim 1-20$,
scattering rates in the relevant $\tilde{\omega}$ regime are \it indeed on
the order of $\sim 0.1-1$ of the bandwidth $t_{*}$\rm. We also add that while
a temperature range of this order certainly encompasses that relevant to
experiment (\S7),
the essential point is not dependent on it, since for $\tilde{T} \gg 1$
we find the Fermi level scattering rate to decay slowly with temperature
(specifically $\tau^{-1}(\omega =0; T) \propto 1/ln^{2}(\tilde{T})$, see
equation (5.7) below).

  The issue of scattering rates was considered in [21] using 
 IPT [43], explicitly 
so for a particular choice of bare parameters $U/t_{*} =3$ and $V/t_{*} = 0.25$;
and at a temperature $T=0.1t_{*}$, a significant fraction of the free conduction
bandwidth and some 50 times the corresponding IPT gap $\Delta_{g} \equiv
ZV^{2}/t_{*}$.
The resultant scattering rate in the vicinity of the Fermi
level was found to be $\tau^{-1}(0;T) \sim 10^{-2}t_{*}$, with which the 
authors of [21] support their view mentioned above. We have also performed 
IPT calculations
for the same $U$ and $V$; and indeed for $T=0.1t_{*}$ recover
the results of [21]. We have further investigated the $T$-dependence of
IPT over a wide
$\tilde{T} = T/\Delta_{g}$ range (as well as a broad $(U,V)$ range).
Significantly, we find that for temperatures $\tilde{T}$ up to $\sim 5$,
the behaviour of the resultant $\tau^{-1}(\omega; T)$ is qualitatively similar
to that shown in figure 6: in particular, scattering rates in the vicinity of the
Fermi level  are again found to lie in the range $0.1t_{*} -t_{*}$. For
temperatures $\tilde{T} \gtrsim 3-5$, the IPT scattering rates decay much more
rapidly with increasing $\tilde{T}$ than those arising from the LMA. This reflects
the inability of IPT to capture the logarithmically slow decays in
$\tilde{\omega}$ and/or $\tilde{T}$ that are characteristic of the model in
the strong coupling/scaling regime, as illustrated below (figure 9); and is in turn
related to the fact that IPT leads to an algebraically rather than an
exponentially small gap scale in strong coupling, and hence does not produce
a `clean' separation of low/high energy scales. Nonetheless if it was 
indeed the case
that non-universal temperature scales on the order of {\it e.g.\ }$0.1t_{*}$
were pertinent in relation to experiment, then the resultant scattering rates
would in general be too small to explain observed conductivities. Our 
view is that
this is {\it not} the case, but rather that the relevant thermal
scale for comparison to experiment is the gap $\Delta_{g}$ and multiples of it;
and on which scales the transport rates readily attain values on the
order of $\sim 0.1t_{*} - t_{*}$. Further support for this view will be provided
in the following sections. 

  We turn now to more detailed consideration of the conduction electron
scattering rate in strong coupling, in particular its high-frequency `tail' 
behaviour evident in figure 6 (inset) and its relation to the $f$-electron
single-particle spectrum. We begin with the latter. It is straightforward
to show generally (using equations (2.5) together with the definition equation (2.8) of 
$\gamma (\omega)$ and equation (3.4)), that the asymptotic behaviour of the
dimensionless scattering rate $\tilde{\tau}^{-1}(\omega; T)$ (equation (3.4)) 
in fact coincides with the $f$-spectrum, {\it viz}
\begin{equation}
\frac{1}{\tilde{\tau}(\omega;T)} \sim \pi\Delta_{0}D^{f}(\omega).
\end{equation}
This holds asymptotically for $\tilde{\tau}^{-1}(\omega;T) \ll 1$; which in
practice means $(|\omega|/\Delta_{g} =)$ $|\tilde{\omega}| \gg 1$ for any
$\tilde{T}$ (the `tails') or $\tilde{T} \gg max(1,|\tilde{\omega}|)$ for 
any $|\tilde{\omega}|$. This is illustrated in figure 7(a) for
$\tilde{T}=0$, where the $\tilde{\omega}$-dependence of 
$\pi\Delta_{0}D^{f}(\omega)$ and $\tilde{\tau}^{-1}(\omega;T)$ are compared
(omitting the $\delta (\tilde{\omega})$ contribution for clarity):
they naturally differ very significantly at low frequencies, but for 
$|\tilde{\omega}| \gtrsim 10$ or so their tails rapidly become coincident. 
The same holds at finite $\tilde{T}$ as illustrated in figure 7(b) for 
$\tilde{T} = 2.5$ and $10$; and for $\tilde{T} \gg 1$ the two coincide
asymptotically for all $\tilde{\omega}$.
\begin{figure}[h]
\epsfxsize=400pt
\centering
\epsffile{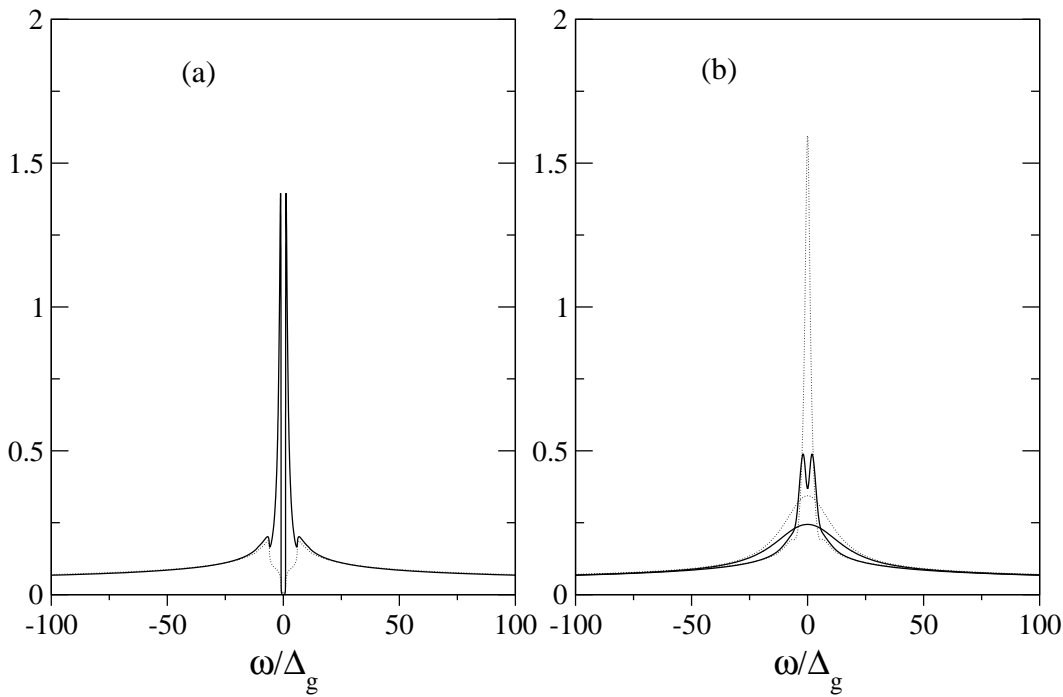}
\caption{$\pi\Delta_{0}D^{f}(\omega)$ (solid lines) and scattering 
rates $\tilde{\tau}^{-1}(\omega;T)$ (dotted lines) {\it vs}
$\om/\Delta_g$ (for the BL). Figure 7(a) is for $\tilde{T}=
T/\Delta_g=0$, and figure 7(b) for $\tilde{T}= 2.5$ and 10 (in an
obvious sequence).}
\label{fig:scatdf}
\end{figure}

\begin{figure}[t]
\epsfxsize=370pt
\begin{center}
\epsffile{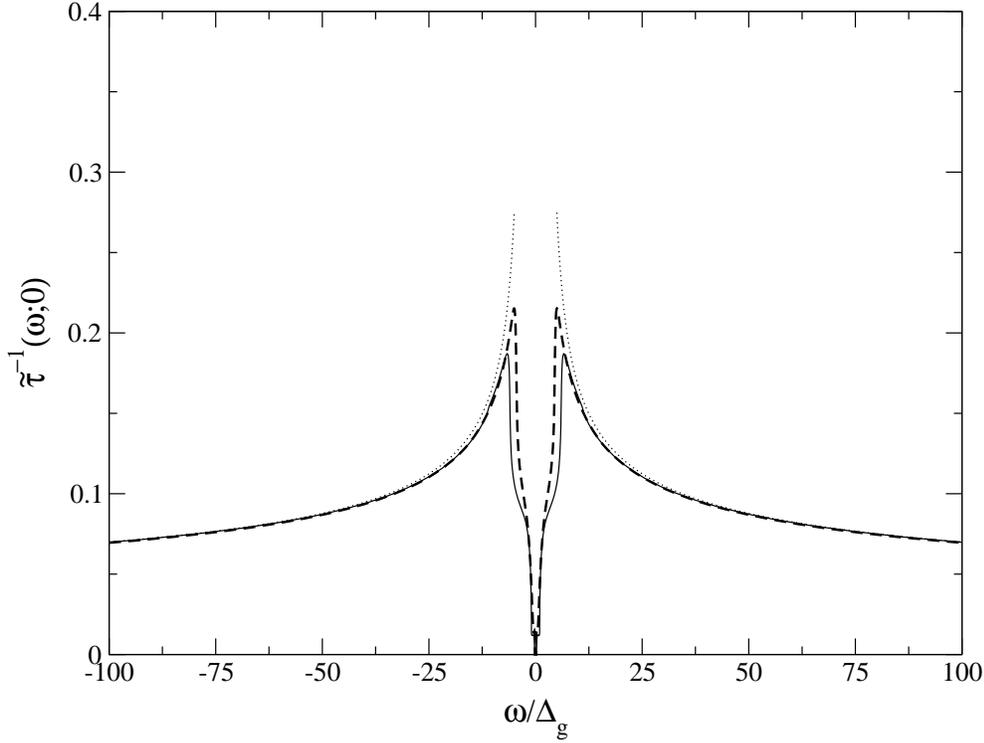}
\end{center}
\caption{$T=0$ scattering rates $\tilde{\tau}^{-1}(\omega;0)$
{\it vs} $\om/\Delta_g$ for the BL (solid line) and HCL (dashed line).
Explicit comparison is also made to the asymptotic behaviour equation
 (4.4) (dotted line).}
\label{fig:asymp}
\end{figure}

  An obvious question arises: what is the functional form of the 
$|\tilde{\omega}| \gg 1$ `tails'
in $\tilde{\tau}^{-1}(\omega;T)$ (or equivalently $\pi\Delta_{0}D^{f}(\omega))$,
which as seen in figure 6 (inset) arise at sufficiently high frequencies
for any temperature $\tilde{T}$? For large $|\tilde{\omega}|$ and/or $\tilde{T}$
one expects [1,2] on physical grounds that properties of the Kondo lattice 
such as the $f$-electron spectrum or the resistivity $\rho (T)$ (\S5),
should asymptotically approach those of an Anderson impurity model (AIM)
({\it i.e.} the pure Kondo model in strong coupling). The high-frequency behaviour of 
the impurity single-particle scaling spectrum $D_{imp}(\omega)$ for the AIM has only
recently been uncovered, using the LMA [31] (which gives excellent agreement
with NRG
results [34] for that problem). On the testable assumption that the high-frequency
behaviour of the $f$-electron self-energy for the PAM has the same scaling
form as that for the AIM, we thus expect $\Delta_{0}^{-1}\Sigma_{f}(\omega)$
to behave as [31]
\numparts
\begin{equation}
\Delta_{0}^{-1}\Sigma_{f}^{I}(\omega) \sim \frac{2}{3}[1 +
\frac{8}{\pi^{2}}ln^{2}(a|\tilde{\omega}|)]
\end{equation}
\begin{equation}
\Delta_{0}^{-1}\Sigma_{f}^{R}(\omega) \sim
-sgn(\omega)\frac{16}{3\pi}ln(a|\tilde{\omega}|)
\end{equation}
\endnumparts
for $|\tilde{\omega}| \gg 1$ (or $\gg max(1,\tilde{T})$ at finite-$\tilde{T}$),
with $a$ a pure constant $\cal{O}$$(1)$; and hence from equations (3.4) and (4.2)
that
\begin{equation}
\pi\Delta_{0}D_{f}(\omega) \sim \frac{1}{\tilde{\tau}(\omega; T)}
\stackrel{|\tilde{\omega}| \gg 1}{\sim} \frac{3\pi^{2}}{16}
\frac{[ln^{2}(a|\tilde{\omega}|) + \frac{\pi^{2}}{8}]}
{[ln^{2}(a|\tilde{\omega}|) + \frac{\pi^{2}}{8}]^{2} +
[\pi ln(a|\tilde{\omega}|)]^{2}}.
\end{equation}
In figure 8 for $T=0$, the $\tilde{\omega}$-dependence of $\tilde{\tau}^{-1}(\omega;0)$
is shown, for both
the BL and HCL. It is also compared directly to equation (4.4). The asymptotics for 
the two lattices coincide in practice for $|\tilde{\omega}| \gtrsim 5-10$ (as
expected {\it e.g.\ }from figure 2). And the `high'-energy behaviour of the tails in
$\tilde{\tau}^{-1}(\omega;T)$ (or equivalently $\pi\Delta_{0}D^{f}(\omega)$)
is indeed seen to be that of equation (4.4) (with the constant $a \sim 0.55$ determined
numerically). We also add in passing that the full form equation (4.4) is required
for the very good agreement over a wide $\tilde{\omega}$-range evident in figure 8,
{\it i.e.} it is not exclusively
dominated by the ultimate large-$\tilde{\omega}$ asymptotic behaviour
$\pi\Delta_{0}D^{f}(\omega) \sim \tilde{\tau}^{-1}(\omega; T)
\sim 3\pi^{2}/[16ln^{2}(|\tilde{\omega}|)]$.
\vspace{1.0cm}
\begin{figure}[h]
\epsfxsize=350pt
\centering
\epsffile{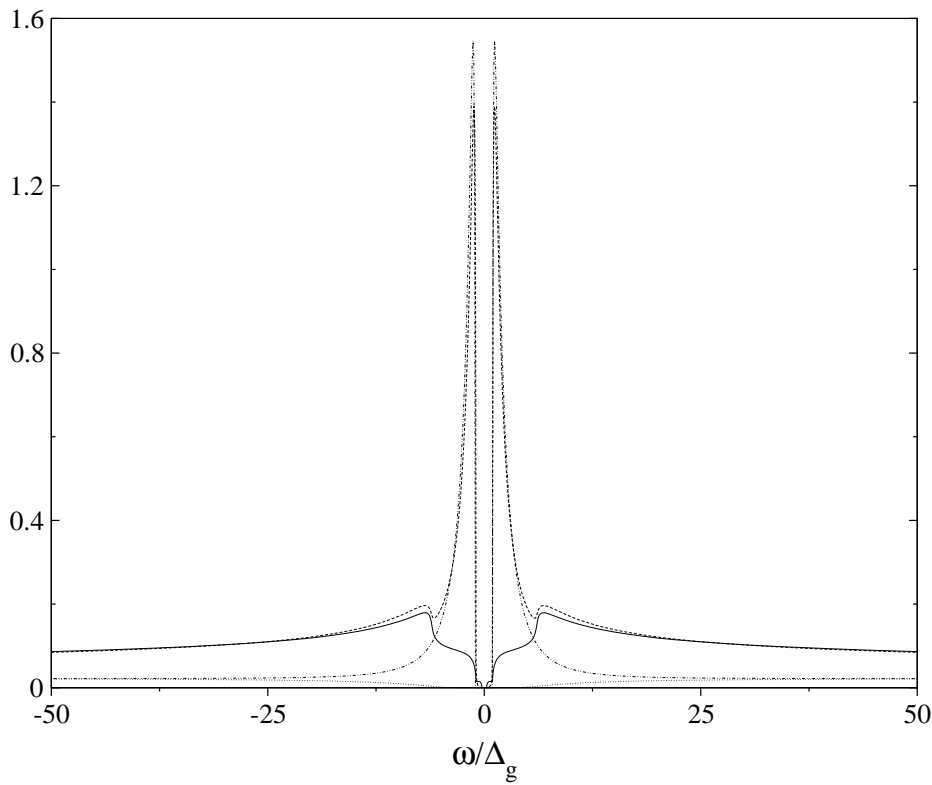}
\caption{Comparing the $\om/\Delta_g$-dependence of scattering rates
and $f$-electron spectra (BL) arising from the LMA and IPT:
$\pi \Delta_0 D^f(\om)$ (LMA: dashed line; IPT: double point-dash line)
and $\tilde{\tau}^{-1}(\omega;0$) (LMA: solid line; IPT: dotted line).
The LMA results are naturally universal; IPT results were obtained for
$U/t_*=3$ and $V/t_*=0.25$.}
\label{fig:iptlma}
\end{figure}

  Finally, explicit comparison to IPT [21,10] is made in figure 9, for the
$\tilde{\omega}$-dependence of both the $T=0$ $f$-electron spectrum
$\pi\Delta_{0}D^{f}(\omega)$ and the scattering rate $\tilde{\tau}^{-1}(\omega;0)$
(again omitting the $\delta(\tilde{\omega})$ contribution). The LMA results
shown are naturally the universal scaling forms, independent of the bare model 
parameters. Corresponding IPT results are again obtained for
$U/t_{*}=3$ and $V/t_{*}=0.25$. The figure illustrates three points.
(i) That IPT does not capture the important logarithmic tails (and hence
{\it e.g.\ }produces much reduced scattering rates, even for $T=0$ considered).
(ii) Its $f$-electron spectrum recovers, but amounts to little more than, the
limiting low-frequency quasiparticle form (as seen by comparison to figure 2 above).
(iii) The approach does not produce an exponentially small gap scale, and hence
a `clean' separation between low (meaning universal) and high  
energy scales; {\it e.g.\ }non-universal scales on the order of $\omega \sim 0.1t_{*}$
are reached by $\tilde{\omega} \sim 50$ in the IPT results shown in figure 9.
We also note that the IPT transport rate $\tilde{\tau}^{-1}(\omega; 0)$
actually vanishes for $|\tilde{\omega}| \leq 3\Delta_{g}$, rather than 
$\Delta_{g}$ itself. This in turn is related to the spurious behaviour
in the $T$-dependence of the resistivity arising from IPT, that is
discussed further in the following section.
\section{ D.C.\ transport. }

  We now consider d.c.\ transport properties of the PAM. Our aim here
is to understand the temperature dependence of the static conductivity/resistivity,
in particular over the full $\tilde{T}$-range in the strong coupling/Kondo
lattice scaling regime; as well as the connection between the high-$\tilde{T}$ 
behaviour of the PAM and that of the single-impurity Anderson model.

  As discussed in \S3 the d.c.\ conductivity $\sigma(0;T) =
\frac{1}{3}\sigma_{0}F_{\alpha}(0;T)$, with the dimensionless $F_{\alpha}(\omega;T)$
given respectively for the hypercubic and Bethe lattices by equations (3.7a,b). In the
strong coupling regime of interest the static conductivity naturally
exhibits scaling in $\tilde{T} =T/\Delta_{g}$, with no dependence on the bare system
parameters $U$, $V$, $t_{*}$. The resultant scaling form does of course
depend on the lattice type (cf \S4 for single-particle dynamics), and
$F_{\alpha}(0;T)$ for the two lattices is shown in figure 10. For the HCL the figure 
also shows (dotted) the $\tilde{T}$-dependence arising from the approximate but 
physically intuitive form equation (3.10), {\it viz} $F_{HCL}(0;T) \propto
-\int^{+\infty}_{-\infty}d\omega \frac{\partial f(\omega)}{\partial \omega}
 \tilde{\tau}(\omega;T)$ with $\tilde{\tau}(\omega;T)$ the (dimensionless) scattering 
time. Two initial points should be noted here. First that the
latter approximation, while qualitatively inadequate as $\tilde{T} \rightarrow 0$
where it fails to capture the insulating ground state, is very
good for $\tilde{T} \gtrsim 2$ or so; in fact as further evident from the inset to
figure 10 where comparison is made up to $\tilde{T} =200$, it coincides
asymptotically, and rapidly, with $F_{HCL}(0;T)$ obtained from the exact 
equation (3.7a). Second, recalling (\S3) that the constant $\sigma_{0}$ is
realistically on the order of $\sim 10^{4} - 10^{5} (\Omega cm)^{-1}$, we
note that the results of figure 10 readily encompass static conductivities
$\sigma(0;T)$ on the order of $\sim 10^{3} - 10^{4} (\Omega cm)^{-1}$ 
that are typical of Kondo insulating materials around room temperature [3-7]
(see also \S7).

  The insulating nature of the ground state 
is self-evident in figure 10; and for sufficiently low $T$ one expects 
and finds activated transport of form
\begin{equation}
\tilde{\sigma}(0; T) \propto \exp[-\Delta_{tr}/T]
\end{equation}
which defines the transport gap $\Delta_{tr}$. For the BL we find $\Delta_{tr}
= \Delta_{g}$,
while for the HCL one naturally finds
$\Delta_{tr} \propto \Delta_{g}$, but with a proportionality constant differing
from unity ($\sim 0.4$). 

  We turn now to consider the $\tilde{T}$-dependence of
the static conductivity/resistivity over the full $\tilde{T}$-range, for both the
HCL and BL; and in addition to compare the resultant behaviour, at 
high-$\tilde{T}$ in particular, to that arising for the AIM. The resistivity
for the pure impurity model, $\rho^{'}_{imp}(T) = \rho_{imp}(T)/\rho_{imp}(0)$
with $\rho_{imp}(T)$ the change in resistivity due to addition of the impurity,
is given from equation (3.11). To compare like with like in the following, comparison
of equation (3.10) for the HCL with equation (3.11) for the AIM suggests that for the HCL
we consider the dimensionless resistivity
\begin{equation}
\rho^{'}_{HCL}(T)
 = \frac{\frac{1}{2}[\rho_{0}t_{*}]^{2}}{F_{HCL}(0;T)}
\equiv \frac{1}{2\pi F_{HCL}(0;T)}
\end{equation}
(the BL counterpart of which is given below, equation (5.4)).
\begin{figure}[h]
\epsfxsize=320pt
\begin{center}
\epsffile{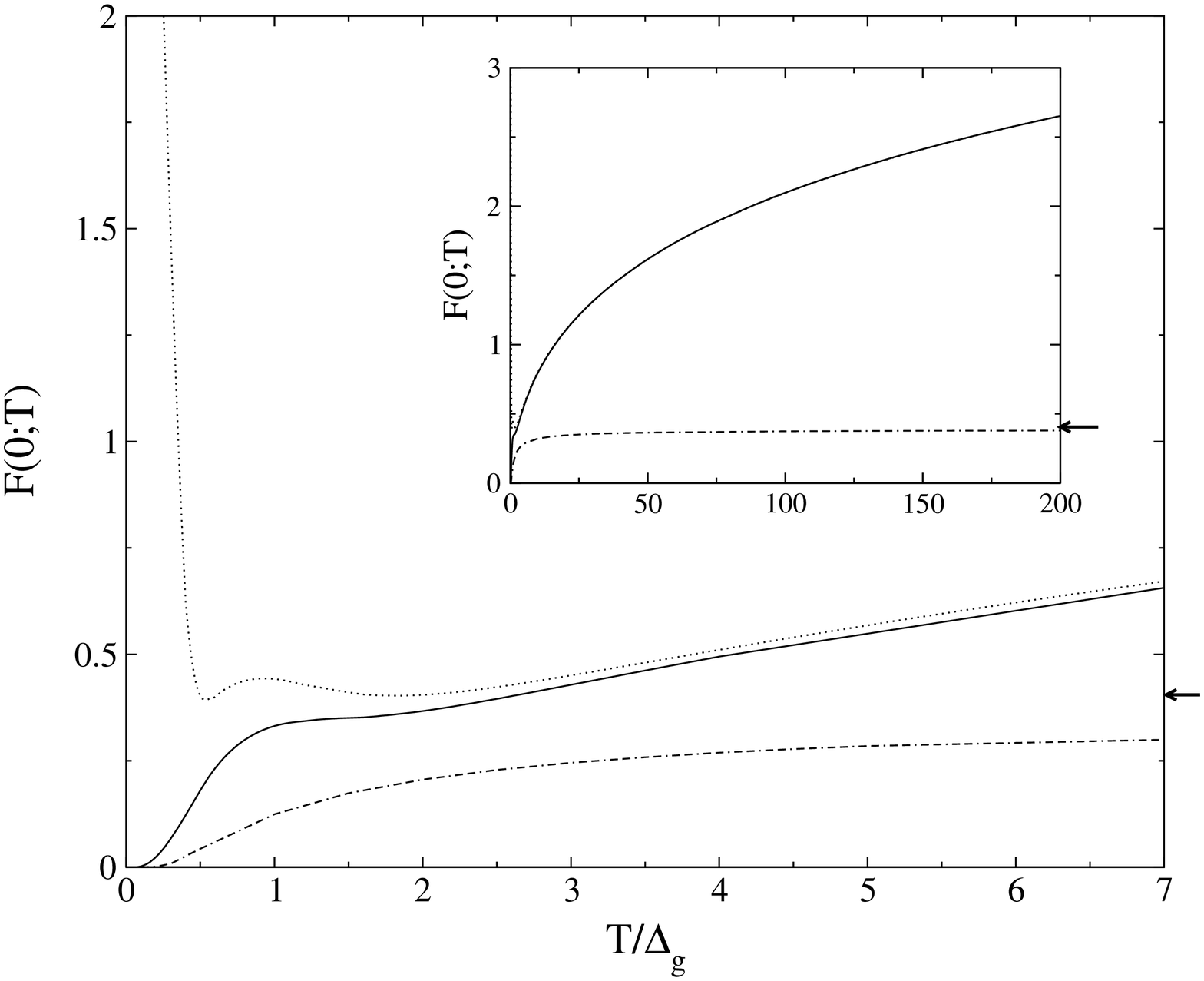}
\end{center}
\caption{Temperature dependence of the static conductivities in the 
Kondo lattice scaling regime: $F_\alpha(\om=0;T)$ {\it vs} $\tilde{T}=
T/\Delta_g$ for the HCL (solid line) and BL (point-dash line); 
the approximate equation (3.10) for the HCL is also shown (dotted line).
The high-temperature asymptote for the BL ($4/\pi^2 \simeq 0.4$) is indicated
by an arrow. Inset: results on an expanded scale out to $\tilde{T}=200$.}
\label{fig:dcc_gl}
\end{figure}

  First let us consider the important differences between the
hypercubic and Bethe lattices; which in turn reveals equally important
similarities between the two. From figure 10 it appears that with increasing
$\tilde{T}$, $F_{HCL}(0;T)$ is growing apparently unboundedly, albeit
relatively slowly; while $F_{BL}(0;T)$ is slowly saturating to a constant value 
$F_{BL}(0;T) \approx 0.4$.
This is indeed the case, and stems from the fact that in the limit
of vanishing  hybridization $V=0$ --- where the $f$-levels decouple from the
conduction band --- the BL has a non-zero residual ($T=0$) conductivity/resistivity
(in contrast of course to \it any \rm $V>0$, for which the $T=0$ conductivity
vanishes); specifically, as follows straightforwardly using equation (3.7b), that:
\begin{equation}
F_{BL}^{V=0}(0; T=0) = [\rho_{0}t_{*}]^{2} = \frac{4}{\pi^{2}}
\end{equation}
It is this, the residual conductivity of the free ($V=0$) conduction band, 
that corresponds to the ultimate high-$\tilde{T}$ constant asymptote
seen in the inset to figure 10 for the BL ($4/\pi^{2} = 0.405..$, marked by an arrow
in the Figure). 
This is further evident if one considers the temperature dependence of 
$F_{BL}(0;T)$ on `all scales', {\it i.e.\ }{\it vs }$T/t_{*}$;
as shown in figure 11 for a sequence of increasing $U/t_{*}$ in strong coupling.
Here the $T/t_{*}$-dependence of $F_{BL}^{V=0}(0;T)$ is also
shown, and is seen to constitute an upper bound to the $T/t_{*}$-dependence
of $F_{BL}(0;T)$ itself.
Since the gap $\Delta_{g}$ is exponentially small in strong coupling, the scaling
behaviour of  $F_{BL}(0;T)$ illustrated in figure 10 for finite 
$\tilde{T} = T/\Delta_{g}$ corresponds in figure 11 to exponentially small values of 
$T/t_{*} = \tilde{T}\frac{\Delta_{g}}{t_{*}}$; 
whence as seen from figure 11, 
$F_{BL}(0;T)$ is bounded above by  $F_{BL}^{V=0}(0; T=0)$ for large-$\tilde{T}$
in the scaling regime (which corresponds formally to any finite $\tilde{T}$ in the
limit $\Delta_{g} \rightarrow 0$).
\begin{figure}[h]
\epsfxsize=350pt
\centering
\epsffile{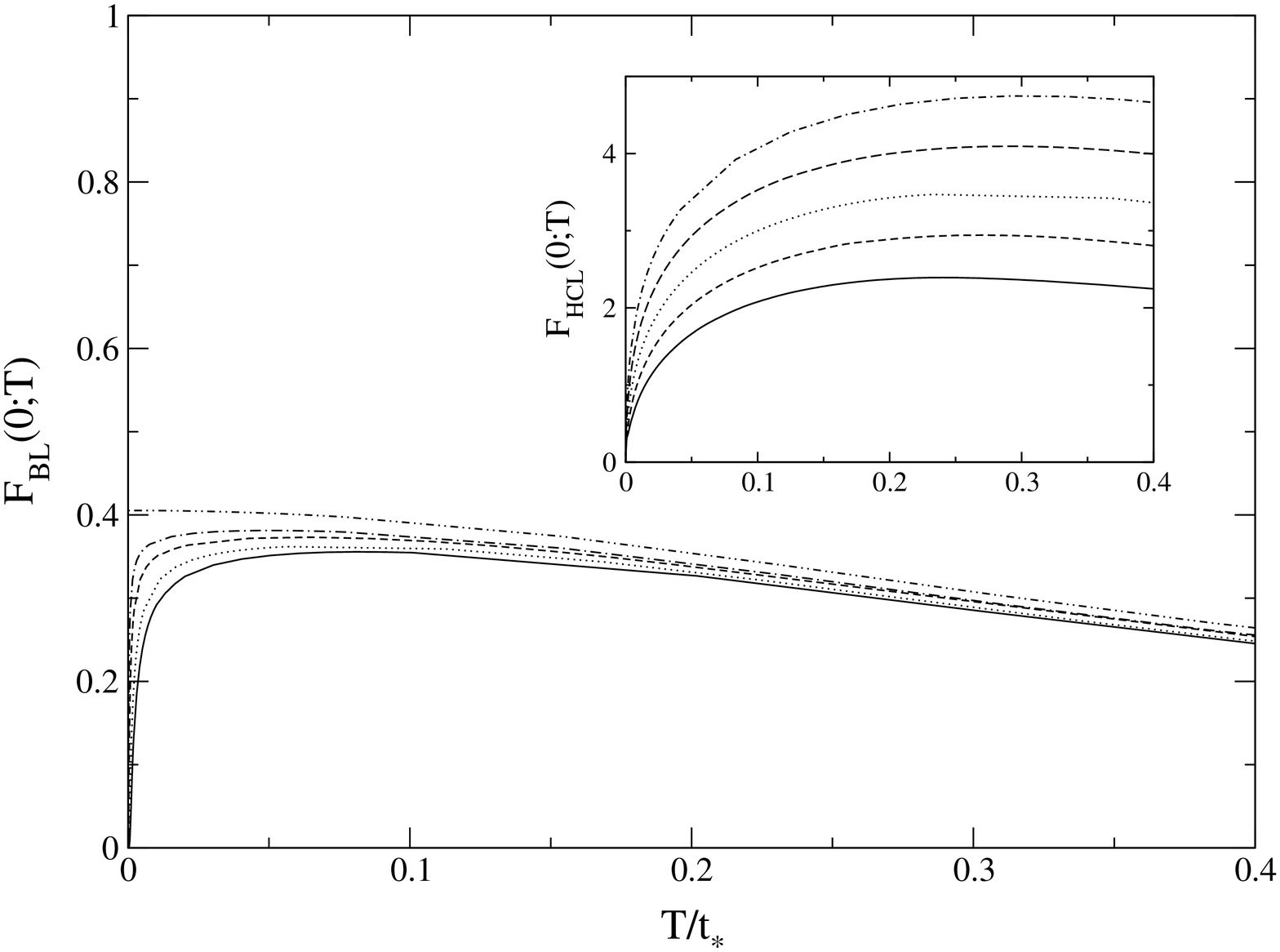}
\caption{Temperature dependence of static conductivities on `all scales',
{\it vs} $T/t_*$: $F_{BL}(0;T)$ for $V^2/t_*^2=0.2$ and $U/t_*$=4.6 (solid),
5.1 (dotted), 6.1 (dashed) and 7 (point-dash). The $V=0$ limit
$F_{BL}^{V=0}(0;T)$ is also shown (double point-dash). Inset: Corresponding
results for $F_{HCL}(0;T)$ for  $U/t_*$=4.6 (solid), 5.1 (short dash), 5.6
(dotted), 6.1 (long dash) and 6.6 (point dash); in this case, 
$F_{HCL}^{V=0}(0;T) = \infty$.}
\label{fig:dcc_gl_allsc}
\end{figure}

  The latter behaviour is readily understood in physical terms, since 
in the scaling regime the gap in the conduction band $D^{c}(\omega)$ rapidly 
fills in with increasing $\tilde{T}$ and approaches the free $V=0$ conduction
band (see {\it e.g.\ }figures 4b,5b). For $\tilde{T} \gg 1$ the gap $\Delta_{g}$ is 
thus in essence irrelevant, {\it i.e.\ }might as well be zero --- which is of
course just the $V=0$ limit, with its characteristic residual 
resistivity for the BL. For the HCL by contrast the residual resistivity for $V=0$ 
is precisely zero; reflecting the infinite conductivity associated with the coherent 
Bloch states that in this case arise when $V=0$. So for the HCL by the same
reasoning, one does not therefore expect the conductivity \it in the scaling
regime \rm to saturate for $\tilde{T} \gg 1$, as indeed evident in figure 10
(see also figure 11 (inset) where the $T/t_{*}$-dependence of $F_{HCL}(0;T)$
is shown for a sequence of increasing $U/t_{*}$, as well as figure 12 
below). We believe this conclusion to be rather
general: if for whatever reason the free ($V=0$) conduction band is characterised
by a non-vanishing residual conductivity/resistivity, be it due {\it e.g.\ }to the
intrinsic nature of the free conduction band or to the presence of disorder,
then we expect the conductivity in the scaling regime to approach asymptotically
this limiting value for $\tilde{T} \gg 1$.

  To take the above into account when comparing the two lattices it is natural
to subtract from the resistivity any finite high-$\tilde{T}$ asymptote, as
embodied in $1/F_{\alpha}(0;T) - 1/F_{\alpha}^{V=0}(0;0)$ (a redundant operation
for the HCL where $F_{HCL}^{V=0}(0;0) = \infty$);
considering therefore --- in obvious parallel to the AIM ---  the `change in
resistivity due to coupling the $f$-levels to the host conduction band'.
Specifically, for the BL we consider below the following dimensionless
resistivity:
\begin{equation}
\rho_{BL}^{'}(T) = \frac{F_{BL}^{V=0}(0;0)}{F_{BL}(0;T)} -1 ~~~
\equiv \frac{4}{\pi^{2}F_{BL}(0;T)} -1
\end{equation}

\begin{figure}[h]
\epsfxsize=370pt
\centering
\epsffile{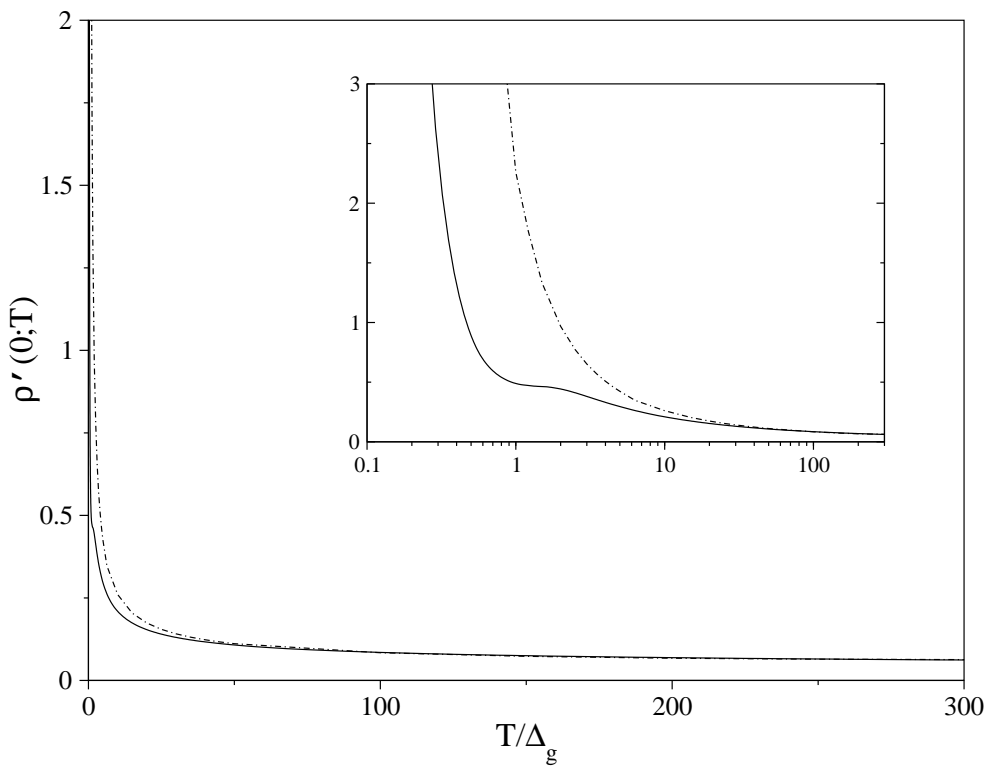}
\caption{ Resistivities in the Kondo lattice scaling regime:
$\rho_{HCL}^\prime(T)$ (equation (5.2), solid line) and $\rho_{BL}^\prime(T)$
(equation (5.4), point-dash line) {\it vs} $\tilde{T}=T/\Delta_g$. Inset:
same results on a logarithmic $\tilde{T}$ scale.}
\label{fig:scres}
\end{figure}

  We have just highlighted the differences between the HCL and BL. But their
important similarities are evident in figure 12, where we show the resultant 
scaling behaviour of the resistivities $\rho^{'}_{HCL}(T)$ (equation (5.2)) and
$\rho^{'}_{BL}(T)$ (equation (5.4)) as a function of $\tilde{T} = T/\Delta_{g}$
out to $\tilde{T} \simeq 300$; the inset shows the same results on a log scale,
indicating that the $\rho^{'}(T)$'s ultimately vanish as $\tilde{T} \rightarrow
\infty$.
Two immediate features are apparent. First that the high-$\tilde{T}$ asymptotics 
of the $\rho^{'}(T)$'s coincide (whence re figure 10, the rate at which the BL
conductivity $F_{BL}(0;T)$ asymptotically approaches its saturation value is the 
same as that with which its HCL counterpart grows unboundedly); the leading asymptotics
is obtained explicitly below. In fact for $\tilde{T} \gtrsim 5-10$ or so,
the behaviour of the two lattices is barely distinguishable. Second, cursory
inspection of figure 12 might suggest that the resistivities $\rho^{'}(T)$
are themselves plateauing with increasing $\tilde{T}$, since they 
change only slightly over a large temperature range out to many multiples 
of the gap. This interpretation, while rather natural at first sight, is not 
correct since  the $\rho^{'}(T) \rightarrow 0$ as 
$\tilde{T} \rightarrow \infty$. But it is indicative of the very slow
$\tilde{T}$-decays involved; and for which, see \S7, we believe 
there is experimental evidence in Kondo insulating materials.

  We now obtain the leading high-$\tilde{T}$ behaviour of the $\rho^{'}(T)$,
here considering explicitly $\rho^{'}_{HCL}(T)$. Precisely the same result
(equation (5.8) below) is obtained for $\rho^{'}_{BL}(T)$, which separate calculation
is outlined in the Appendix. From equation (5.2) together with equation (3.10) (which
is asymptotically valid at large-$\tilde{T}$ as illustrated in figure 10),
\begin{equation}
\frac{1}{\rho^{'}_{HCL}(T)} \sim
-\int^{\infty}_{-\infty} d\omega ~ \frac{\partial f(\omega;T)}{\partial\omega} ~
\tilde{\tau}(\omega;T)
\end{equation}
where the scattering time $\tilde{\tau}(\omega;T)$ is given by equation (3.4) in
terms of the $f$-electron self-energy $\Delta_{0}^{-1}\Sigma_{f}(\omega;T)$.
The latter is given for $|\tilde{\omega}| \gg max(1,\tilde{T})$ by equation (4.3);
the generalization of which to $\tilde{T} \gg 1$ and any $|\tilde{\omega}|$
(holding also for the BL) is [32]
\begin{equation}
\Delta_{0}^{-1}\Sigma_{f}^{I}(\omega;T) \sim
\frac{16}{3\pi^{2}} ln^{2}[a\sqrt{|\tilde{\omega}|^{2} + (\pi\tilde{T})^{2}}]
\end{equation}
with $\Delta_{0}^{-1}\Sigma_{f}^{R}(\omega;T) \propto
[\Delta_{0}^{-1}\Sigma_{f}^{I}(\omega;T)]^{1/2}$ (such that
it may be neglected in equation (3.4) for
the scattering time). Equation (3.4) thus yields
\begin{equation}
\tilde{\tau}(\omega;T) \sim \Delta_{0}^{-1}\Sigma_{f}^{I}(\omega;T) \sim
\frac{16}{3\pi^{2}}ln^{2}(\tilde{T})L(y;\tilde{T})
\end{equation}
where $y= \tilde{\omega}/\tilde{T}$, and $L(y;\tilde{T}) =
[1+ln[a\sqrt{\pi^{2}+y^{2}}]/ln(\tilde{T})]^{2}$ such that 
$L(y;\tilde{T}) \rightarrow 1$ as $\tilde{T} \rightarrow \infty$ for 
finite $y$. Using equation (5.7) in equation (5.5), and changing variables therein
from $\omega$ to $y$, gives directly the leading high-$\tilde{T}$ behaviour
of $\rho^{'}_{HCL}(T)$ as
\begin{equation}
\rho^{'}(T) \stackrel{\tilde{T} \gg 1}{\sim}
\frac{3\pi^{2}}{16ln^{2}(\tilde{T})}
\end{equation}
(which as above also holds for $\rho^{'}_{BL}(T)$, see Appendix).

  Equation (5.8) reflects the anticipated connection between the PAM at
high-$\tilde{T}$ and the Anderson impurity model; being also the exact
high-temperature asymptote for the strong coupling AIM (with Kondo temperature
$\propto \Delta_{g}$) [2], first obtained
[44] for the Kondo model from the leading logarithmic sum of parquet diagrams.
That connection is seen more generally in figure 13, where the full temperature
dependence of the HCL resistivity is compared directly to corresponding LMA
results for the AIM (analogous comparison for
the BL is clear from figure 12). Specifically we show $\rho^{'}_{HCL}(T)$
\it vs \rm $\tilde{T} = T/\Delta_{g}$ (with $\Delta_{g} = ZV^{2}/t_{*}
\equiv \Delta_{0}Z/\sqrt{\pi}$); compared to the impurity resistivity
$\rho^{'}_{imp}(T) = \rho_{imp}(T)/\rho_{imp}(0)$
{\it vs} $\tilde{T} = T/\Delta_{g}^{imp}$ (where $\Delta_{g}^{imp} \equiv
\Delta_{0}Z_{imp}/\sqrt{\pi}$ with $Z_{imp}$ the AIM quasiparticle weight).
The LMA $\rho^{'}_{imp}(T)$ is detailed in [32] where it is shown to agree rather well
with NRG calculations for the AIM [45], to be asymptotically exact at high-$\tilde{T}$
(equation (5.8) being recovered); to agree well with the Hamann 
approximation [46] (obtained
by further parquet resummation) in the latter's regime of applicability down
to $\tilde{T} \sim 1$, and to cross over smoothly to the AIM Fermi liquid form
$\rho^{'}_{imp}(T) - 1 \propto -\tilde{T}^{2}$ as $\tilde{T} \rightarrow 0$.
As evident in figure 13 the PAM resistivity $\rho^{'}_{HCL}(T)$, which exhibits activated
insulating behaviour (equation (5.1)) for $\tilde{T} \lesssim 0.5$ or so,  
progressively diminishes with increasing $\tilde{T}$ and for $\tilde{T} \gtrsim 1-2$
essentially coincides with that for the AIM.
\begin{figure}[h]
\epsfxsize=350pt
\centering
\epsffile{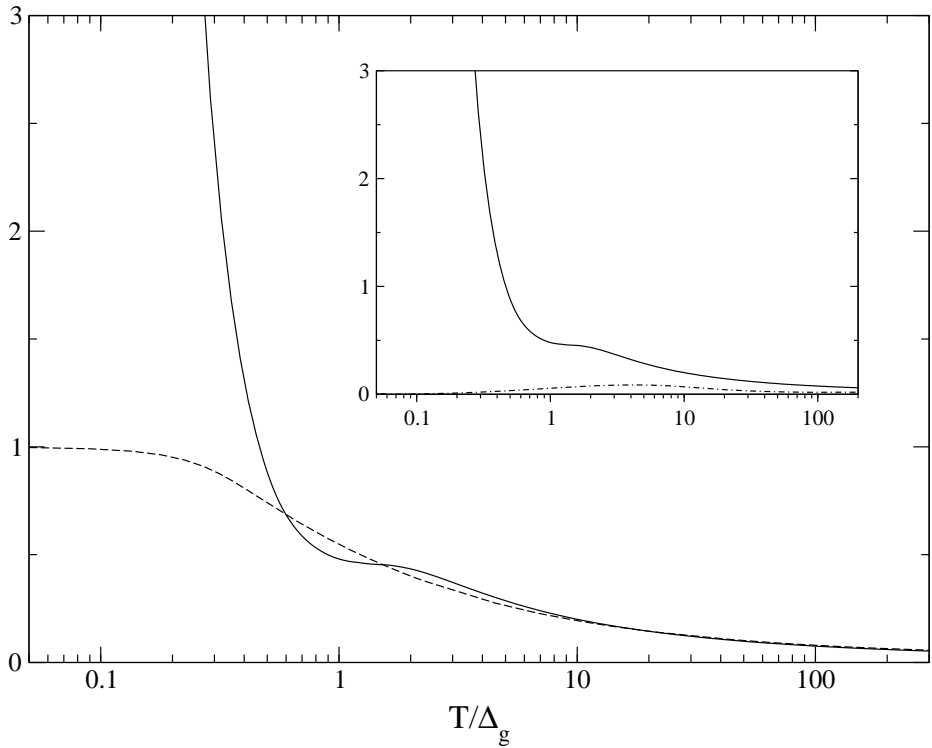}
\caption{ Comparison of resistivities for the HCL and Anderson impurity
model: $\rho^\prime_{HCL}(T)$ (solid line) and $\rho^\prime_{imp}(T)$
(dashed line) {\it vs} $\tilde{T}=T/\Delta_g$ on a log-scale.
Inset: comparison of $\rho^\prime_{HCL}(T)$ arising from the local
moment approach (solid line) and IPT (point-dash line).}
\label{fig:dc_si_pam}
\end{figure}

  The above results are in marked contrast to those arising from iterated
perturbation theory [21,10], the behaviour of which is qualitatively
wrong.  IPT results for $\rho^{'}_{HCL}(T)$ \it vs \rm
$\tilde{T}$ are compared to the LMA in the inset to figure 13. 
The IPT calculations were performed for $U/t_{*}=3$ and $V/t_{*} =0.25$ (as 
employed for the BL in \S4.1); and the inability of IPT to 
recover an exponentially small gap scale means that non-universal temperatures
on the order of $T \sim 0.1t_{*}$ are reached by $\tilde{T} \sim 60$. 
It is evident
from figure 13 that the IPT resistivities are generally much less than their
LMA counterparts; and no hint of logarithmic $\tilde{T}$-decays are evident
(which mirrors the absence of logarithmic tails in the 
$\tilde{\omega}$-dependence of the IPT single-particle spectra and scattering
rates, \S4). Most significantly however, we see that the IPT resistivity
actually \it vanishes \rm as $\tilde{T} \rightarrow 0$; rather than exhibiting 
the correct divergence symptomatic of the insulating ground state. At first sight
this is surprising, since IPT is qualitatively correct in predicting a $T=0$ gap 
in the single-particle spectra $D^{c/f}(\omega)$ [21,10]; we have thus examined the
problem in some detail. Its physical origins stem from the fact that 
the imaginary part of
the IPT $T=0$ self-energy, $\Sigma_{f}^{I}(\omega;0)$, is zero for $|\omega| <
3\Delta_{g}$ rather than for $|\omega| \lesssim \Delta_{g}$ as one might expect 
(where `zero' strictly means exponentially small for the HCL with its soft-gap, 
although that is irrelevant in the following); and in consequence the scattering 
rate $\tau^{-1}(\omega;0)$ (see equation (3.4)) likewise vanishes for
$0 < |\omega| < 3\Delta_{g}$. But for 
$|\omega| \gtrsim \Delta_{g}$, there is a high density of conduction electron 
states embodied in $D^{c}(\omega)$ (see {\it e.g.\ }figures 2 or 4). 
Hence states in the interval 
$\Delta_{g} \lesssim |\omega| \lesssim 3\Delta_{g}$ are
essentially  \it unscattered\rm; and it is this
that, on initially increasing $T$ from zero (and thus accessing states
in this interval), leads to an arbitrarily large conductivity and hence vanishing 
of $\rho^{'}_{HCL}(T)$ as $T \rightarrow 0$. This can be demonstrated numerically
in several ways ({\it e.g.\ }by `cutting off' the Fermi functions 
$\partial f(\omega;T)/\partial \omega$ entering $\rho^{'}_{HCL}(T)$
in the above $|\omega|$ interval); but the physical argument above is the basic 
origin of the problem. This spurious behaviour does not arise
within the LMA, for which $\Sigma_{f}^{I}(\omega;0)$ and the single-particle
spectra are non-vanishing in the \it same \rm $|\omega|$ intervals. The problem
can be circumvented within IPT by \it ad hoc \rm addition of a 
finite-$\eta$ in frequency factors $\omega^{+} = \omega + i\eta$ (to mimic
an additional white-noise inelastic scattering rate); as has been employed
in a weak coupling IPT study [47]. In this case we find the ultimate low-$T$ behaviour
of the resistivity to be of form $\rho^{'}_{HCL}(T) \sim \eta exp(b\Delta_{g}/T)$
(with $b \sim 0.4$); but with a crossover temperature scale to such
activated behaviour that is entirely determined by $\eta$ and vanishes as
$\eta \rightarrow 0$.

\section{Optical conductivities.}

  The natural progression is now to consider the optical conductivities
$\sigma(\omega;T) =\frac{1}{3}\sigma_{0} F_{\alpha}(\omega;T)$),
with the $F_{\alpha}(\omega;T)$ given explicitly for the hypercubic and
Bethe lattices by equations (3.7). Before proceeding to specific results arising
from the LMA, we consider a basic question regarding the optical gap:
what is it? While strictly an issue at $T=0$ since all gaps are technically
destroyed for $T>0$, this of course has major ramifications for 
both the frequency and temperature dependence of the optical conductivity.

  We note at the outset that our most basic conclusion  here
differs qualitatively from the work of [21,15]. It has been argued
hitherto (see {\it e.g.\ }figure 2 of [21]) that the optical gap in
$\sigma(\omega;0)$
corresponds to the minimum \it direct gap \rm $\Delta_{dir}$ of the renormalized band 
structure, as opposed to the \it indirect gap \rm
$\Delta_{ind} \propto \Delta_{g} =ZV^{2}/t_{*}$
that is manifest in the single-particle spectra $D^{c}(\omega)$ or $D^{f}(\omega)$;
a conclusion that in turn underpins the interpretation of experiment given in
[21] (see also [6]). We do not believe this to be correct on general grounds,
neither do we find it supported by the present theory. By contrast we show,
in agreement with the qualitative conclusions of [7], that : 
(i) the optical gap corresponds to the indirect gap (albeit that the direct 
gap scale is obviously also manifest in the $\omega$-dependence of the optical 
conductivity); and (ii) that it is the indirect gap which sets the temperature
scale for the thermal evolution of the optical conductivity in the Kondo lattice 
regime, in particular the `filling' of the optical gap with increasing 
temperature (which
we regard as entirely natural, since in strong coupling $\Delta_{g}$ is 
the characteristic low-energy scale of the system).

  It is first necessary to explain why we disagree in particular with the conclusions
of [21]. As discussed in \S2.1 (see also [2,10,36]) the limiting low-frequency behaviour
of the single-particle Green functions amounts to a renormalization of the
non-interacting limit; arising  from the exact equations (2.5) by neglecting the imaginary
part of the $f$-electron self-energy $\Sigma_{f}^{I}(\omega;0)$ and replacing
($Re\Sigma_{f}(\omega;0) \equiv$)
$\Sigma_{f}^{R}(\omega;0)$ by its leading low-$\omega$ behaviour, {\it viz}
\begin{equation}
\Sigma_{f}(\omega;0) \sim \Sigma_{f}^{R}(\omega;0) \sim - (\frac{1}{Z} -1)\omega
\end{equation}
with $Z$ the quasiparticle weight. This leads directly to the quasiparticle
behaviour embodied in equations (2.20), {\it e.g.\ }$G^{c}(\omega) \sim g_{0}^{c}(\omega;ZV^{2})$
with $g_{0}^{c}(\omega;V^{2})$ referring to the $U=0$ limit; whence the
lowest-$\omega$ behaviour of $D^{c}(\omega)$ (and likewise $D^{f}(\omega)$)
is a simple quasiparticle renormalization of the non-interacting limit, with
$V^{2} \rightarrow ZV^{2}$. In particular (\S2.1) the gap in the
single-particle spectra is generically preserved. This is the indirect 
gap scale defined by,
\begin{equation}
\Delta_{ind} = 2\Delta_{g} = 2Z\frac{V^{2}}{t_{*}}
\end{equation}
which is indeed manifest in the single-particle spectra (the 2 simply reflecting
the `full' gap, as opposed to that measured from the Fermi level at mid-gap,
$\omega =0$). The quasiparticle form for the single-particle propagators is not
approximate: it must be satisfied asymptotically at
sufficiently low frequencies, reflecting as it does Fermi liquid behaviour in
the sense of adiabatic continuity to the non-interacting limit; and indeed it
is, as illustrated in figure 2 (see also [36]). We emphasise however that this
quasiparticle behaviour is confined to the lowest frequency scales 
$|\tilde{\omega}| = |\omega|/\Delta_{g}$, up to at most a few times the
(indirect) gap $\Delta_{g} \propto Z$; beyond which, as detailed in \S4 and
[36], this simple picture no longer holds.

  The arguments above may be extended --- but with a wholly different
validity --- to the optical conductivity; it is this that underlies the
conclusion of [21]. From equations (3.7), the optical conductivities are determined
by the $D_{c}(\epsilon;\omega) = -\frac{1}{\pi}ImG^{c}(\epsilon;\omega)$; where
(\S3) $G^{c}(\epsilon;\omega) = [\gamma(\omega) - \epsilon]^{-1}$ with
$\gamma (\omega) = \omega^{+} -V^{2}[\omega^{+}-\Sigma_{f}(\omega;T)]^{-1}$.
If the asymptotic form equation (6.1) for $\Sigma_{f}(\omega;0)$ is again employed
here, then $\gamma(\omega) \approx \omega^{+}-\frac{ZV^{2}}{\omega^{+}}$ and hence
$G^{c}(\epsilon;\omega) \approx [\omega^{+}- \frac{ZV{2}}{\omega^{+}}
-\epsilon]^{-1}$ which again amounts to a renormalization of the non-interacting
limit. The two branches characteristic of this renormalized band structure then
follow from the zeros of the approximate $[G^{c}(\epsilon;\omega)]^{-1}$; being
given by $\omega_{\pm}(\epsilon) = \frac{1}{2}[\epsilon \pm
\sqrt{\epsilon^{2} + 4ZV^{2}}]$, with the $\epsilon$-dependent
gap function $\Delta_{d}(\epsilon) = \omega_{+}(\epsilon) -
\omega_{-}(\epsilon)$. The minimum such gap occurs for $\epsilon =0$; this is the
(approximate) direct gap,
\begin{equation}
\Delta_{dir} \simeq 2\sqrt{Z}V
\end{equation}
(which we add in passing is radically different from the indirect gap equation (6.2)
in strong coupling where $Z \ll 1$, since $\Delta_{dir}/\Delta_{ind}
\propto Z^{-1/2}$).
If now one interprets the optical conductivity of the hypercubic lattice (given
from equation (3.7a)) in terms of a naive picture of renormalized interband transitions,
then since $\Delta_{dir}$ above corresponds to the minimum energy for such, one
would clearly expect it to vanish for $\omega < \Delta_{dir}$; concluding [21] therefore
that the optical gap coincides with the direct gap, while the indirect gap is by
contrast manifest only in the single-particle spectra, see {\it e.g.\ }figure 2 of [21]. 
And indeed, if the above approximation $G^{c}(\epsilon;\omega) \approx
[\omega^{+} - \frac{ZV^{2}}{\om^{+}} -\epsilon]^{-1}$ is used explicitly in equation (3.7a),
one finds directly that $F_{HCL}(\omega;0) =0$  for all $\omega \leq 
\Delta_{dir} = 2\sqrt{Z}V$ (and in fact likewise for $F_{HCL}(\omega;T)$ at 
any $T$).

  The above argument is certainly correct for the \it non-interacting \rm 
limit of the PAM, which is characterised strictly by simple one-electron states.  
Despite its superficial appeal however, it is incorrect beyond the
confines of this limit: for it neglects completely [7] the effects 
of scattering arising
from electron interactions, as embodied in the scattering rates 
$\tau^{-1}(\omega;T) \equiv \gamma_{I}(\omega)$ (equation (3.3)) considered 
explicitly in \S4.1; or equivalently (see equation (3.4)) in the imaginary part of the
$f$-electron self-energy $\Sigma_{f}^{I}(\omega;T)$, which also controls the
single-particle dynamics (\S2,4). The effect of this many-body scattering on the 
optical conductivity may be inferred from equation (3.7a), from which it is straightforward
to prove that if $\Sigma_{f}^{I}(\omega;0)$ is non-zero for $|\omega| > \lambda$,
then $F_{HCL}(\omega;0)$ is non-zero for $\omega > 2\lambda$. 
But the scale $\lambda$ above which $\Sigma_{f}^{I}(\omega;0)$ 
effectively (HCL) becomes non-zero is of course the single-particle gap scale
$\Delta_{g}$, whence (equation (6.2)) the optical gap corresponds to the \it indirect \rm gap 
$\Delta_{ind}$. Four further points should be noted here. (a) We emphasize that this
qualitative conclusion is not dependent on the LMA, although the latter indeed
gives $\lambda \sim \Delta_{g}$. In fact it arises also within IPT for which,
as discussed in \S5, $\Sigma_{f}(\omega;0)$ is non-vanishing for
$|\omega| \gtrsim 3\Delta_{g}$ (the central point again being that 
$\lambda \propto \Delta_{g}$).
(b) The above conclusion is certainly
consistent (\S5) with a transport gap (equation (5.1)) $\Delta_{tr} \sim \Delta_{g}$ in the
static  conductivity: the latter is simply the $\omega =0$ limit of the optical
conductivity, and at the very least it is natural to expect the optical gap to be
proportional to the low-temperature gap scale for static transport. 
(c) Recall that the scaling regime characteristic of the strong coupling Kondo
lattice (where the quasiparticle weight $Z \rightarrow 0$) corresponds to any 
finite $\tilde{\omega} = \omega/\Delta_{g}$ in the formal limit 
$\Delta_{g} \propto Z \rightarrow 0$. But if the optical gap corresponded to the 
direct gap equation (6.3), this would require $\Sigma_{f}^{I}(\omega;0) =0$ for all 
$|\omega| \leq \Delta_{dir} \propto Z^{1/2}$. Since such a frequency range 
clearly encompasses (and goes well beyond) the $\tilde{\omega}$-scaling regime,
this would entail \it vanishing \rm
scattering rates throughout the scaling regime; which is physically untenable.
(d) Our discussion of the optical gap has in large part focussed implicitly on 
$F_{HCL}(\omega;0)$ for the hypercubic lattice. But the same conclusion
arises (essentially trivially) for the Bethe lattice $F_{BL}(\omega;0)$: from
equation (3.7b)
\begin{equation}
F_{BL}(\omega;0) = \frac{t_{*}^{2}}{\omega}
 \int^{0}_{-\omega} d\omega_{1} ~
D^{c}(\omega_{1})D^{c}(\omega_{1}+\omega)
\end{equation}
where, since (\S4) the conduction electron spectrum has a gap $\Delta_{g}$ (measured
from the Fermi level),
the optical gap follows immediately as $\Delta_{ind} = 2\Delta_{g}$. This
suggests in addition that we should expect the optical properties of the HCL
and BL to be qualitatively similar, at least for sufficiently low frequencies
and/or temperatures. That is indeed the case, as we show below.
\begin{figure}[h]
\epsfxsize=400pt
\centering
\epsffile{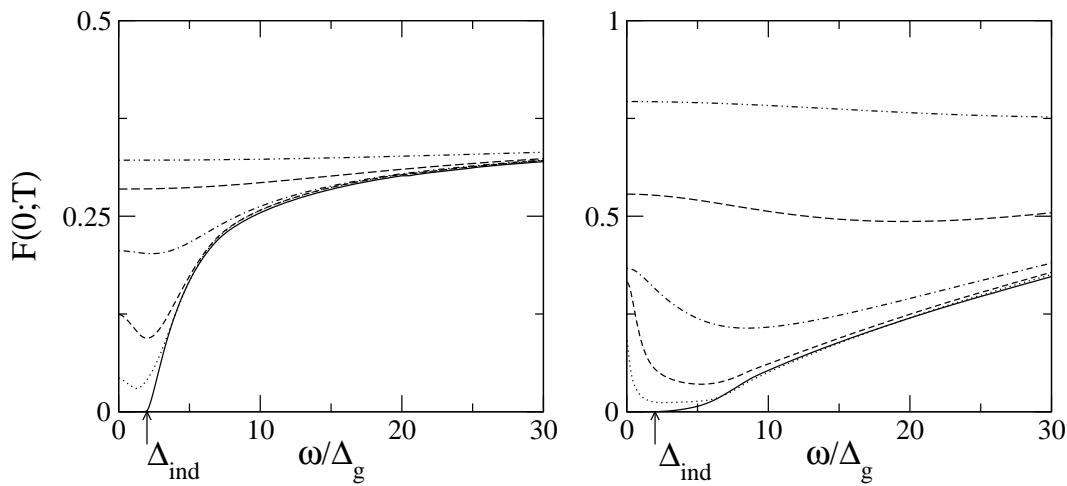}
\vspace{0.1cm}
\caption{Optical conductivities in the scaling regime: $F_{BL}(\omega;T)$
(left) and $F_{HCL}(\omega;T)$ (right) {\it vs} $\om/\Delta_g$ for
$\tilde{T}=T/\Delta_g$=0 (solid), 0.5 (dotted), 1 (short dash),
2 (point-dash), 5 (long dash) and 10 (double point-dash). The optical gap
$\Delta_{ind}=2\Delta_g$ is indicated in each case.}
\label{fig:opt_gl}
\end{figure}

  Figure 14 shows the resultant LMA optical conductivities $F_{\alpha}(\omega;T)$
(equations (3.7)) for both the BL and HCL, as a function of $\tilde{\omega} =
\omega/\Delta_{g}$ and for a range of temperatures $\tilde{T} = T/\Delta_{g}$.
These are the universal forms, scaling in terms of 
$\tilde{\omega}$ and $\tilde{T}$ with no dependence on bare material parameters
[and by way of orientation we add that for gaps $\Delta_{g}$
in the range $\sim 10K - 300K$, $\tilde{\omega} = 30$
would correspond to frequencies $\omega$ in the range $\sim 200 - 6000 cm^{-1}$].
The indirect gap $\Delta_{ind}$ ($ = 2\Delta_{g} = 2ZV^{2}/t_{*}$) is marked 
on the figures and is indeed seen to be the ($T=0$) optical gap.
The $T=0$ optical conductivity is `sharper' in an obvious sense for the BL
than the HCL, but in either case it is again the indirect gap that sets the
scale for thermal destruction of the gap: by $\tilde{T} = T/\Delta_{g} \gtrsim 5$
or so, the gap is well filled in and the optical conductivity essentially
constant over the $\tilde{\omega}$-range shown. These are of course characteristic
features of experimental Kondo insulators [3-7], as discussed for specific materials
in \S7. Here we simply add that $F_{\alpha}(\omega;T)$'s on the order of
$\sim 0.25 - 0.75$ lead to absolute conductivities $\sigma(\omega;T)
= \frac{1}{3}\sigma_{0}F_{\alpha}(\omega;T)$ in the range $\sim 2500 -7500
(\Omega cm)^{-1}$ (taking a typical $\sigma_{0}$ (\S3) of $\sim 3.10^{4}
(\Omega cm)^{-1}$); values that are typical for Kondo insulators around room
temperature [3-7].

  Figure 14 also shows that on initially increasing $\tilde{T}$ from $0$, the
optical conductivity acquires a Drude-like peak centred on $\omega =0$, which
broadens with increasing $\tilde{T}$ and is subsequently destroyed as the gap
is filled in; as also seen in a QMC study of the HCL [15] (discussed
further below). This is more clearly evident in figure 15, where the thermal
evolution of the Drude peak for the HCL is shown for $\tilde{T} = 0.4, 0.8$
and $1.2$. The figure also makes comparison to a Lorentzian form for the Drude
peaks; the quality and range of which fit (and it is merely that) are self-evident.
\begin{figure}[h]
\epsfxsize=350pt
\centering
\epsffile{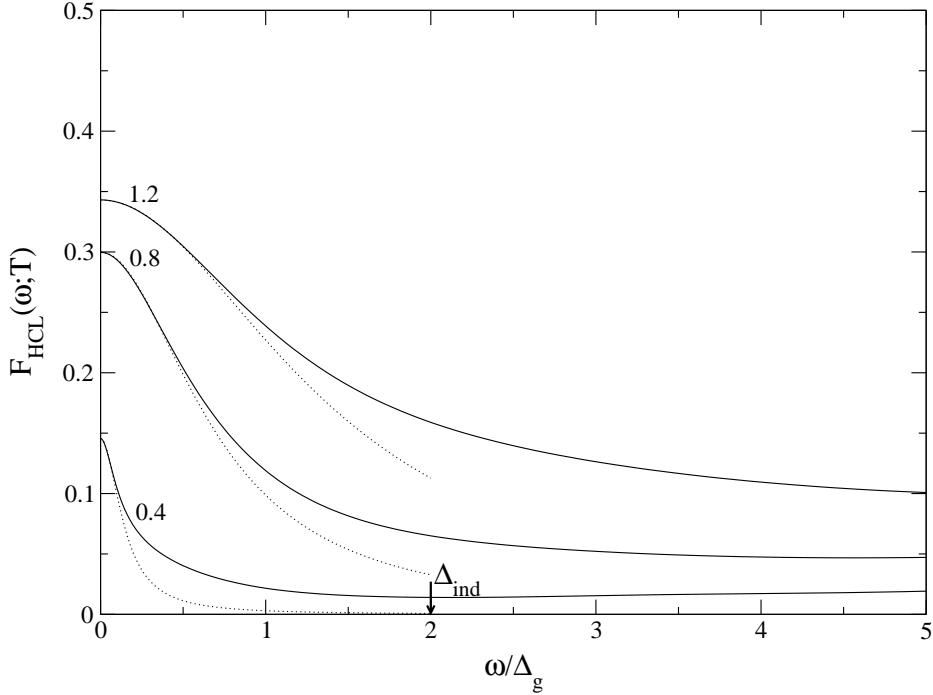}
\caption{Drude-like peaks in the low-$(\tilde{\om},\tilde{T})$ optical
conductivity: $F_{HCL}(\om;T)$ {\it vs} $\om/\Delta_g$ for $\tilde{T}
=T/\Delta_g$=0.4, 0.8 and 1.2. Lorentzian fits to the 
data (dotted lines) are also shown.}
\label{fig:drude}
\end{figure}

\begin{figure}[t]
\epsfxsize=290pt
\centering
\epsffile{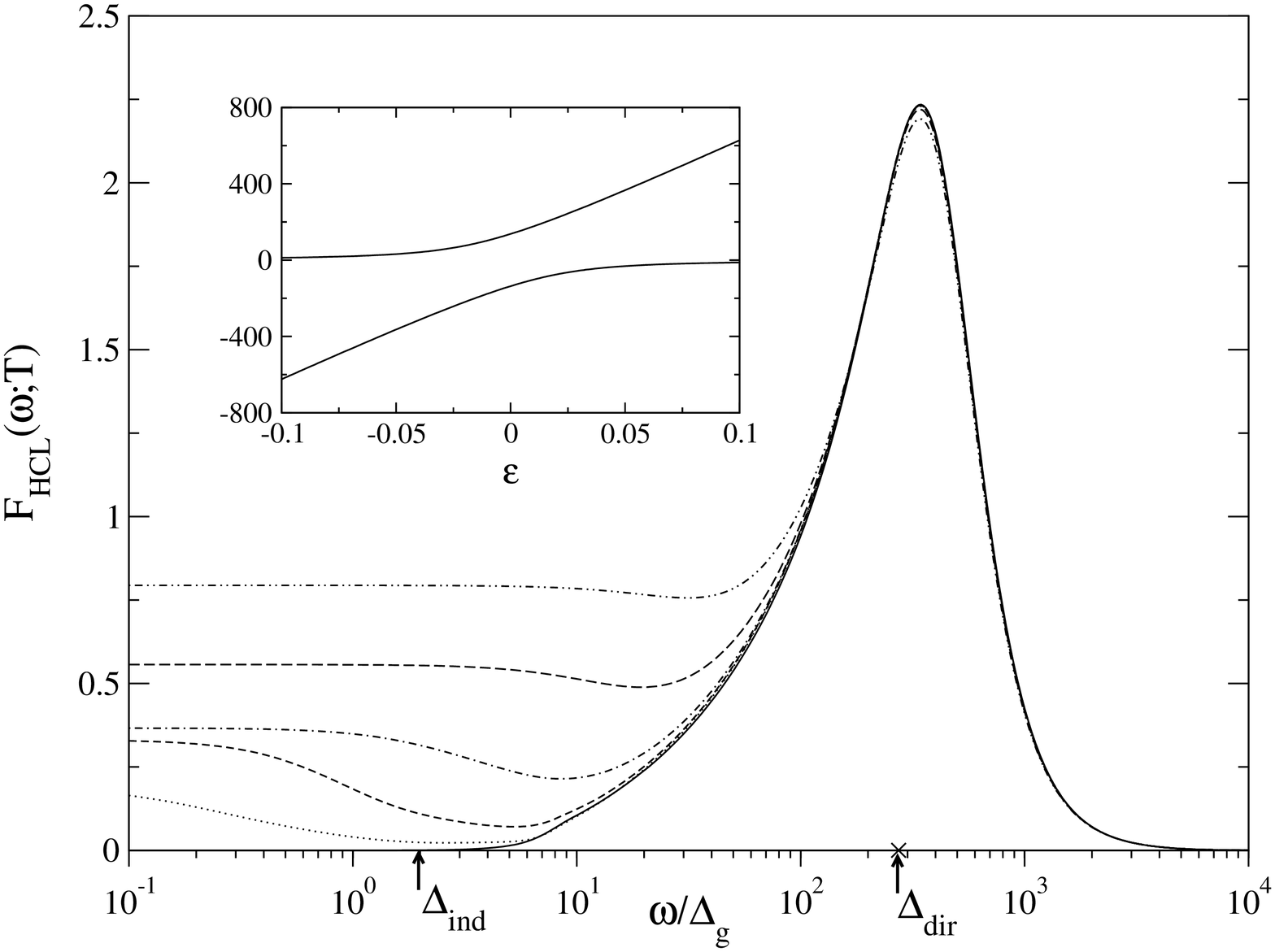}
\epsfxsize=290pt
\centering
\epsffile{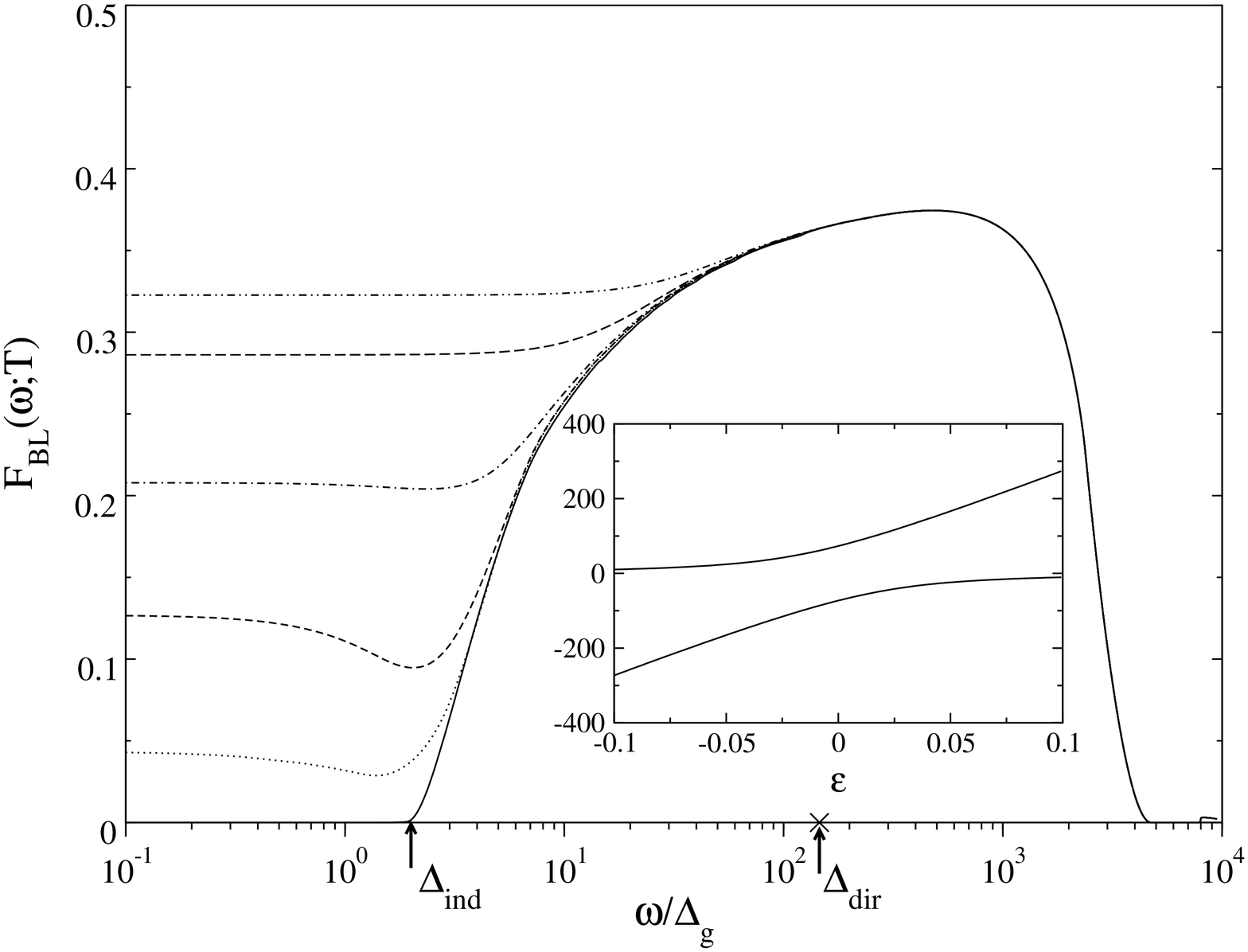}
\caption{Optical conductivities on all $\om$-scales: $F_{HCL}(\om;T)$
 (top panel) and $F_{BL}(\om;T)$ (lower panel) {\it vs} $\om/\Delta_g$
 on a log-scale, for $\tilde{T}=T/\Delta_g$=0 (solid), 0.5 (dotted),
 1 (short dash), 2 (point-dash), 5 (long dash) and 10 (double point-dash).
 The indirect and direct gap scales are also indicated. Temperatures in
 the range shown produce essentially no effect at $\om$'s on the order
 of the direct gap scale $\Delta_{dir}$. 
 Insets: renormalized  band structure $\om_\pm(\ep)/\Delta_g$ {\it vs} the
 free ($V=0$) conduction band energies $\ep/t_*$.
}
\label{fig:16B}
\end{figure}

  The optical conductivities shown in figure 14 refer to the universal
$\tilde{\omega}$-regime. To consider all $\omega$-scales, encompassing 
in particular
frequencies on the order of the direct gap and beyond, bare model 
parameters must of 
course be specified. Figure 16 provides an example, for $U/t_{*} = 6$ and 
$V^{2}/t_{*}^{2} =0.2$. For both the HCL and BL the optical conductivities
$F_{\alpha}(\omega;T)$ are shown for all $\omega$ over which
$F_{\alpha}(\omega;T)$ is non-zero, as a function of
$\tilde{\omega} = \omega/\Delta_{g}$ on a log scale from $10^{-1}$ to $10^{4}$; 
and for the same range of temperatures $\tilde{T} = T/\Delta_{g}$ (up to $10$)
employed in figure 14. The insets in each case show the renormalized band structure
$\tilde{\omega}_{\pm}(\epsilon) = \omega_{\pm}(\epsilon)/\Delta_{g}$ \it vs \rm
the free ($V=0$) conduction band energies $\epsilon/t_{*}$; with the branches
$\omega_{\pm}(\epsilon)$ obtained from solution of 
($Re\gamma(\omega) \equiv$) $\gamma_{R}(\omega) = \epsilon$. From this the
fiducial direct gap $\Delta_{dir}$ is obtained (occurring as expected for
$\epsilon =0$); it is indicated on the frequency axis, and for the particular
chosen bare parameters is seen in either case
to be $\sim 10^{2}$ times the indirect gap $\Delta_{ind}$ ($=
2\Delta_{g}$), or $\sim 0.05t_{*}$. Two features are immediately apparent in
figure 16. First, unsurprisingly, that the optical conductivity 
is large on frequency
scales on the order of the direct gap; particularly for the HCL 
where, as expected
physically and known from previous work [15,21], $F_{HCL}(\omega;T)$ is 
strongly peaked around $\omega \sim \Delta_{dir}$.

  The second point concerns the thermal evolution of the optical conductivity.
The $\tilde{T} =T/\Delta_{g}$  range shown in figure 16 corresponds to temperatures
up to 5 times the \it indirect \rm gap  $\Delta_{ind} =2\Delta_{g}$; over which
range, as shown in figure 14, the optical gap `fills in'. As seen from
figure 16 however, temperatures of this order have essentially no effect on the
optical conductivity at frequency scales on the order of the direct gap
$\Delta_{dir} \sim 10^{2}\Delta_{ind}$, which for all practical purposes retain
their $T=0$ values. This is not surprising, for one should expect the optical
conductivity on frequency scales $\omega \sim \Delta_{dir}$ to be thermally
eroded only on temperature scales of the same order.
This is indeed the case, as illustrated in figure 17 which (for the same bare
parameters) shows the thermal evolution of $F_{HCL}(\omega;T)$ up to
temperatures $\tilde{T} = 200$ ({\it i.e.\ }$T \sim 0.7\Delta_{dir}$):
significant thermal erosion on the direct gap scale sets by around
$T/\Delta_{dir} \sim 0.1-0.3$, and is well established at the highest
temperature shown in the figure. 
\begin{figure}[h]
\epsfxsize=350pt
\centering
\epsffile{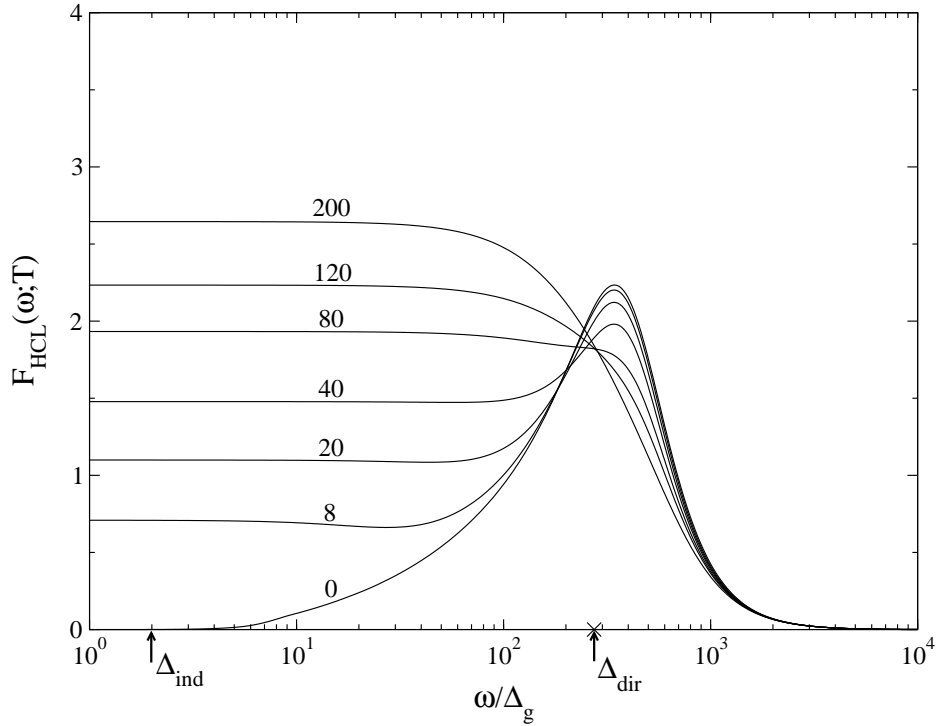}
\caption{For the same strong coupling parameters as figure 16, 
$F_{HCL}(\om;T)$ {\it vs} $\om/\Delta_g$ for a much wider range of 
temperatures $\tilde{T}=T/\Delta_g$ up to 200 ({\it i.e.\ } 
$T/\Delta_{dir}\sim {\cal{O}}(1)$); significant thermal erosion on
the direct gap $\om$-scale occurs only for temperatures of the
same order.
}
\label{fig:17}
\end{figure}

  It is clear from the above that the indirect and direct gap frequency scales,
each determined by but with their different dependences 
upon the quasiparticle weight $Z$ (equations (6.2,3)),
are qualitatively distinct in the strong coupling Kondo lattice regime where 
$Z \ll 1$; as too in consequence are the corresponding temperatures for which dynamics 
on these respective scales evolve. But this `clean' separation of scales will not
of course be evident in the $(\omega, T)$-dependences of the optical conductivity
if one is {\it e.g.\ }restricted technique-wise either to low interaction strengths (where 
$Z \sim{\cal{O}}(1)$) or to high temperatures. This is the case with Quantum
Monte Carlo, and is we believe the reason why the above scale separation is not
apparent in the QMC work of [15]. This lack of scale separation is also intrinsic 
to IPT [21], since its resultant quasiparticle weight decays algebraically rather
than exponentially in the interaction strength. Numerical renormalization group 
calculations [12,13] by contrast are not constrained in the above sense, although NRG
calculations of finite-$T$ dynamics and transport properties of the PAM have 
not to our knowledge been performed thus far.

\section{Experiment}

  Experimentally, Kondo insulators have  been widely studied 
via an impressive range of techniques (for reviews see {\it e.g.\ }[3-7]).
Here we
consider briefly three prototypical materials among those for which the most
extensive and reliable data is available, {\it viz} $Ce_{3}Bi_{4}Pt_{3}$, $SmB_{6}$
and $YbB_{12}$. Each naturally possesses features specific to itself, but the
well known commonality of behaviour between the different materials is 
of course the dominant theme [3-7]. Our aim here is simple: to compare the 
present 
theory directly to experimental results for d.c.\ transport and optical
conductivities. We also emphasise that subsequent comparison to experiment does
\it not \rm involve multi-parameter fits, or a detailed knowledge of the bare
material parameters; and in these terms is minimalist. The essential 
point of scaling, as detailed in the preceding
sections, is that in strong coupling the temperature dependence of transport/optics
is encoded in $\tilde{T} = T/\Delta_{g}$ alone, independently of the bare model 
parameters. The gap $\Delta_{g}$ ($\propto \Delta_{tr}$) may thus itself be obtained
from experiment (as below); given which, the theory then predicts the full
$T$-dependence. The same comment applies to the $\omega$-dependence of 
the optical conductivity, which in the scaling regime depends only on 
$\tilde{\omega} = \omega/\Delta_{g}$ (albeit of course that `non-universal' 
$\;\omega$'s  (\S6) are also accessed experimentally). All this is naturally
based on the assumption that the experimental materials are indeed strongly
correlated, which we take for granted unless there appear to be experimental
hints to the contrary (as we suggest below may be the case for $YbB_{12}$).

  Granted the above there is one unknown in the theory, {\it viz } 
the constant $\sigma_{0}$ entering the conductivity $\sigma(\omega;T)
= \frac{1}{3}\sigma_{0}F_{\alpha}(\omega;T)$ as an overall scale factor (see {\it e.g.\ }
equation (3.6)); and which, as noted in \S3, should realistically lie in the
range $\sim 10^{4} - 10^{5} (\Omega cm)^{-1}$.
In comparing to experiment we take
$\sigma_{0}$ as a parameter, which in practice is indeed found to lie
in the above range; although even this is not strictly necessary (knowledge
of $\sigma_{0}$ may be bypassed entirely by taking {\it e.g.\ }the room temperature d.c.\
conductivity as a reference and considering
$\sigma(\omega;T)/\sigma(0; 300K)$).
One must of course choose whether to compare experiment to theoretical results 
arising from the HCL or BL, the two host lattices we have 
considered explicitly. Here we simply make comparison to whichever of 
the two appears optimal, which in practice means the hypercubic lattice
for all but $Ce_{3}Bi_{4}Pt_{3}$; a more realistic free ($V=0$)
conduction band density of states $\rho_{0}(\epsilon)$ could readily be 
employed, but the comparisons below suggest this would be barely necessary.

\subsection{ $ \bf Ce_{3}Bi_{4}Pt_{3}.$ }

  Experimental results for the  resistivity $\rho(T)$ of this classic Kondo
insulator are shown in figure 18. These are compared to results arising
for $\rho (T)$ ($= 1/\sigma (0;T)$) from the present
theory for the BL, taking $\sigma_{0} \simeq 4.2 \times 10^{4} (\Omega cm)^{-1}$.
The transport gap (equation (5.1)) inferred experimentally is $\Delta_{tr} \simeq 35K$ [48],
from which (\S5) $\Delta_{g} = \Delta_{tr}$ follows.

  The quantitative agreement between experiment and theory is 
clear from figure 18, being excellent for $T \gtrsim 50K$ or so, as seen
in particular from the d.c.\ conductivity shown directly in the inset.
The agreement extends up to $T=300K$ or $\tilde{T} = T/\Delta_{g} \sim 10$,
an appreciable multiple of the gap where (see {\it e.g.\ }figure  10)
the high-temperature logarithmic asymptotics of the Kondo regime are being
approached; so that the temperature range shown spans essentially the full range
of expected physical behaviour. The activated insulating nature of the low-$T$
transport (equation (5.1)) is likewise evident in the figure, although transport in 
$Ce_{3}Bi_{4}Pt_{3}$ at the lowest temperatures is dominated by variable range
hopping between extrinsic states in the gap [48] which the theory does not of
course seek to include.
\begin{figure}[h]
\epsfxsize=350pt
\centering
\epsffile{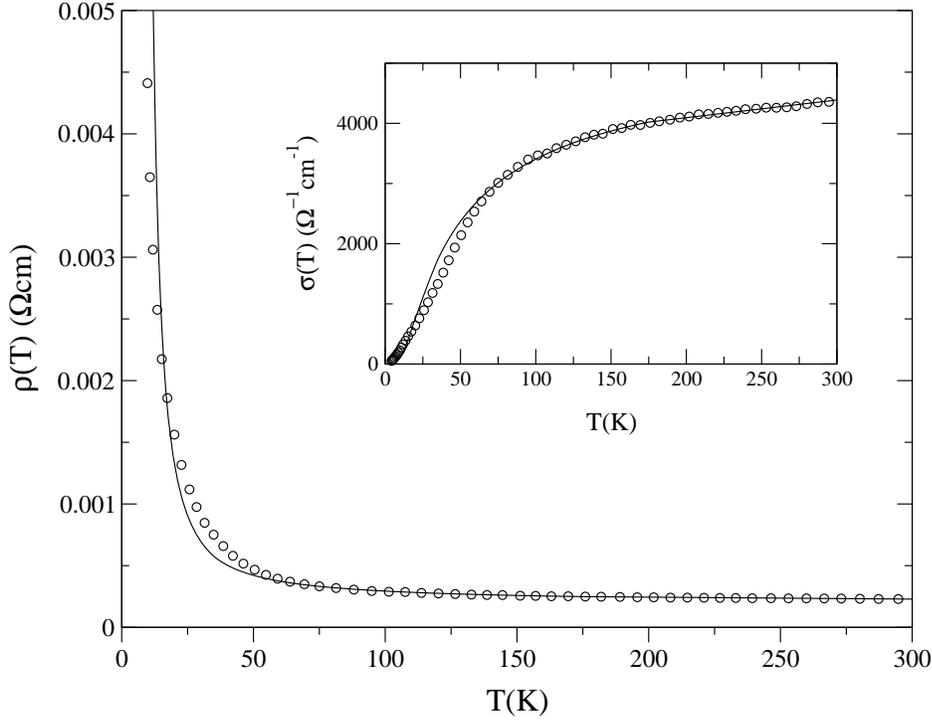}
\caption{$Ce_{3}Bi_{4}Pt_{3}$ resistivity $\rho(T)$ (in $\Omega cm$)
{\it vs} $T$ up to 300K. Experimental results [48], open circles; theory,
solid line. Inset: corresponding results for the d.c.\ conductivity
$\sigma(0;T)$.
}
\end{figure}

  As discussed in \S6, the d.c.\ transport and optics should be consistent
in the sense that the ($T=0$) gap in the optical conductivity should
correspond to the  indirect gap, $\Delta_{ind} = 2\Delta_{g}$;
and the thermal evolution of the optical conductivity should likewise be 
controlled by the indirect gap scale. Taking 
$\Delta_{ind} \simeq 70 K \simeq 50 cm^{-1}$ from the d.c.\ experiments as above, 
figure 19 (top panel) shows the theoretical optical conductivity $\sigma(\omega;T) = 
\frac{1}{3}\sigma_{0}F_{BL}(\omega;T)$ \it vs \rm $\omega$ up to $1000 cm^{-1}$, 
for temperatures $T= 0, 25, 50, 75, 100$ and $300K$ (each curve thus corresponding to
a particular `realisation' of the universal BL optical conductivities shown 
in figure 10). The lower panel in figure 19 shows corresponding experimental results [49] 
for $\omega > 50cm^{-1}$, obtained from Kramers-Kronig transformation of 
reflectivity measurements; we have also indicated the indirect gap on the
$\omega$-axis, and have marked experimental values of the d.c.\ conductivity 
[48]
on the vertical axis.
The level of agreement between experiment/theory for the optics is self-evident
from figure 19, and we regard it as rather good. The relevant temperature scale 
indeed appears to be the indirect gap [7,50]: as noted in [49], for $T$
between $100K
\sim 1.4\Delta_{ind}$ and $300K$ the optical gap is well filled in and
the optical conductivity nearly constant in the far infrared; and the gap begins to
develop markedly only below $T \sim 75K \sim \Delta_{ind}$ (which we add is
naturally consistent with thermal evolution of the single-particle dynamics, see
{\it e.g.\ }figure 4).
\begin{figure}[t]
\epsfxsize=290pt
\centering
\epsffile{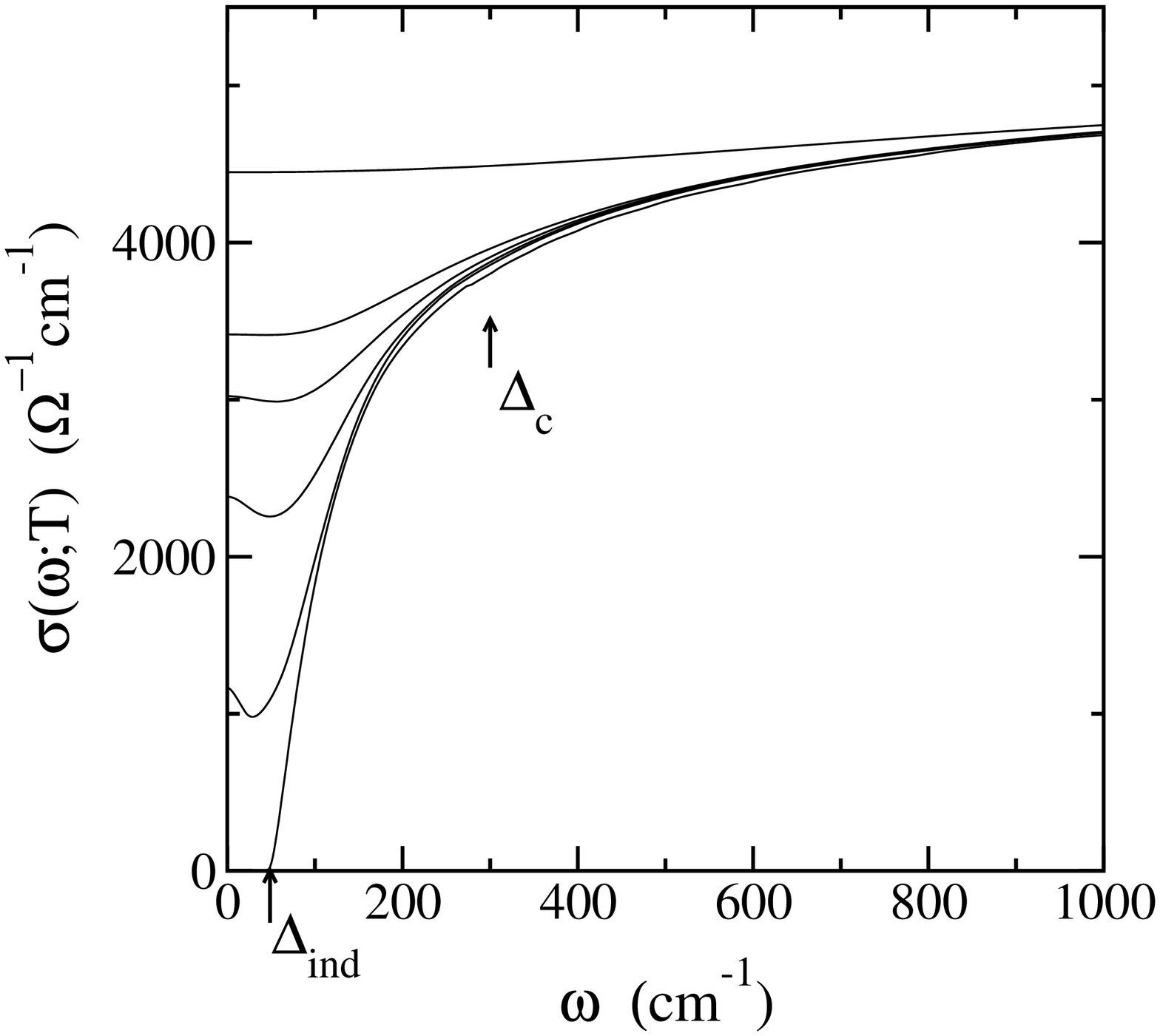}
\epsfxsize=310pt
\centering
\epsffile{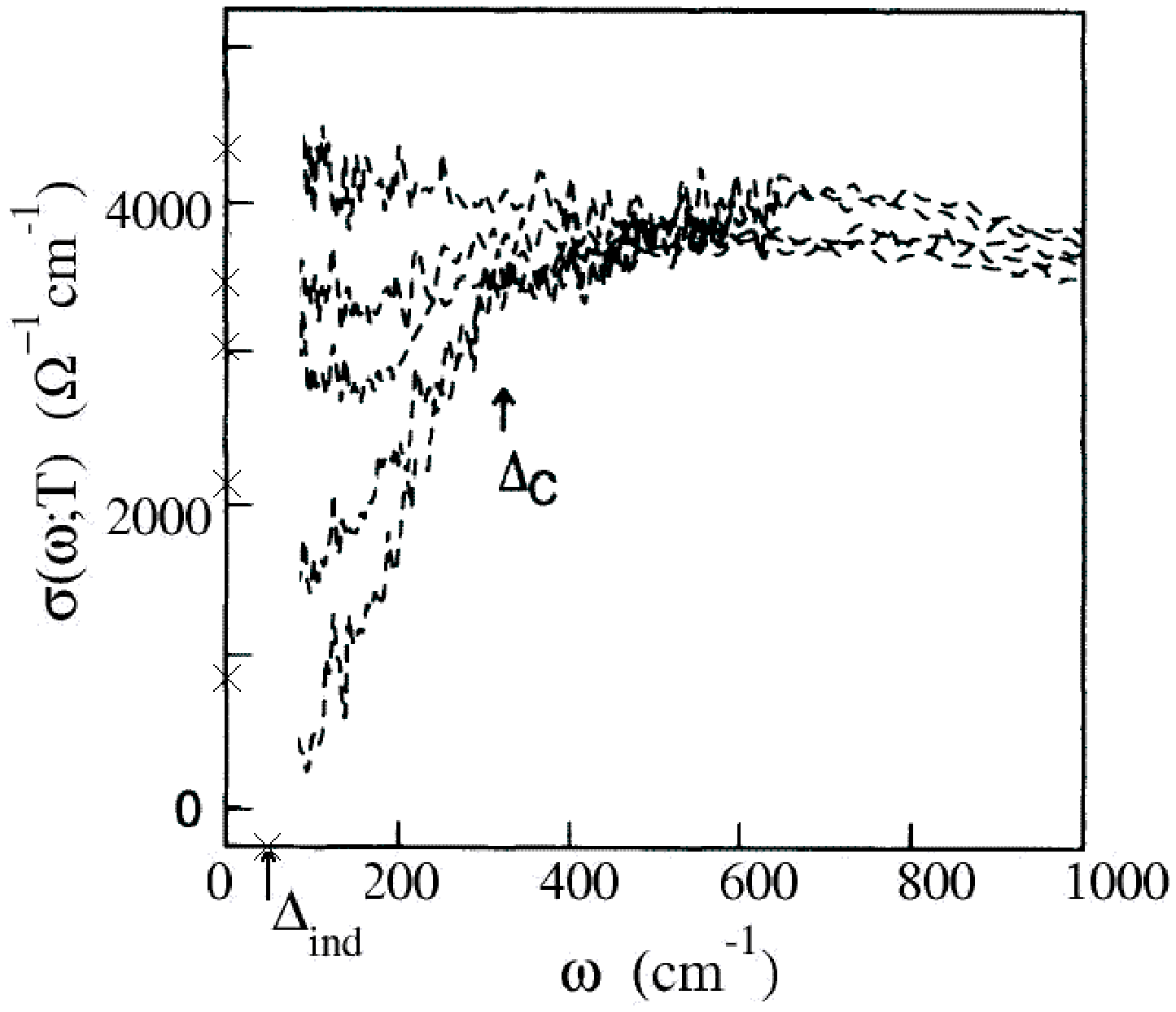}
\caption{ $Ce_{3}Bi_{4}Pt_{3}$ optical conductivity $\sigma(\om;T)$
 (in $\Omega^{-1} cm^{-1}$) {\it vs} $\om$ up to 1000 $cm^{-1}$, and
 for temperatures $T$=25,50,75,100 and 300K (in obvious sequence).
 Top panel: theory (including $T=0$); bottom panel: experimental
 results for $\om > 50 cm^{-1}$ [49], with experimental d.c.\ 
 conductivities marked by crosses on the vertical axis. 
 The optical/indirect gap 
 $\Delta_{ind}$ is indicated; for `$\Delta_c$', see text.
}
\end{figure}

  As seen from the experimental data in figure 19, the low-temperature d.c.\ 
conductivities are larger than their a.c. counterparts at the lowest
$\omega = 50 cm^{-1}$, the difference diminishing with increasing $T$ and
being barely perceptible by $T \sim 75K \simeq \Delta_{ind}$. This we believe to
be symptomatic of the low-$\omega$ Drude-like peak discussed in \S6 (figures 14,15)
that lies below the detection limit of $50cm^{-1}$ but is apparent (albeit weakly)
in the theoretical $\sigma (\omega;T)$; and further evidence for which
arises in the systems discussed below (see figures  21,23). Finally, a remark on
the optical gap itself. As noted in [49], linear extrapolation to zero of the
steep part of the experimental $\sigma (\omega; T=25K)$ would suggest
a gap value on the order of $100K$. This is certainly consistent with
$\Delta_{ind} \simeq 70K \simeq 50cm^{-1}$, inferred as above from d.c.\ transport.
On the other hand a charge gap $\Delta_{c} \simeq 300 cm^{-1}$ has been
identified in [49], for natural reasons evident in the experimental data shown
in figure 19 (and with $\Delta_{c}$ indicated on both panels in figure 19). Our point
here is simply that $\Delta_{ind}$ and $\Delta_{c}$ are fundamentally 
equivalent scales (each being proportional to the quasiparticle
weight $Z$). The former is `correct' insofar as the optical gap is strictly 
a $T=0$ notion; while the latter is natural if one wishes instead to focus on
the incipient development of a gap coming from the `high'-temperature regime 
$T \gtrsim \Delta_{ind}$.

\subsection{ $\bf SmB_{6}. $}

  Samarium hexaboride provides another long studied [51], prime example
of a Kondo insulator. Here we refer to a recent comprehensive study [52] of
both static transport and low-energy electrodynamics, performed on a high
quality single crystal sample; with the optical conductivity obtained directly via
sub-millimeter spectroscopy in the frequency range $5-36 cm^{-1}$ [52], and by 
Kramers-Kronig analysis of reflectivity in the infrared [52,53]. 
We restrict our 
considerations mainly to temperatures $T \gtrsim 8K$, since for lower 
temperatures variable range hopping again arises.  

  Experimental results for the resistivity $\rho(T)$ [52], spanning five orders of
magnitude, are shown in figure 20;
and compared to theoretical results for the hypercubic lattice (as one might
anticipate to be appropriate for a clean sample), taking $\sigma_{0} =
4.7 \times 10^{4} (\Omega cm)^{-1}$ and $\Delta_{g} = 101K$. Between 
$T=8K$ and $25K$ the experimental d.c.\ conductivity/resistivity has the activated
form equation (5.1), with a transport gap $\Delta_{tr} \simeq 3.5meV$ 
$( \simeq 41K$) [52]. As discussed in \S5, for the hypercubic lattice we find 
$\Delta_{tr} = b \Delta_{g}$ with the constant $b =0.40$; hence 
$\Delta_{g} \simeq 8.75meV \simeq 101K$ as above. The main figure shows
$\rho(T)$ on a log-scale \it vs \rm $T$; while the insets show the corresponding
d.c.\ conductivity \it vs \rm both $1/T$ (to exemplify in particular the activated
regime) and temperature on a log-scale. The agreement between theory/experiment
for $T \gtrsim 8K$ is rather striking; for $T \lesssim 8K$ the experimental variable
range hopping behaviour is of course evident from the right inset to the figure.
\begin{figure}[h]
\epsfxsize=350pt
\centering
\epsffile{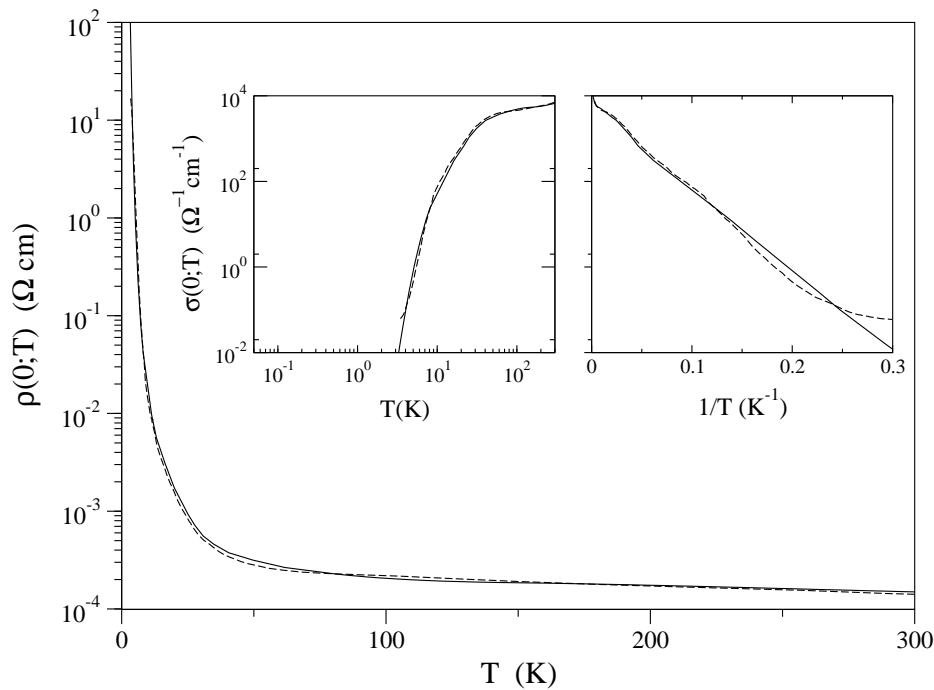}
\caption{$SmB_6$ resistivity $\rho(T)$ (in $\Omega cm$) {\it vs} $T$
up to 300K. Experimental results [52], dashed line; theory, solid line.
Left inset: corresponding d.c.\ conductivity $\sigma(0;T)$ 
{\it vs} $T$ on a log-log scale. Right inset: $\sigma(0;T)$ on a log scale
{\it vs} $1/T$; for $T \lesssim 8K$ variable range hopping arises
experimentally.}
\end{figure}

  The consequent optical conductivity $\sigma (\omega;T)$ for the HCL
is shown \it vs \rm $\omega$ in figure 21 (top panel) on a log-log scale, for
temperatures $T= 3,13,16,18$ and $300K$. Corresponding experimental results for
the same temperatures [52] are shown in the lower panel, including the extrapolated
conductivities (dashed lines) obtained from the phenomenological fit to the data
employed in [52].  The $\omega$-range
shown, up to $\sim 3 \times 10^{4} cm^{-1}$, naturally encompasses non-universal
scales at high-frequencies, and the calculations were performed specifically
for $U/t_{*} = 4.5$ and $V^{2}/t_{*}^{2} =0.2$. The system is however strongly
correlated for these parameters, so we emphasise that the resultant 
`low'-$\omega$ conductivity (up to $\sim 2000 cm^{-1}$ or so in practice) lies
in the $\tilde{\omega} = \omega/\Delta_{g}$ scaling regime that is actually
independent of the bare parameters (\S 's 4-6): the choice of bare parameters
`matters' only at the high frequency end, and we have simply chosen those above
as illustrative. We would however add that for the chosen parameters the quasiparticle
weight $Z \sim 5 \times 10^{-3}$, which is in qualitative accord with the 
experimentally inferred effective mass ($1/Z \sim$) $m_{*}/m_{0} \sim 10^{2}$
(from [52]) and $\sim 10^{3}$ from a previous study [54] (and which values themselves
attest to the correlated nature of $SmB_{6}$).
\begin{figure}[t]
\epsfxsize=295pt
\centering
\epsffile{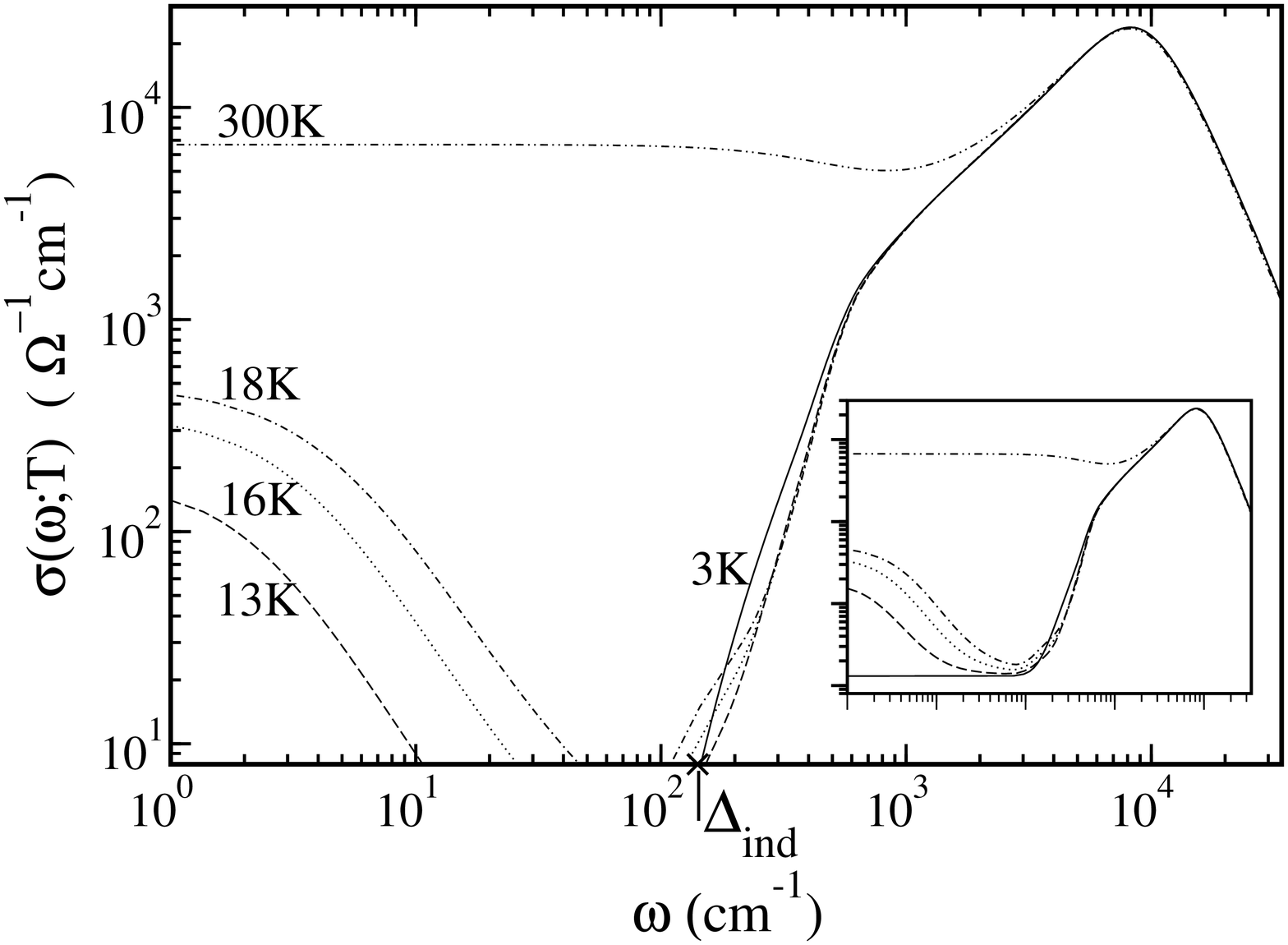}
\epsfxsize=320pt
\centering
\epsffile{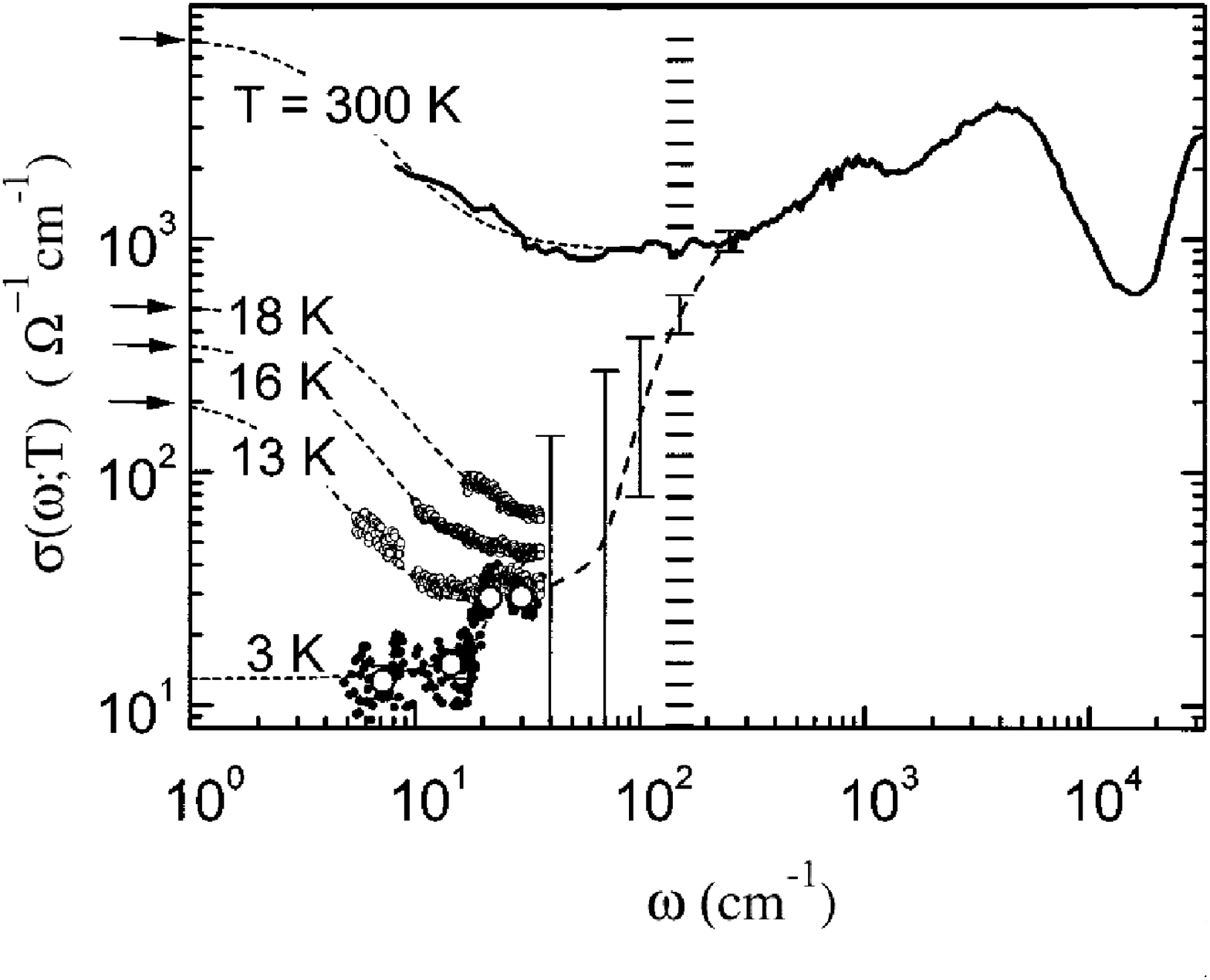}
\caption{$SmB_6$ optical conductivity $\sigma(\om;T)$ (in $\Omega^{-1}
cm^{-1}$) {\it vs} $\omega$ on a log-log scale, for temperatures
$T=$ 3,13,16,18 and 300K. Top panel: theory (with the theoretical
optical/indirect gap indicated). Inset: corresponding results when
a constant $12 \Omega^{-1}cm^{-1}$ is added. Bottom panel: experimental
results [52]. Circles are from sub-mm data [52], solid lines from
reflectivity spectrum via Kramers-Kronig (KK) [53]. Error
bars refer to IR conductivity obtained from KK analysis of the
$T=3$K reflectivity spectrum assuming 0.5\% uncertainty [52]. The shaded
area corresponds to the experimental optical gap [52]. The short
dash lines show the extrapolated fit used in [52], and the arrows indicate 
the experimental d.c.\ conductivity.}
\end{figure}

  The first point to note here is that the theoretical optical gap 
$\Delta_{ind} = 2\Delta_{g}$, indicated on figure 21, is 
$\Delta_{ind}  \simeq 17.5meV \simeq 200K$ (from $\Delta_{g}$ obtained as above).
This accords remarkably well with the gap of $19 \pm 2 meV$ inferred experimentally
from the optical conductivity [52], and we emphasise that there is no \it a
priori \rm connection between these two ways of obtaining the optical gap
--- as above, the theoretical optical gap is inferred directly from the much 
smaller transport gap $\Delta_{tr}$. There is in consequence no conundrum
between a transport gap of $\simeq 40K$ and an optical gap of $\simeq 200K$. 
The level of agreement between theory/experiment is self-evident in figure 21, and
encouraging both in terms of its $\omega$-dependence and
thermal evolution. One small twist may also be added. For the phenomenological fitting
employed in [52], it was found that for a complete description of the
conductivity spectra an additional parameter (termed `$\sigma_{min}$') had to
be introduced in the form of an additive, frequency-independent contribution to 
the optical  conductivity; with  $\sigma_{min} \sim 12 (\Omega cm)^{-1}$ in the 
low-temperature regime [52].  Without wishing to
speculate here on the origin of the $\sigma_{min}$, the inset to figure 21 
(top panel) shows the effect of simply adding a constant $12 (\Omega cm)^{-1}$ 
contribution to our theoretical optical conductivity. For low-temperatures its 
effect is  noticeable in the $\sim 10-100 cm^{-1}$ range, particularly
at the lowest $T=3K$; and the improved agreement with experiment is clear.

  Finally we mention that we have taken no consideration above of the presence,
in the Kondo insulating gap, of an additional narrow donor-type band which
is known to occur experimentally [52] in $SmB_{6}$ (and may possibly be due to 
impurities). While certainly of interest in itself, the present theory has of 
course nothing to say about it; and it plays no role in the extent to which, 
as above, theory concurs with experiment.

\subsection{$\bf YbB_{12}. $}

  Resistivity [55] and optical measurements [56] have likewise been performed
on single crystal $YbB_{12}$, the only known $Yb$-based Kondo insulator.
Experimental results for $\rho(T)$ up to $\sim 350K$ are shown in figure 22,
and for $15 K < T < 40K$ exhibit activated behaviour (equation (5.1)) with a transport
gap $\Delta_{tr} \simeq 68 K$. Optical conductivity results [56], again obtained via 
Kramers-Kronig from reflectivity spectra combined with a Hagen-Rubens
extrapolation at low-$\omega$, are shown in the top panel of figure 23. The 
experimental optical gap is determined as $\Delta_{ind} 
\simeq 25meV = 290K$ [56];
while the strong IR peak around $\sim 0.2-0.25 eV$ as naturally 
interpretable [7] in terms of direct gap excitations
(see {\it e.g.\ }figures 16,17 for the HCL).
\begin{figure}[ht]
\epsfxsize=350pt
\centering
\epsffile{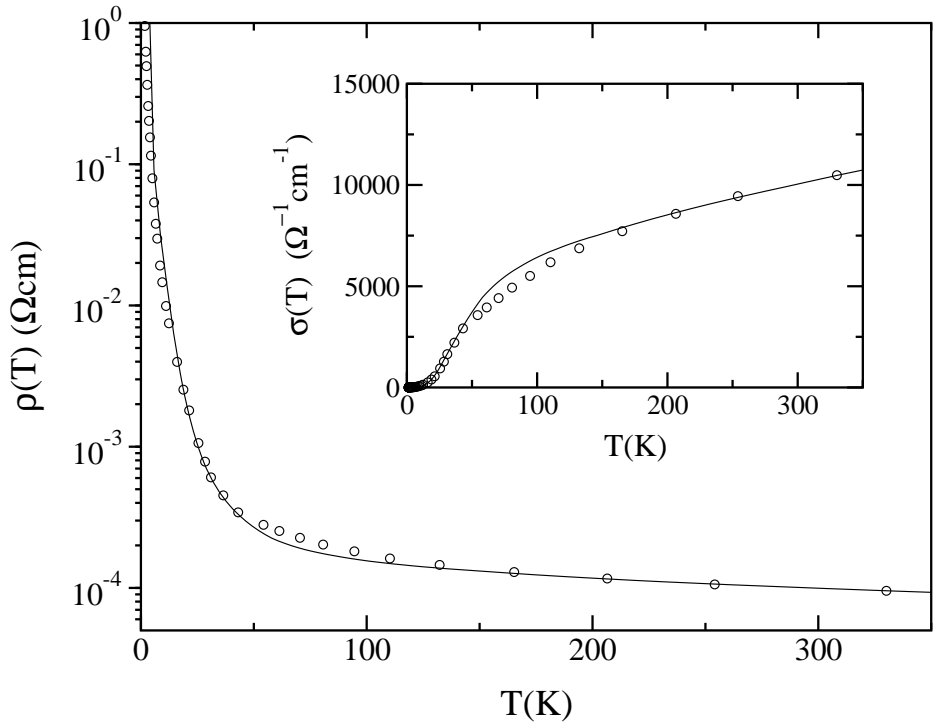}
\caption{$YbB_{12}$ resistivity $\rho(T)$ (in $\Omega cm$) {\it vs} $T$.
Experimental results [55], open circles; theory, solid line (for parameters
specified in text). Inset: corresponding d.c.\ conductivities {\it vs} 
$T$.}
\end{figure}

  It is the temperature dependence of the experimental optical data which suggests
to us that $YbB_{12}$ may be in an intermediate-weak coupling regime. As seen from
figure 23 [56], increasing temperatures up to $T = 290K \equiv \Delta_{ind}$ leads to
significant redistribution of spectral weight at much higher energy scales on
the order of the direct gap and beyond. This is not behaviour characteristic of 
strong coupling, as evident from the discussion of \S 6 (figures 16,17). It is
however typical of intermediate-weak coupling interactions, theoretical 
consideration of which thus requires specification of bare model parameters.
In the following we consider specifically $U/t_{*} = 1.65$ and
$V^{2}/t_{*}^{2} =0.2$: these values should not of course be taken too
seriously {\it per se}, but they lead within our approach to behaviour 
representative of intermediate-weak coupling and should be viewed simply
as such.
\begin{figure}[t]
\epsfxsize=300pt
\centering
\epsffile{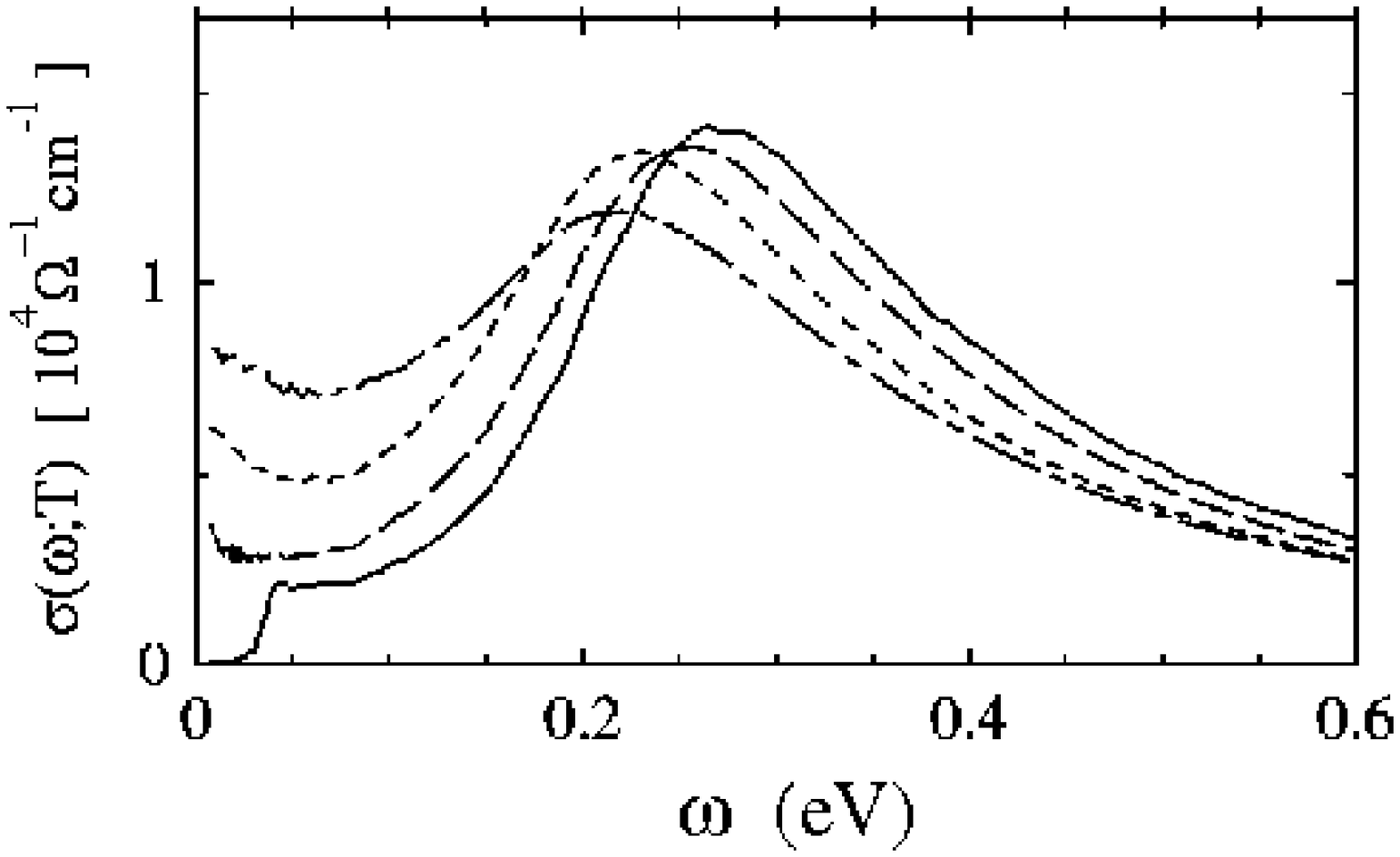}
\epsfxsize=300pt
\centering
\epsffile{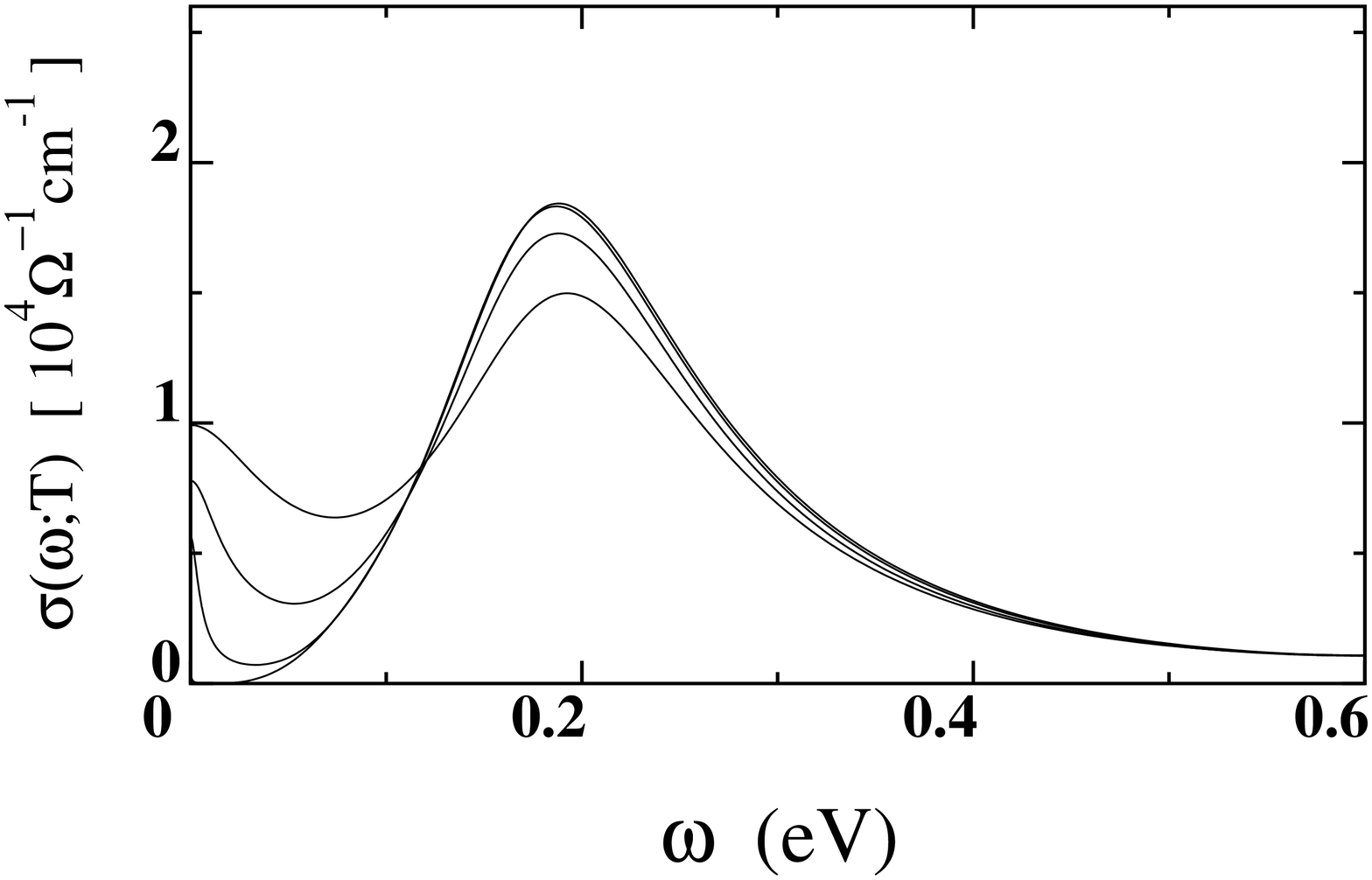}
\caption{$YbB_{12}$ optical conductivity $\sigma(\om;T)$ (in $\Omega^{-1}
cm^{-1}$) {\it vs} $\om$ (in $eV$), for $T$=20,78,160 and 290K (in obvious
sequence). Top panel: experimental results [56]. Bottom panel: theory
 (for same parameters as figure 22).}
\end{figure}

  We find with the latter that the low-temperature transport has as expected 
the activated form equation (5.1), with $\Delta_{tr} = b\Delta_{g}$
and the constant $b = 0.47$ (as opposed to $b=0.40$ for the HCL in
strong coupling).  Equating the theoretical $\Delta_{tr}$ with the experimental
$\Delta_{tr} \simeq 68 K$, and taking $\sigma_{0} \simeq 10^{5} (\Omega
cm)^{-1}$, direct comparison between the $T$-dependence of the theoretical and 
experimental $\rho (T)$'s is shown in figure 22; and the agreement with experiment 
is seen to be rather good across essentially the full temperature range.
  Does this lead to a consistent description of the optical conductivity? The
answer is yes in two senses. First, the theoretical estimate of the optical
gap $\Delta_{ind} =2\Delta_{g} = (2/b)\Delta_{tr}$ follows directly as
$\Delta_{ind} \simeq 290K = 25meV$ using only the experimental transport gap
above; which value coincides with $\Delta_{ind}$ obtained directly from the
optical experiments (a situation analogous to that for $SmB_{6}$ discussed
above). Second, the resultant theoretical optical conductivity is shown in the
bottom panel of figure 23. It is seen to accord well qualitatively with the experimental
results (top panel), both in terms of its overall $\omega$-dependence and thermal
evolution; including the redistribution of spectral weight on energy
scales beyond the direct gap, for temperatures up to $T =290K$ that merely
corresponds to the indirect gap itself. (Improved agreement with experiment
could no doubt be obtained by playing with the bare parameters, but would 
add little new.)

\section{Summary.}

  We have developed in this paper a non-perturbative local moment approach
to dynamics and transport properties of the symmetric periodic Anderson model,
the basic microscopic model for understanding small-gap Kondo insulator materials
[3-7]. Our primary focus has naturally been the strong coupling, Kondo lattice
regime. Here the system is characterised by the low-energy scale $\Delta_{g}$
which, being exponentially small in strong coupling, leads to a pristine separation
between low- and high-energy scales; and hence to `universal scaling' of
dynamics/transport in terms of $\tilde{\omega} = \omega/\Delta_{g}$ and
$\tilde{T} = T/\Delta_{g}$ alone, with no explicit dependence on bare model
parameters. It is this single, indirect gap scale $\Delta_{g}$ that is of paramount
importance in controlling the physical properties of the system that we have 
investigated systematically here; for it determines 
the single-particle spectral gap, the transport gap for d.c.\ conductivity and the 
optical gap in the dynamical conductivity, all of which are simply proportional to 
each other. It sets the scale for thermal evolution of single-particle dynamics 
and d.c. transport, from the gapped/activated behaviour symptomatic of the 
low-temperature insulator through to the incoherent single-impurity physics that 
is found to arise naturally for $\tilde{T} \gg 1$. And likewise it is $\Delta_{g}$ 
that sets the thermal scale for `filling' the optical gap with increasing 
temperature; the much higher direct gap scale naturally also being apparent in the
$\omega$-dependence of the optical conductivity, but in strong coupling being essentially
irrelevant as a thermal scale for its evolution.

Notwithstanding the innate simplicity of the PAM itself, and the range of
material-specific factors it naturally omits, the present theory also
appears to account well for experiments on materials such
as $Ce_{3}Bi_{4}Pt_{3}$, $SmB_{6}$ and $YbB_{12}$; with many characteristic
features arising theoretically apparent in experiment, and a mutually consistent
picture of d.c.\ transport and optics arising. We believe this lends further
support to the essential veracity of both the underlying model (the PAM within DMFT)
as well as the present theory, the local moment approach. Further development
of the LMA to encompass the asymmetric PAM, and hence heavy fermion metals, will
be reported in subsequent work.

\ack

We are most grateful to Z. Fisk, A. Loidl and
T. Takabatake for permission to use their
experimental results.
  It is a pleasure to acknowledge support from the EPSRC
and Leverhulme Trust, as well as the Royal Society and 
Indian National Science Academy.

\appendix

\section*{Appendix.}
  Here we sketch the steps leading to equation (5.8) for the leading high-$\tilde{T}$
behaviour of $\rho^{'}_{BL}(T)$ for the Bethe lattice. Using equation (3.7b) for
$\omega =0$, together with $<D_{c}(\epsilon;\omega)>_{\epsilon} = D^{c}(\omega)$
(the $c$-electron spectrum), $F_{BL}(0;T)$ is given generally by
\begin{displaymath}
F_{BL}(0;T) = t_{*}^{2}\int^{\infty}_{-\infty} d\omega
-\frac{\partial f(\omega;T)}{\partial \omega} ~ [D^{c}(\omega)]^{2}.
\hspace{4.6cm} (1)
\end{displaymath}
For the BL, the Feenberg self-energy $S(\omega) = \frac{1}{4}t_{*}^{2}G^{c}(\omega)$;
so equation (2.5a) implies
\begin{displaymath}
G^{c}(\omega) = [\gamma (\omega) - \case{1}{4} t_{*}^{2}G^{c}(\omega)]^{-1}
\hspace{7.35cm} (2)
\end{displaymath}
which determines the $\gamma$-dependence
of $G^{c} \equiv G^{c}[\gamma]$. For $\tilde{T} \gg 1$ in the scaling regime,
$|\gamma| \ll t_{*}$; which corresponds physically to scattering rates
$(\tilde{\gamma}_{I}(\omega) = \pi\rho_{0}\gamma_{I}(\omega) \equiv)$ 
$ \tilde{\tau}^{-1}(\omega;T) \ll 1$ 
for (all) finite $|\tilde{\omega}|$. Using equation (2), the leading asymptotics of
$D^{c}(\omega) = -\frac{1}{\pi} ImG^{c}(\omega)$ is given for 
$|\tilde{\gamma}| \ll 1$ by
\begin{displaymath}
D^{c}(\omega) \sim \rho_{0}[1 -\case{\pi}{2}\rho_{0}\gamma_{I}(\omega) 
+{\cal{O}}(\tilde{\gamma}_{I}^{2})] =
 \rho_{0}[1- \case{1}{2}\tilde{\tau}^{-1}(\omega;T)
+ {\cal{O}}(\tilde{\tau}^{-2})]
\hspace{0.7cm} (3)
\end{displaymath}
whence from equation (1): 
\begin{displaymath}
F_{BL}(0;T) \sim [\rho_{0}t_{*}]^{2} \{1 +
\int^{\infty}_{-\infty} d\omega ~ \frac{\partial f(\omega;T)}{\partial \omega} ~
\tilde{\tau}^{-1}(\omega;T)\}
\hspace{3.0cm} (4)
\end{displaymath}
Equations (5.6,7) for $\tilde{\tau}^{-1}(\omega;T)$ then lead to
\begin{displaymath}
F_{BL}(0;T) \sim [\rho_{0}t_{*}]^{2} \{1 + \frac{3\pi^{2}}{16ln^{2}(\tilde{T})}
\int^{\infty}_{-\infty} dy ~ \frac{\partial f(y)}{\partial y}
\frac{1}{L(y;\tilde{T})} \}
\hspace{2.35cm} (5)
\end{displaymath}
where $f(y) = [e^{y}+1]^{-1}$; and using $L(y;\tilde{T}) \rightarrow 1$ as
$\tilde{T} \rightarrow \infty$ (\S5) gives
$F_{BL}(0;T) \sim [\rho_{0}t_{*}]^{2} \{1 - \frac{3\pi^{2}}{16ln^{2}(\tilde{T})} \}$.
Using this in equation (5.4) for $\rho^{'}_{BL}(T)$ (together with $[\rho_{0}t_{*}]^{2} 
= \frac{4}{\pi^{2}}$), gives directly the leading large-$\tilde{T}$ behaviour
\begin{displaymath}
\rho^{'}_{BL}(T) \stackrel{\tilde{T} \gg 1}{\sim} \frac{3\pi^{2}}{16ln^{2}(\tilde{T})}
\hspace{8.7cm} (6)
\end{displaymath}
as sought.

\end{document}